\documentclass[12pt]{article}
\usepackage{epsfig}
\usepackage{amsmath}
\usepackage{hhline}
\usepackage{amssymb}
\usepackage{times}
\usepackage{cite}
\usepackage{rotate}

\newlength{\dinwidth}
\newlength{\dinmargin}
\begin{document}  
\newcommand{\pom}{{I\!\!P}}
\newcommand{\reg}{{I\!\!R}}
\newcommand{\slowpi}{\pi_{\mathit{slow}}}
\newcommand{\fiidiii}{F_2^{D(3)}}
\newcommand{\fiidiiiarg}{\fiidiii\,(\beta,\,Q^2,\,x)}
\newcommand{\n}{1.19\pm 0.06 (stat.) \pm0.07 (syst.)}
\newcommand{\nz}{1.30\pm 0.08 (stat.)^{+0.08}_{-0.14} (syst.)}
\newcommand{\fiidiiiful}{F_2^{D(4)}\,(\beta,\,Q^2,\,x,\,t)}
\newcommand{\fiipom}{\tilde F_2^D}
\newcommand{\ALPHA}{1.10\pm0.03 (stat.) \pm0.04 (syst.)}
\newcommand{\ALPHAZ}{1.15\pm0.04 (stat.)^{+0.04}_{-0.07} (syst.)}
\newcommand{\fiipomarg}{\fiipom\,(\beta,\,Q^2)}
\newcommand{\pomflux}{f_{\pom / p}}
\newcommand{\nxpom}{1.19\pm 0.06 (stat.) \pm0.07 (syst.)}
\newcommand {\gapprox}
   {\raisebox{-0.7ex}{$\stackrel {\textstyle>}{\sim}$}}
\newcommand {\lapprox}
   {\raisebox{-0.7ex}{$\stackrel {\textstyle<}{\sim}$}}
\def\gsim{\,\lower.25ex\hbox{$\scriptstyle\sim$}\kern-1.30ex%
\raise 0.55ex\hbox{$\scriptstyle >$}\,}
\def\lsim{\,\lower.25ex\hbox{$\scriptstyle\sim$}\kern-1.30ex%
\raise 0.55ex\hbox{$\scriptstyle <$}\,}
\newcommand{\pomfluxarg}{f_{\pom / p}\,(x_\pom)}
\newcommand{\dsf}{\mbox{$F_2^{D(3)}$}}
\newcommand{\dsfva}{\mbox{$F_2^{D(3)}(\beta,Q^2,x_{I\!\!P})$}}
\newcommand{\dsfvb}{\mbox{$F_2^{D(3)}(\beta,Q^2,x)$}}
\newcommand{\dsfpom}{$F_2^{I\!\!P}$}
\newcommand{\gap}{\stackrel{>}{\sim}}
\newcommand{\lap}{\stackrel{<}{\sim}}
\newcommand{\fem}{$F_2^{em}$}
\newcommand{\tsnmp}{$\tilde{\sigma}_{NC}(e^{\mp})$}
\newcommand{\tsnm}{$\tilde{\sigma}_{NC}(e^-)$}
\newcommand{\tsnp}{$\tilde{\sigma}_{NC}(e^+)$}
\newcommand{\st}{$\star$}
\newcommand{\sst}{$\star \star$}
\newcommand{\ssst}{$\star \star \star$}
\newcommand{\sssst}{$\star \star \star \star$}
\newcommand{\tw}{\theta_W}
\newcommand{\sw}{\sin{\theta_W}}
\newcommand{\cw}{\cos{\theta_W}}
\newcommand{\sww}{\sin^2{\theta_W}}
\newcommand{\cww}{\cos^2{\theta_W}}
\newcommand{\trm}{m_{\perp}}
\newcommand{\trp}{p_{\perp}}
\newcommand{\trmm}{m_{\perp}^2}
\newcommand{\trpp}{p_{\perp}^2}
\newcommand{\alp}{\alpha_s}

\newcommand{\alps}{\alpha_s}
\newcommand{\sqrts}{$\sqrt{s}$}
\newcommand{\LO}{$O(\alpha_s^0)$}
\newcommand{\Oa}{$O(\alpha_s)$}
\newcommand{\Oaa}{$O(\alpha_s^2)$}
\newcommand{\PT}{p_{\perp}}
\newcommand{\JPSI}{J/\psi}
\newcommand{\sh}{\hat{s}}
\newcommand{\uh}{\hat{u}}
\newcommand{\MP}{m_{J/\psi}}
\newcommand{\PO}{I\!\!P}
\newcommand{\xbj}{x}
\newcommand{\xpom}{x_{\PO}}
\newcommand{\ttbs}{\char'134}
\newcommand{\xpomlo}{3\times10^{-4}}  
\newcommand{\xpomup}{0.05}  
\newcommand{\dgr}{^\circ}
\newcommand{\pbarnt}{\,\mbox{{\rm pb$^{-1}$}}}
\newcommand{\gev}{\,\mbox{GeV}}
\newcommand{\WBoson}{\mbox{$W$}}
\newcommand{\fbarn}{\,\mbox{{\rm fb}}}
\newcommand{\fbarnt}{\,\mbox{{\rm fb$^{-1}$}}}

\newcommand{\vtab}{\rule[-1mm]{0mm}{4mm}}
\newcommand{\htab}{\rule[-1mm]{0mm}{6mm}}
\newcommand{\photoproduction}{$\gamma p$}
\newcommand{\ptmiss}{$P_{T}^{\rm miss}$}
\newcommand{\epz} {$E{\rm-}p_z$}
\newcommand{\vap}{$V_{ap}/V_p$}
\newcommand{\Zero}   {\mbox{$Z^{\circ}$}}
\newcommand{\Ftwo}   {\mbox{$\tilde{F}_2$}}
\newcommand{\Ftwoz}   {\mbox{$\tilde{F}_{2,3}$}}
\newcommand{\Fz}   {\mbox{$\tilde{F}_3$}}
\newcommand{\FL}   {\mbox{$\tilde{F}_{_{L}}$}}
\newcommand{\wtwogen} {W_2}
\newcommand{\wlgen} {W_L}
\newcommand{\xwthreegen} {xW_3}
\newcommand{\Wtwo}   {\mbox{$W_2$}}
\newcommand{\Wz}   {\mbox{$W_3$}}
\newcommand{\WL}   {\mbox{$W_{_{L}}$}}
\newcommand{\Fem}  {\mbox{$F_2$}}
\newcommand{\Fgam}  {\mbox{$F_2^{\gamma}$}}
\newcommand{\Fint} {\mbox{$F_2^{\gamma Z}$}}
\newcommand{\Fwk}  {\mbox{$F_2^{Z}$}}
\newcommand{\Ftwos} {\mbox{$F_2^{\gamma Z, Z}$}}
\newcommand{\Fzz} {\mbox{$F_3^{\gamma Z, Z}$}}
\newcommand{\Fintz} {\mbox{$F_{2,3}^{\gamma Z}$}}
\newcommand{\Fwkz}  {\mbox{$F_{2,3}^{Z}$}}
\newcommand{\Fzint} {\mbox{$F_3^{\gamma Z}$}}
\newcommand{\Fzwk}  {\mbox{$F_3^{Z}$}}
\newcommand{\Gev}  {\mbox{${\rm GeV}$}}
\newcommand{\Gevv}{\mbox{${\rm GeV}^2$}}
\newcommand{\QQ}  {\mbox{${Q^2}$}}
\newcommand{\gv}{GeV$^2\,$}
\newcommand{\bs}{\overline{s}}
\newcommand{\bc}{\overline{c}}
\newcommand{\bu}{\overline{u}}
\newcommand{\bb}{\overline{b}}
\newcommand{\bU}{\overline{U}}
\newcommand{\bD}{\overline{D}}
\newcommand{\bd}{\overline{d}}
\newcommand{\bq}{\overline{q}}
\newcommand{\FLc}{$ F_{L}\,$} 
\newcommand{\xg}{$xg(x,Q^2)\,$}
\newcommand{\xgc}{$xg\,$}
\newcommand{\ipb}{pb$^{-1}\,$}               
\newcommand{\TOSS}{x_{{i}/{\PO}}}                                              
\newcommand{\un}[1]{\mbox{\rm #1}}
\newcommand{\pdsi}{$(\partial \sigma_r / \partial \ln y)_{Q^2}\,$}
\newcommand{\pdff}{$(\partial F_{2} / \partial \ln  Q^{2})_x\,$ }
\newcommand{\Fc}{$ F_{2}~$}
\newcommand{\amz}{$\alpha_s(M_Z^2)\,$} 

%
%
\newcommand{\qsq}{\ensuremath{Q^2} }
\newcommand{\gevsq}{\ensuremath{\mathrm{GeV}^2} }
\newcommand{\et}{\ensuremath{E_t^*} }
\newcommand{\rap}{\ensuremath{\eta^*} }
\newcommand{\gp}{\ensuremath{\gamma^*}p }
\newcommand{\dsiget}{\ensuremath{{\rm d}\sigma_{ep}/{\rm d}E_t^*} }
\newcommand{\dsigrap}{\ensuremath{{\rm d}\sigma_{ep}/{\rm d}\eta^*} }
\def\Journal#1#2#3#4{{#1} {\bf #2} (#3) #4}
\def\NCA{\em Nuovo Cimento}
\def\NIM{\em Nucl. Instrum. Methods}
\def\NIMA{{\em Nucl. Instrum. Methods} {\bf A}}
\def\NPB{{\em Nucl. Phys.}   {\bf B}}
\def\PLB{{\em Phys. Lett.}   {\bf B}}
\def\PRL{\em Phys. Rev. Lett.}
\def\PRD{{\em Phys. Rev.}    {\bf D}}
\def\ZPC{{\em Z. Phys.}      {\bf C}}
\def\EJC{{\em Eur. Phys. J.} {\bf C}}
\def\CPC{\em Comp. Phys. Commun.}

\begin{titlepage}

\noindent
DESY 03-038 \hfill ISSN 0418-9833\\
March 2003 \hfill

\vspace{2cm}

\begin{center}
\begin{Large}

{\bf\boldmath Measurement and QCD Analysis of Neutral and Charged \\
Current Cross Sections at HERA}
\vspace{2cm}

H1 Collaboration

\end{Large}
\end{center}

\vspace{2cm}

\begin{abstract}
\noindent
The inclusive $e^+ p$ single and double differential cross sections for
neutral and charged current processes are measured with the H1 detector
at HERA. The data were taken in $1999$ and $2000$ at a centre-of-mass
energy of $\sqrt{s} = 319{\rm\,GeV}$ and correspond to an integrated
luminosity of $65.2$\,pb$^{-1}$.  The cross sections are measured
in the range of four-momentum transfer squared $Q^2$ between $100$ and
$30\,000{\rm\,GeV}^2$ and Bjorken $x$ between $0.0013$ and $0.65$. 
The neutral current analysis for the new $e^+p$ data and the earlier 
$e^-p$ data taken in $1998$ and $1999$ is extended to small energies of 
the scattered electron and therefore to higher values of inelasticity $y$,
allowing a determination of the longitudinal structure function $F_L$ 
at high $Q^2$ ($110-700{\rm\,GeV}^2$).
A new measurement of the structure function $x\tilde{F}_3$ is obtained
using the new $e^+p$ and previously published $e^\pm p$ neutral current 
cross section data at high $Q^2$.
These data together with H1 low $Q^2$ precision data are further used to 
perform new next-to-leading order QCD analyses in the framework of 
the Standard Model to extract flavour separated parton 
distributions in the proton.
\end{abstract}

\vspace{1.5cm}

\begin{center}
Submitted to Eur. Phys. J. C.
\end{center}

\end{titlepage}

\begin{flushleft}

C.~Adloff$^{33}$,              
V.~Andreev$^{24}$,             
B.~Andrieu$^{27}$,             
T.~Anthonis$^{4}$,             
A.~Astvatsatourov$^{35}$,      
A.~Babaev$^{23}$,              
J.~B\"ahr$^{35}$,              
P.~Baranov$^{24}$,             
E.~Barrelet$^{28}$,            
W.~Bartel$^{10}$
% ...

\bigskip{\it
 $ ^{1}$ I. Physikalisches Institut der RWTH, Aachen, Germany$^{ a}$ \\
% ...
$ ^{4}$ Inter-University Institute for High Energies ULB-VUB, Brussels;
          Universiteit Antwerpen (UIA), Antwerpen; Belgium$^{ c}$ \\
% ...
$ ^{10}$ DESY, Hamburg, Germany \\
% ...
$ ^{23}$ Institute for Theoretical and Experimental Physics,
          Moscow, Russia$^{ l}$ \\
 $ ^{24}$ Lebedev Physical Institute, Moscow, Russia$^{ e}$ \\
% ...
$ ^{27}$ LPNHE, Ecole Polytechnique, IN2P3-CNRS, Palaiseau, France \\
 $ ^{28}$ LPNHE, Universit\'{e}s Paris VI and VII, IN2P3-CNRS,
          Paris, France \\
% ...
$ ^{33}$ Fachbereich Physik, Bergische Universit\"at Gesamthochschule
          Wuppertal, Wuppertal, Germany \\
% ...
$ ^{35}$ DESY, Zeuthen, Germany \\
% ...
$ ^a$ Supported by the Bundesministerium f\"ur Bildung und Forschung, FRG,
      under contract numbers 05 H1 1GUA /1, 05 H1 1PAA /1, 05 H1 1PAB /9,
      05 H1 1PEA /6, 05 H1 1VHA /7 and 05 H1 1VHB /5 \\
% ...
$ ^c$ Supported by FNRS-FWO-Vlaanderen, IISN-IIKW and IWT \\
% ...
$ ^e$ Supported by the Deutsche Forschungsgemeinschaft \\
% ...
$ ^l$ Partially Supported by Russian Foundation
      for Basic Research, grant    no. 00-15-96584 \\
}

\end{flushleft}

\newpage
\section{Introduction}
\label{intro}
In $1992$ the HERA accelerator began operation, colliding lepton and proton
beams within the H1 and ZEUS experiments.
The phase space covered by HERA inclusive deep inelastic scattering (DIS)
cross section measurements ranges from small Bjorken $x$ at low $Q^2$,
the four-momentum transfer squared, to large $x$ at $Q^2$ values larger 
than the squared masses of the $W$ and $Z$ gauge bosons.
These measurements provide an insight into the partonic structure of matter
and the dynamics of strong interactions
and test quantum chromodynamics (QCD) over a huge kinematic range.

Both neutral current (NC) interactions, \mbox{$ep \rightarrow eX$} via
$\gamma$ or $Z^0$ exchange, and charged current (CC) interactions,
\mbox{$ep \rightarrow \nu X$} via $W$ exchange, can be observed at
HERA, yielding complementary information on the QCD and electroweak (EW)
parts of the Standard Model. The cross sections are defined in terms
of three kinematic variables $Q^2$, $x$ and $y$, where $y$ quantifies
the inelasticity of the interaction. The kinematic variables are
related via $Q^2=sxy$, where $s$ is the $ep$ centre-of-mass energy squared.

Measurements of the NC and CC cross sections in $e^+p$ scattering have
been made by H1 and ZEUS based on $\simeq 40$\,pb$^{-1}$ data sets taken
between $1994$ and $1997$~\cite{h1hiq2,zeushiq2} with protons of energy
$820{\rm\,GeV}$ and positrons of energy $27.6{\rm\,GeV}$, leading to a
centre-of-mass energy $\sqrt{s}=301{\rm\,GeV}$.
Here, new $e^+p$ NC and CC cross section measurements, based on data
taken at $\sqrt{s}=319{\rm\,GeV}$ in $1999$ and $2000$, are
presented with improved precision using a luminosity of $65.2$\,pb$^{-1}$.
The increased centre-of-mass energy stems from the change
in the proton beam energy from $820{\rm\,GeV}$ to $920{\rm\,GeV}$ since 1998.
These data sets together provide the most accurate neutral and charged
current cross sections measured by H1 at high $Q^2$ ($\geq 100{\rm\,GeV}^2$)
in the first phase of HERA operation (HERA-I).

The NC analysis is extended to higher $y$ up to $0.9$ for
$100{\rm\,GeV}^2 \leq Q^2 \leq 800{\rm\,GeV}^2$. This extension
of the kinematic range allows a
determination of the longitudinal structure function, $F_L(x,Q^2)$, to
be made at high $Q^2$ for the first time.
This analysis is performed on both the $99-00$
$e^+p$ data and the $e^-p$ data, taken in $1998$ and
$1999$ with a luminosity of $16.4$\,pb$^{-1}$ at $\sqrt{s}=319{\rm\,GeV}$.
The extended high-$y$ $e^{-}p$ analysis and $F_L$ extraction complement
the inclusive cross section measurements published in~\cite{h1elec}.
The difference in NC cross sections between $e^+p$ and $e^-p$ scattering
at high $Q^2$ is employed to update the measurement of $x\tilde{F}_3$,
superseding the earlier measurement~\cite{h1elec}. 

The accuracy and kinematic coverage of the H1 neutral and charged
current cross section data enable dedicated QCD analyses, which test
the prediction of logarithmic scaling violations over four orders of
magnitude in $Q^2$ and allow parton distribution functions (PDFs) 
of the proton to be deduced. This in turn allows predictions to be made 
for future facilities such as the LHC,
deviations from which may be due to exotic phenomena beyond 
the Standard Model. 

A next-to-leading order (NLO) QCD analysis of the H1 data alone is
performed, using a novel decomposition of the quark species into the
up- and down-type quark distributions to which the NC and CC
cross section data are sensitive.
The fit parameter space is narrowed using
theoretical constraints adapted to the new ansatz and the experimental
and phenomenological uncertainties are systematically approached.
This leads to a description of the complete set of NC and CC data 
as well as to new determinations of the PDFs and their uncertainties.
For comparison, the QCD analysis is further extended to include the accurate 
proton and deuteron data from the BCDMS muon scattering 
experiment~\cite{bcdms}.

This paper is organised as follows. In section~\ref{sec:theory} the
definitions of the inclusive NC and CC cross sections are given. In
section~\ref{expt} the detector, simulation and measurement procedures are
described.
The QCD analysis method is explained in section~\ref{qcdana}, followed
by the measurements and the QCD analysis results in section~\ref{results}. 
The paper is summarised in section~\ref{summary}.

\section{Neutral and Charged Current Cross Sections}
\label{sec:theory}

After correction for QED radiative effects, the measured NC cross section for
the process $e^{\pm}p\rightarrow e^{\pm}X$ with unpolarised beams is given by
\begin{eqnarray}
 \label{Snc1}
 \frac{{\rm d}^2\sigma_{NC}^{\pm}}{{\rm d}x\;{\rm d}\QQ}
 & = & \frac{2\pi \alpha^2}{xQ^4} 
 \;\phi^\pm_{NC}\;
 (1+\Delta^{\pm,weak}_{NC})\,,\\
 \label{eq:phinc}
 \mbox{with} \hspace{1cm} \phi_{NC}^{\pm} & = & 
 Y_+ \Ftwo \mp Y_{-}x\Fz  - y^2 \FL\,,
\end{eqnarray}
where $\alpha \equiv \alpha(Q^2=0)$ is the fine structure constant.
The $\Delta^{\pm,weak}_{NC}$ corrections are defined in~\cite{workshop},
with $\alpha$ and the $Z$ and $W$ boson masses (taken here as in~\cite{h1elec}
to be $M_Z=91.187{\rm\,GeV}$ and $M_W=80.41{\rm\,GeV}$) as the main 
electroweak inputs.  The weak corrections
are typically less than $1\%$ and never more than $3\%$.
The NC structure function term $\phi^\pm_{NC}$ was introduced in~\cite{h1hiq2}
and is expressed in terms of the generalised structure functions 
$\Ftwo$, $x\Fz$ and $\tilde{F}_L$.
The helicity dependences of the electroweak interaction are contained in 
\mbox{$Y_{\pm} \equiv 1 \pm (1-y)^2$}.
The generalised structure functions $\Ftwo$ and $x\Fz$ can be further
decomposed as~\cite{klein}
\begin{eqnarray}
 \label{f2p}
 \Ftwo  \equiv & \Fem & - \ v_e  \ \frac{\kappa  \QQ}{(\QQ + M_Z^2)}
  \Fint  \,\,\, + (v_e^2+a_e^2)  
 \left(\frac{\kappa  Q^2}{\QQ + M_Z^2}\right)^2 \Fwk\,, \\
 \label{f3p}
 x\Fz    \equiv &      & - \ a_e  \ \frac{\kappa  \QQ}{(\QQ + M_Z^2)} 
 x\Fzint + \,\, (2 v_e a_e) \,\,
 \left(\frac{\kappa  Q^2}{\QQ + M_Z^2}\right)^2  x\Fzwk\,,
\end{eqnarray} 
with $\kappa^{-1}=4\frac{M_W^2}{M_Z^2}(1-\frac{M_W^2}{M_Z^2})$ in the
on-mass-shell scheme~\cite{pdg}.  The quantities $v_e$ and $a_e$ are the
vector and axial-vector weak couplings of the electron\footnote{In this paper
  ``electron'' refers generically to both electrons and positrons. Where
  distinction is required the symbols $e^+$ and $e^-$ are used.} to the
$Z^{0}$~\cite{pdg}.  The electromagnetic structure function $\Fem$
originates from photon exchange only. The functions $\Fwk$ and $x \Fzwk$
are the contributions to $\Ftwo$ and $x\Fz$ from $Z^0$ exchange and the
functions $\Fint$ and $x\Fzint$ are the contributions from $\gamma Z$
interference. The longitudinal structure function $\FL$ may be
decomposed in a manner similar to $\Ftwo$. Its contribution is
significant only at high $y$.

Over most of the kinematic domain at HERA the dominant contribution to
the cross section comes from pure photon exchange via
$F_2$. The contributions due to $Z^0$ boson exchange only become important 
at large values of $Q^2$.  For longitudinally unpolarised lepton
beams the $\Ftwo$ contribution is the same for $e^-$ and for $e^+$ scattering,
while the $x \Fz$ contribution changes sign as can be seen in
eq.~\ref{eq:phinc}.

In the quark parton model (QPM) the structure functions $F_2$,
$F_2^{\gamma Z}$ and $F_2^Z$ are related to the sum of the quark
and anti-quark momentum distributions, $xq(x,Q^2)$ and $x\overline{q}(x,Q^2)$, 
\begin{equation}
 \label{eq:f2}
 [F_2,F_2^{\gamma Z},F_2^{Z}] = x \sum_q 
 [e_q^2, 2 e_q v_q, v_q^2+a_q^2] 
 \{q+\bq\} 
\end{equation}
and the structure functions $xF_3^{\gamma Z}$ and $xF_3^Z$ to their
difference, which determines the valence quark distributions, $xq_v(x,Q^2)$,
\begin{equation}
 \label{eq:xf3}
 [ x F_3^{\gamma Z},x F_3^{Z} ] = 2x \sum_q 
 [e_q a_q, v_q a_q]
 \{q -\bq \} = 2 x \sum_{q=u,d} [e_q a_q, v_q a_q] q_v\,.
\end{equation}
In eqs.~\ref{eq:f2} and \ref{eq:xf3}, $e_q$ is the electric charge of quark $q$
and $v_q$ and $a_q$ are respectively the vector and
axial-vector weak coupling constants of the quarks to the $Z^0$. 
In the QPM the longitudinal structure function $\FL \equiv 0$. 

For CC interactions the measured unpolarised $ep$ scattering cross section 
corrected for QED radiative effects may be expressed as
\begin{eqnarray}
 \label{eq:cccross}
 \frac{{\rm d} ^2 \sigma_{CC}^{\pm}}{{\rm d} x\; {\rm d} Q^2} & = &
 \frac{G_F^2 }{2 \pi x} \left[ \frac{M_W^2}{Q^2+M_W^2} \right]^2 
 \;\phi_{CC} ^\pm \; (1+\Delta^{\pm,weak}_{CC})\,,\\
 \mbox{with } \hspace{1cm} \phi_{CC}^\pm & =& 
 \frac{1}{2}(Y_+ \wtwogen^\pm   \mp Y_- \xwthreegen^\pm - y^2 \wlgen^\pm)\,,
\end{eqnarray}
where $\Delta^{\pm,weak}_{CC}$ represents the CC weak radiative corrections.
In this analysis $G_F$ is defined~\cite{hector} using the weak boson
masses and is in very good agreement with $G_F$ determined
from the measurement of the muon lifetime~\cite{pdg}.  The CC
structure function term $\phi_{CC}^{\pm}$~\cite{h1hiq2} is expressed 
in terms of the CC structure functions $\wlgen^\pm$, $\wtwogen^\pm$ and 
$\xwthreegen^\pm$,
defined in a similar manner to the NC structure functions~\cite{hector}. 
In the QPM (where $\wlgen^\pm \equiv 0$), they
may be interpreted as lepton beam-charge dependent sums and differences
of quark and anti-quark distributions and are given for an unpolarised
lepton beam by
\begin{eqnarray}
 \label{ccstf}
    \wtwogen^{+}  =  x (\bU+D)\hspace{0.1cm}\mbox{,}\hspace{0.3cm}
 \xwthreegen^{+}  =  x (D-\bU)\hspace{0.1cm}\mbox{,}\hspace{0.3cm}
    \wtwogen^{-}  =  x (U+\bD)\hspace{0.1cm}\mbox{,}\hspace{0.3cm}
 \xwthreegen^{-}  =  x (U-\bD)\,.
\end{eqnarray}
Below the $b$ quark mass threshold, $xU$, $xD$, $x\bU$ and $x\bD$ are 
defined respectively as the sum of up-type, of down-type and of their 
anti-quark-type distributions
\begin{eqnarray}
\label{ud}
  xU  &=& x(u + c)     \nonumber\\
 x\bU &=& x(\bu + \bc) \nonumber\\
  xD  &=& x(d + s)     \nonumber\\
 x\bD &=& x(\bd + \bs)\,. 
\end{eqnarray}

For the presentation of the subsequent measurements it is convenient
to define the NC and CC ``reduced cross sections'' as
\begin{equation}
 \label{Rnc}
 \tilde{\sigma}_{NC}(x,Q^2) \equiv  \frac{1}{Y_+} \ 
 \frac{ Q^4 \ x  }{2 \pi \alpha^2}
 \          \frac{{\rm d}^2 \sigma_{NC}}{{\rm d}x{\rm d}Q^2}\,,\;\;\;
 \tilde{\sigma}_{CC}(x,Q^2) \equiv  
 \frac{2 \pi  x}{ G_F^2}
 \left[ \frac {M_W^2+Q^2} {M_W^2} \right]^2
          \frac{{\rm d}^2 \sigma_{CC}}{{\rm d}x{\rm d}Q^2}\,.
\end{equation}

\section{Experimental Technique}
\label{expt}

\subsection{H1 Apparatus and Trigger}
\label{det}

The H1 co-ordinate system is defined such that the positive $z$
axis is in the direction of the outgoing proton beam (forward
direction). The polar angle $\theta$ is then defined with respect to
this axis. A full description of the H1 detector can be found
in~\cite{h1detector, lar, spacal}. The detector
components most relevant to this analysis are the Liquid Argon (LAr)
calorimeter, which measures the positions and energies of particles
over the range $4^\circ<\theta<154^\circ$, a lead-fibre calorimeter
(SPACAL) covering the range $153^\circ<\theta<177^\circ$, the Plug
calorimeter covering the range $0.7^\circ<\theta<3.3^\circ$ and the
inner tracking detectors, which measure the angles and momenta of
charged particles over the range $7^\circ<\theta<165^\circ$.  In the
central region, $25^{\circ}\lapprox\theta\lapprox165^{\circ}$, the
central jet chamber (CJC) measures charged track trajectories in the 
$(r, \phi)$ plane and is supplemented by two $z$ drift chambers to
improve the $\theta$ measurement of reconstructed tracks. The forward
tracking detector, $\theta\lapprox30^{\circ}$, is used to determine
the vertex position of events when no reconstructed CJC track is found.

The $ep$ luminosity is determined by
measuring the QED bremsstrahlung ($ep \rightarrow ep\gamma$) event rate 
by tagging the photon in a photon detector located at $z=-103$\,m.  
An electron tagger is
placed at $z=-33$\,m adjacent to the beam-pipe. It is used to check the
luminosity measurement and to provide information on $ep\rightarrow
eX$ events at very low $Q^2$ (photoproduction) where the electron
scatters through a small angle ($\pi - \theta < 5$\,mrad). 

NC events are triggered mainly using information from the LAr
calorimeter. The calorimeter has a finely segmented geometry
allowing the trigger to select localised energy deposits in the
electromagnetic section of the calorimeter.
For electrons with energy above $11{\rm\,GeV}$ this is $100\%$ efficient as
determined using an independently triggered sample of events.  At
lower energies the triggers based on LAr information are supplemented by
using additional information from the tracking detectors.  In $1998$ the
LAr calorimeter electronics were upgraded in order to trigger scattered
electrons with energies as low as $6{\rm\,GeV}$, 
the minimum value considered in this analysis. 
This gives access to the high $y$ kinematic region. 
For electron energies of $6{\rm\,GeV}$, the overall trigger efficiency is 
$96\%$ for the $e^+p$ data set and $90\%$ for the earlier $e^-p$ data set.

The characteristic feature of CC events is a large missing transverse
momentum $P_T^{miss}$, which is identified at the trigger level using
the LAr calorimeter vector sum of ``trigger towers'', i.e. groups of trigger
regions with a projective geometry pointing to the nominal interaction
vertex. At low $P_T^{miss}$, the efficiency is enhanced by making use of
an additional trigger requiring calorimeter energy 
in association with track information from the inner tracking chambers. 
For the minimum $P_T^{miss}$ of $12{\rm\,GeV}$ considered in the analysis
the efficiency is $60\%$, rising to $90\%$ for $P_T^{miss}$ of $25{\rm\,GeV}$. 
In terms of $Q^2$, the efficiency is $79\%$ at $300{\rm\,GeV}^2$ and 
increases to $98\%$ at $3\,000{\rm\,GeV}^2$.  
These efficiencies are determined from the data using a sample of NC events
in which all information from the scattered lepton is suppressed, 
the so-called {\em pseudo-CC} sample. 
The trigger energy sums are then recalculated for
the remaining hadronic final state. This sample also provides a useful
high statistics cross check of further aspects of the CC analysis.

\subsection{Simulation Programs}
\label{mc}

Simulated DIS events are used in order to determine acceptance corrections.
DIS processes are generated using the DJANGO~\cite{django} Monte Carlo (MC) 
simulation program, which is based on LEPTO~\cite{lepto} for 
the hard interaction and HERACLES~\cite{heracles} for single photon
emission off the lepton line and virtual EW corrections.
LEPTO combines ${\cal O}(\alpha_s)$ matrix elements with higher order QCD
effects using the colour dipole model as implemented in ARIADNE~\cite{cdm}.
The JETSET program is used to simulate 
the hadronisation process~\cite{jetset}.  
In the event generation the DIS cross section is calculated with the PDFs
of~\cite{mrsh}. The simulated cross section is reweighted using 
a NLO QCD fit (H1 $97$ PDF fit) to previous data~\cite{h1hiq2}.

The detector response to events produced by the generation
programs is simulated in detail using a program based on
GEANT~\cite{GEANT}.  These simulated events are then subjected to the
same reconstruction and analysis chain as the data.

The dominant photoproduction background processes are
simulated using the PYTHIA~\cite{pythia} generator with leading order
PDFs for the proton and photon taken from~\cite{ggrv}. 
Further background from QED-Compton scattering, lepton pair
production via two-photon interactions, prompt photon production and
heavy gauge boson ($W^{\pm},Z^0$) production are included in the
background simulation.  
Further details are given in~\cite{h1hiq2}.

\subsection{Polar Angle Measurement and Energy Calibration}
\label{calib}

In the neutral current analysis the polar angle of the scattered
electron ($\theta_e$) is determined using the position of its energy
deposit (cluster) in the LAr calorimeter, together with the interaction 
vertex reconstructed with tracks from charged particles in the event. 
The relative alignment of the calorimeter and tracking chambers is
determined using a sample of events with a well measured electron track,
using information from both the CJC and the $z$ drift chambers. 
Minimisation of the spatial discrepancy between
the electron track and cluster allows the LAr calorimeter and the inner 
tracking chambers to be aligned. 
The residual discrepancy
then determines the systematic uncertainty on the measurement of
$\theta_e$, which varies from $1$\,mrad for $\theta_e>135^{\circ}$ to 
$3$\,mrad for $\theta_e<120^{\circ}$.

The calibration of the electromagnetic part of the LAr
calorimeter is performed using the method described in~\cite{h1hiq2}.
Briefly, the redundancy of the detector information allows a prediction
of the scattered electron energy ($E_e^{\prime}$) to be made based
on the electron beam energy ($E_e$), the polar angle measurement of 
the scattered electron and the inclusive polar angle 
($\gamma_h$)~\cite{h1hiq2} of the hadronic final state.
This prediction of the double angle
(DA) kinematic reconstruction method~\cite{damethod} is then compared with
the measured electromagnetic energy, allowing local calibration
factors to be determined in a finely segmented grid in $z$ and $\phi$.
The calibration procedure is also performed on the simulated data.  
The final calibration is obtained by application of a further small 
correction determined from simulation, which accounts for small biases 
in the reconstruction of $\gamma_h$.
The calibration is cross checked using independent data samples from
QED-Compton scattering and two-photon $e^+e^-$ pair production processes. 
The total systematic uncertainty on the absolute electromagnetic energy scale 
varies from $0.7\%$ in the backward part of the calorimeter to 
$3\%$ in the forward region, where statistics are limited.

The hadronic final state is measured using energy deposits in the
LAr and SPACAL calorimeters supplemented by low momentum tracks.
Isolated low energy calorimetric deposits are classified as noise and
excluded from the analysis. The response of the detector to hadrons is
calibrated by requiring transverse momentum conservation between the
precisely calibrated scattered electron and the hadronic final state in 
NC events as described in~\cite{h1hiq2}. 
The electron transverse momentum is defined as 
$P_{T,e}=\sqrt{p_{x,e}^2+p_{y,e}^2}$.
The hadronic transverse momentum is determined from 
$P_{T,h}=\sqrt{(\sum_i p_{x,i})^2+(\sum_i p_{y,i})^2}$, where
the summation is performed over all hadronic final state particles $i$,
assuming particles of zero rest mass.

\begin{figure}[tb]   
 \begin{center} 
 \begin{picture}(140,65)(0,0)
 \setlength{\unitlength}{1 mm}
 \put(0,-10){\epsfig{file=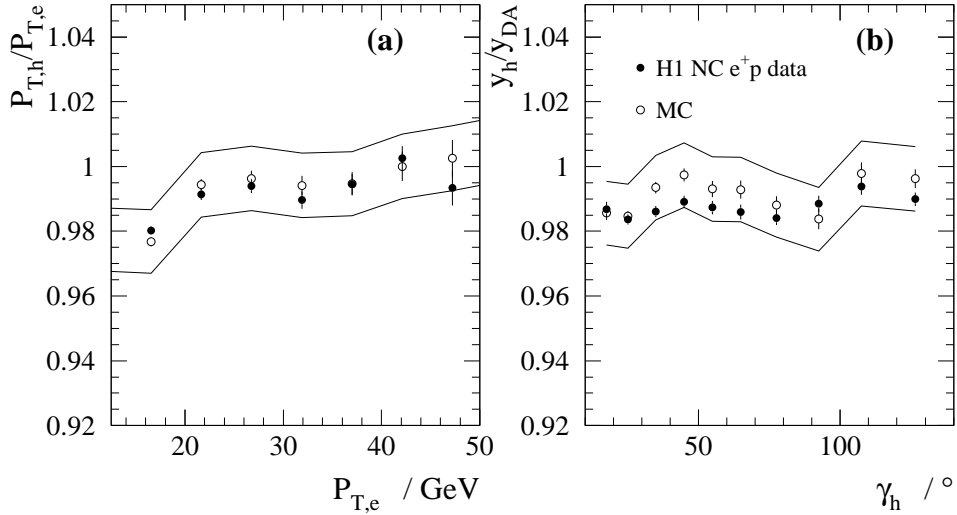,width=14cm}}
 \put(55,54){\bf (a)}
 \put(120,54){\bf (b)}
 \end{picture}
 \end{center}    
 \caption{\sl Mean values of (a) $P_{T,h}/P_{T,e}$ as a function of
 $P_{T,e}$ and (b) $y_{h}/y_{DA}$ as a function of $\gamma_{h}$ for
 neutral current data (solid points) and Monte Carlo (MC) simulation 
 (open points) for
 $\gamma_h>15^{\circ}$ and $12{\rm\,GeV}<P_{T,h}<50{\rm\,GeV}$. 
 The curves correspond to
 a $\pm 1\%$ variation around the simulation.}
 \label{hadscale}
\end{figure}

Detailed studies and cross checks of the hadronic response of the
calorimeter using the enlarged data sample have led to an improved
understanding of the hadronic energy measurement.
The calibration procedure is cross checked by requiring
energy-momentum conservation,
$E-P_z\equiv (E^\prime_e-p_{z,e})+(E_h-P_{z,h})=2E_e$,
with $E_h-P_{z,h}$ being the contribution of all hadronic final state
particles $\sum_i (E_i - p_{z,i})$.
In addition, the reference scale may be taken from the double angle method
prediction rather than from the scattered electron.  
These studies have allowed
the systematic uncertainty on the hadronic scale to be reduced with
respect to previous measurements~\cite{h1hiq2,h1elec} in the region
$12{\rm\,GeV}<P_{T,h}<50{\rm\,GeV}$ and $\gamma_h>15^{\circ}$. The uncorrelated
part (see section~\ref{syserr}) of the hadronic scale uncertainty is 
reduced to $1\%$ from $1.7\%$ previously. 

Fig.~\ref{hadscale} demonstrates the quality of the hadronic calibration 
in the stated region of $P_{T,h}$ and $\gamma_h$, 
showing the level of agreement between data and simulation 
after the calibration procedure. In fig.~\ref{hadscale}(a) the mean value 
of the ratio $P_{T,h}/P_{T,e}$ is shown.  
In fig.~\ref{hadscale}(b) the 
inelasticity $y_h$, defined from the hadron reconstruction 
method~\cite{jbmethod} as $y_h=(E_h-P_{z,h})/2E_e$, 
is compared with the DA variable $y_{DA}$.
In this analysis, it is the relative difference between data and
simulation that is relevant and good agreement is found to within
$1\%$.  In addition a $1\%$ correlated uncertainty is considered,
accounting for possible remaining biases in the reference scale used for
the calibrations.

\subsection{Neutral Current Measurement Procedure}
\label{ncmeas}
Events from inelastic $ep$ interactions are required to have a well defined
interaction vertex to suppress beam-induced background. High \qsq
neutral current events are selected by requiring a compact and 
isolated energy deposit in the electromagnetic
part of the LAr calorimeter\footnote{Local detector regions are removed
  where the cluster of the scattered electron is not expected to be fully 
  contained in the calorimeter, or where the trigger is not fully efficient.}.
The scattered electron is identified as the cluster of highest transverse
momentum. In the central detector region, $\theta>35^{\circ}$, the
cluster has to be associated with a track measured in the inner tracking
chambers.  

As mentioned earlier, energy-momentum conservation requires
$E-P_z=2E_e$. Restricting the measured $E-P_z$ to be greater than
$35{\rm\,GeV}$ thus considerably reduces the radiative corrections 
due to initial state bremsstrahlung,
where photons escape undetected in the backward direction.
It also suppresses photoproduction background in which the scattered
electron is lost in the backward beam-pipe and a hadron fakes
the electron signal in the LAr calorimeter.
Since the photoproduction background contribution increases with $y$,
the analysis is separated into two distinct regions where
different techniques are employed to suppress this background.
The {\em nominal analysis} is restricted to 
\mbox{$y_e< 0.63$} for \mbox{$90{\rm\,GeV}^2<Q^2_e< 890{\rm\,GeV}^2$} and 
$y_e < 0.90$ for $Q^2_e>890{\rm\,GeV}^2$. 
This limits the minimum $E_e^\prime$ to $11{\rm\,GeV}$.
The {\em high-$y$ analysis} is performed for $E^\prime_e>6{\rm\,GeV}$, 
$0.63< y_e < 0.90$ and $90{\rm\,GeV}^2<Q^2_e< 890{\rm\,GeV}^2$.
Here $y_e$ and $Q^2_e$
are reconstructed using the scattered electron energy and angle,
the so-called electron method.

The NC kinematics in the nominal analysis are reconstructed using the
$e\Sigma$ method~\cite{esigma}, which uses $E_e^\prime$, $\theta_e$
and $E_h-P_{z,h}$ and has good resolution and small sensitivity to 
QED radiative corrections over the accessible phase space. 
In the high-$y$ analysis the electron method gives the best resolution
and is used to define the event kinematics.

\begin{figure}[htb]   
 \begin{center} 
 \begin{picture}(140,122.5)(0,0)
 \setlength{\unitlength}{1 mm}
 \put( -15, -10){\epsfig{file=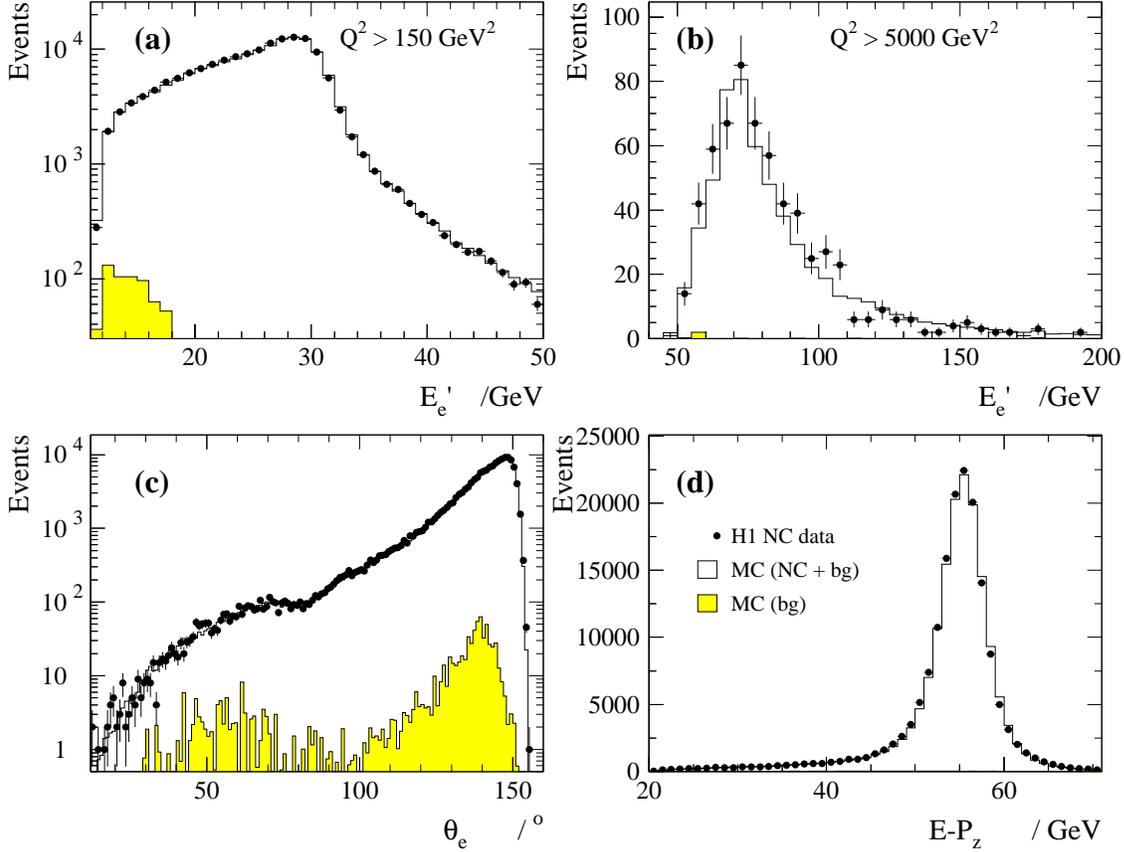,width=16cm}}
 \put(10,99){\bf (a)}
 \put(82,99){\bf (b)}
 \put(10,40) {\bf (c)}
 \put(82,40) {\bf (d)}
 \end{picture}
 \end{center}    
 \caption{\sl Distributions of $E_e^{\prime}$ for (a)
  $Q^2>150{\rm\,GeV}^2$ and (b) $Q^2>5\,000{\rm\,GeV}^2$, 
  (c) $\theta_e$ and (d) $E-P_z$ for $e^+p$ data (solid points) and 
  Monte Carlo (MC) simulation (open histograms) in the nominal analysis.
  The shaded histograms show the simulated background (bg) contribution,
  dominated by photoproduction.}
 \label{nc_cont}
\end{figure}

The nominal data sample consists of about $185\,000$ events. The
comparison of the data and the simulation is shown in
fig.~\ref{nc_cont} for the scattered electron energy and
polar angle spectra and the distribution of $E-P_z$, which are used
in the reconstruction of $x$ and $Q^2$. All distributions
are well described by the simulation, which is normalised to the
luminosity of the data.
In the nominal analysis
the small photoproduction contribution is statistically
subtracted using the background simulation.

In the high-$y$ analysis, the photoproduction background plays 
an increasingly important role, 
as low energies of the scattered electron are accessed.
For this analysis, the calorimeter cluster of the scattered electron
is linked to a well measured track having the same charge as
the electron beam.
This requirement removes a sizeable part of the background, where 
$\pi^0\rightarrow \gamma \gamma$ decays give rise to fake scattered electron 
candidates.
The remaining background from photoproduction was estimated 
from the number of data events in which the detected lepton candidate 
has opposite charge to the beam lepton. 
This background is statistically subtracted assuming charge symmetry.
The charge symmetry is determined to be $0.99\pm 0.07$ by measuring 
the ratio of wrongly charged fake scattered lepton candidates in 
$e^+p$ and $e^-p$ scattering, 
taking into account the difference in luminosity.
The charge symmetry is cross checked using a sample of
data events in which the scattered electron is detected in
the electron tagger and a systematic uncertainty of
$10\%$ on the charge symmetry is assigned. Further details are given
in~\cite{dubak, burkard}.

\begin{figure}[htb]   
 \begin{center} 
\begin{picture}(140,122.5)(0,0)
 \setlength{\unitlength}{1 mm}
 \put( 0, -10){\epsfig{file=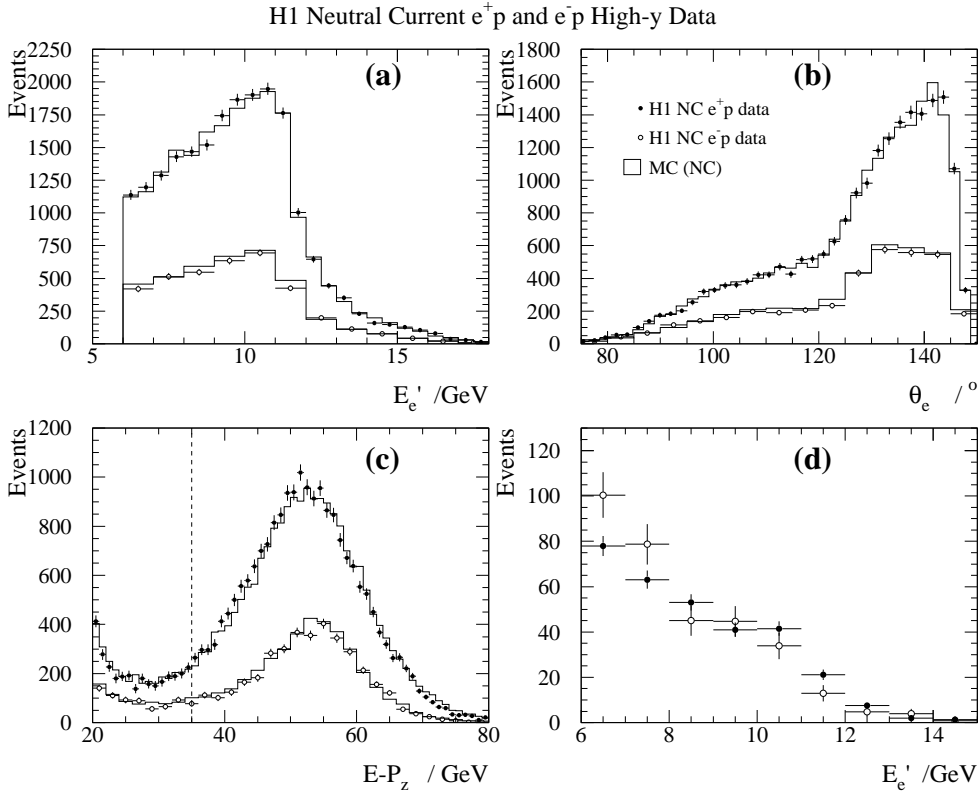,width=14cm}}
 \put(53,86){\bf (a)}
 \put(110,86){\bf (b)}
 \put(53,35){\bf (c)}
 \put(110,35){\bf (d)}
 \end{picture}
 \end{center}    
 \caption{\sl Distributions of (a) $E_e^{\prime}$, (b) $\theta_e$ and (c)
  $E-P_z$ for $e^+p$ data (solid points), $e^-p$ data (open points)
  and Monte Carlo (MC) simulation (histograms) in the high-$y$ analysis
  after background subtraction (see text). In (c) the $E-P_z$ cut is 
  not applied, but is indicated by the dashed vertical line. 
  Shown in (d) is the energy distribution of wrongly charged lepton 
  candidates in background events. In (d) the $e^+p$ data 
  have been normalised to the luminosity of the $e^-p$ data set.}
\label{lowe_cont}
\end{figure}

In total about $24\,000$ $e^+p$ events and $5\,000$ $e^-p$ events are
selected in the high-$y$ analysis. Figs.~\ref{lowe_cont}(a)-(c) show
the scattered lepton energy spectrum, the polar angle distribution and
the $E-P_z$ spectrum for both the $e^+p$ and $e^-p$ data sets
after background subtraction. The simulation, normalised to the luminosity of
the data, provides a good description of these distributions. In
fig.~\ref{lowe_cont}(d) the energy spectra of wrong charge lepton 
candidates in the data sets are shown.
Good agreement is observed when the $e^+p$ data are
normalised to the luminosity of the $e^-p$ data set.

\subsection{Charged Current Measurement Procedure}
\label{ccmeas}

The selection of charged current events requires a large missing
transverse momentum, $P_T^{miss}\equiv P_{T,h}\geq 12{\rm\,GeV}$, assumed to
be carried by an undetected neutrino. In addition the event must have
a well reconstructed vertex as for the NC selection. The
kinematic variables $y_h$ and $Q^2_h$ are determined using the hadron
kinematic reconstruction method~\cite{jbmethod}.  In order to restrict
the measurement to a region with good kinematic resolution the events
are required to have $y_h<0.85$.  In addition the
measurement is confined to the region where the trigger efficiency is
$\gtrsim 50\%$ by demanding $y_h>0.03$.  
The $ep$ background is dominantly due to photoproduction events 
in which the electron escapes undetected in the backward direction
and missing transverse momentum is reconstructed due to fluctuations 
in the detector response or undetected particles. This background is
suppressed using the ratio $V_{ap}/V_p$ and the difference in azimuth between
$\vec{P}_{T,h}$ as measured in the main detector and the Plug
calorimeter, $\Delta \phi_{h,{\rm Plug}}$~\cite{h1elec}.  
The quantities $V_p$ and $V_{ap}$ are respectively the transverse energy
flow parallel and anti-parallel to $\vec{P}_{T,h}$, the transverse
momentum vector of the hadronic final state. The residual $ep$
background is negligible for most of the measured kinematic domain,
though it reaches $15\%$ at the lowest $Q^2$ and the highest $y$. The
simulation is used to estimate this contribution, which is subtracted
statistically from the CC data sample with a systematic uncertainty of
$30\%$ on the number of subtracted events.  The non-$ep$ background is
rejected as described in~\cite{h1hiq2} by searching for event
topologies typical of cosmic ray and beam-induced background.
For further details see~\cite{juergen,zhang}.

\begin{figure}[htb]   
 \begin{center} 
 \begin{picture}(140,65)(0,0)
 \setlength{\unitlength}{1 mm}
 \put(0,-10){\epsfig{file=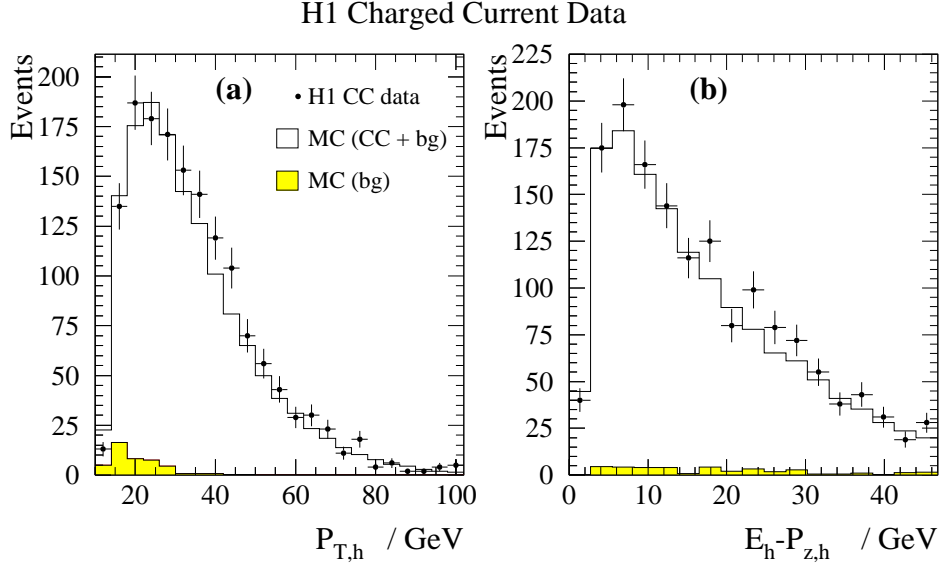,width=14cm}}
 \put(37,54){\bf (a)}
 \put(100,54){\bf (b)}
 \end{picture}
 \end{center}    
 \caption{\sl Distributions of (a) $P_{T,h}$ and (b) $E_h-P_{z,h}$ for CC data 
  (solid points) and Monte Carlo (MC) simulation (open histograms). 
  The shaded histograms show the simulated background (bg) contribution, 
  dominated by photoproduction.}
 \label{cc_cont}
\end{figure}

After all selection criteria are applied, the final CC data sample
contains about $1\,500$ events. The data and simulation are compared in
fig.~\ref{cc_cont} for the $P_{T,h}$ and $E_h-P_{z,h}$ spectra, which are
directly used in the reconstruction of the kinematic variables $y$ and
$Q^2$. In both cases the simulation gives a reasonable
description of the data.

\subsection{Cross Section Measurement and Systematic Uncertainties}
\label{syserr}

For both the NC and CC analyses the selected event samples are corrected for
detector acceptance and migrations 
using the simulation and are converted to bin-centred cross sections. 
The bins in the $(x, Q^2)$ plane are defined as in 
refs.~\cite{h1hiq2,h1elec}, based on the consideration of
the detector resolution and event statistics. 
The bins used in the measurement are required to have values 
of stability and purity\footnote{The stability (purity) was defined 
  in~\cite{h1hiq2} as the number of simulated events which originate 
  from a bin and which are reconstructed in it, divided by the number of 
  generated (reconstructed) events in that bin.}
larger than $30\%$.
This restricts the range of the NC measurements to $y\gtrsim 0.005$.
The QED radiative corrections ($\Delta^{QED}_{NC(CC)}$) are defined 
in~\cite{h1hiq2} and were calculated using the program 
HERACLES~\cite{heracles} as
implemented in DJANGO~\cite{django}.

The systematic uncertainties on the cross section measurements are 
presented in tables~\ref{ncdq2}-\ref{ncextra}. 
They are split into bin-to-bin correlated and uncorrelated parts.  
All the correlated systematic errors are found to be symmetric to 
a good approximation and are assumed so in the following. 
The total systematic error is formed by adding the individual
errors in quadrature. 

The correlated and uncorrelated systematic errors are 
discussed briefly below (see refs.~\cite{dubak,juergen,beate,zhang,burkard} 
for more details).
In addition, there is a global uncertainty of $1.5\%$ and $1.8\%$ 
on the luminosity measurement for the $e^+p$ and $e^-p$ data respectively,
of which $0.5\%$ is common to both (see section~\ref{ansatz}).

\begin{itemize}
\item The total uncertainty on the electron energy scale is $1\%$ if the $z$
  impact position of the electron at the calorimeter surface ($z_{imp}$)
  is in the backward part of the detector ($z_{imp}<-150{\rm\,cm}$), 
  $0.7\%$ in the region $-150{\rm\,cm} < z_{imp}<20{\rm\,cm}$, 
  $ 1.5\%$ for $20< z_{imp}<100{\rm\,cm}$ and
  $3\%$ in the forward part ($z_{imp} >100{\rm\,cm}$). The correlated part of
  the total uncertainty comes mainly from the possible bias of the
  calibration method and is estimated to be $0.5\%$ throughout the LAr
  calorimeter. It results in a correlated systematic error on the NC
  cross section which is typically below $1\%$,
  increasing at low $y$ to $\sim 3\%$ for $Q^2 \lapprox 1\,000{\rm\,GeV}^2$
  and $\sim 8\%$ for larger $Q^2$. 

\item The correlated uncertainty on the electron polar angle is $1$\,mrad,
  $2$\,mrad and $3$\,mrad for $\theta_e>135^{\circ}$,
  $120^{\circ}<\theta_e<135^{\circ}$ and $\theta_e<120^{\circ}$, respectively.
  This leads to a typical uncertainty on the
  NC reduced cross section of less than $1\%$, increasing up to 
  $\sim 5\%$ at high $x$.
  
\item A $0.5\%$ ($1\%$) uncorrelated error originates from the electron 
  identification efficiency in the NC nominal (high-$y$) analysis for
  $z_{imp}<-5{\rm\,cm}$. 
  For $z_{imp}>-5{\rm\,cm}$ the uncertainty is increased to $2\%$. 
  The precision of this efficiency is estimated using an independent 
  track based electron identification algorithm, limited for 
  $z_{imp}> -5{\rm\,cm}$ by statistics. 
  
\item A $0.5\%$ uncorrelated error is assigned on the efficiency of 
  the scattered electron track-cluster link requirement in the NC 
  nominal analysis. In the high-$y$ analysis this is increased to $1\%$. 
  
\item An uncorrelated $1\%$ uncertainty on the hadronic energy
  measured in the LAr calorimeter is assigned for the region
  $12{\rm\,GeV}< P_{T,h}< 50{\rm\,GeV}$ and $\gamma_h>15^{\circ}$.
  Outside this region the
  uncertainty is increased to $1.7\%$. In addition, a $1\%$ correlated
  component to the uncertainty is added in quadrature, originating from
  the calibration method and from the uncertainty on the reference
  scale. This yields a total uncertainty of $1.4\%$ and
  $2\%$ for the two regions respectively.
  
\item The uncertainty on the hadronic energy scale of the SPACAL calorimeter
  is $5\%$. The uncertainty on the hadronic energy measurement due to
  the inclusion of low momentum tracks is obtained by shifting their
  contribution by $3\%$.
  The influence on the cross section measurements from these sources
  is small compared with that from the correlated uncertainty from 
  the LAr calorimeter energy scale. 
  The three contributions (LAr, SPACAL, tracks) are thus combined,
  resulting in a single correlated hadronic error from the hadronic
  energy measurement, which is given in the tables.
  The corresponding error on the NC and CC cross sections is typically 
  $\lapprox \ 1\%$, but increases at low $y$ to $\sim 5\%$.
  
\item A $25\%$ uncertainty is assigned on the amount of energy
  in the LAr calorimeter attributed to noise, which gives rise to a
  sizeable correlated systematic error at low $y$, reaching  $\simeq 10\%$ at
  $x=0.65$ and $Q^2 \lapprox 2\,000{\rm\,GeV}^2$ in the NC measurements.
  
\item In the CC analysis the correlated uncertainties due to the cuts 
  against photoproduction on $V_{ap}/V_p$ and $\Delta \phi_{h,{\rm Plug}}$ 
   are only significant at high $y$, low $Q^2$ and low $P_{T,h}$,
   reaching a maximum of $\sim 7\%$.
  
\item In the CC and the NC nominal analyses the photoproduction
  background is estimated from simulation. A $30\%$ correlated uncertainty 
  on the subtracted photoproduction background is determined from a
  comparison of data and simulation for a phase space region dominated
  by photoproduction background. This results in a 
  systematic error of typically $\lapprox 1\%$.
  
\item In the NC high-$y$ analysis the photoproduction background
  is estimated directly from the data by using wrongly charged (fake) 
  scattered lepton candidates, which leads to a $10\%$ correlated 
  uncertainty on the subtracted photoproduction background. 
  The resulting uncertainty on the measured cross sections is $1\%$ 
  or less.
\item A $0.3\%$ uncorrelated error is considered on the trigger efficiency 
  in the NC nominal analysis and $2-6\%$ in the CC analysis. 
  For the NC high-$y$ analysis the uncertainty on the cross section 
  is $\sim 2\%$ at low $Q^2$, decreasing to $0.6\%$ at the highest 
  $Q^2$ in the analysis.
  
\item An uncorrelated error of $1\%$ (NC) and $3\%$ (CC) is estimated on
  the QED radiative corrections by comparing the radiative corrections
  predicted by the Monte Carlo program (DJANGO) with those calculated from 
  HECTOR and EPRC~\cite{eprc}. The error also accounts for a small missing
  correction in DJANGO due to the exchange of two or more photons between 
  the electron and the quark lines.
  
\item A $3\%$ uncorrelated error is assigned on the event losses due to 
  the non-$ep$ background finders in the CC analysis, 
  estimated from pseudo-CC data (see section~\ref{det}).
  
\item A $2\%$ uncorrelated error ($5\%$ for $y < 0.1$) on the vertex finding
  efficiency for CC events is estimated using pseudo-CC data.

\end{itemize}
Overall a typical total systematic error of about $3\%$ ($6\%$) is
reached for the NC (CC) double differential cross section.
This precision has been achieved through 
detector improvements for triggering and a better understanding of 
the hadronic response of the detector, the electron identification and 
its angular measurement. 

\section{QCD Analysis}
\label{qcdana}
The cross section data presented here, together with the low $Q^2$ precision
data~\cite{h1lowq2} and high $Q^2$ $e^\pm p$ data~\cite{h1hiq2,h1elec}
previously published by the H1 Collaboration, cover a huge range in $Q^2$ 
and $x$. The improved accuracy now available allows the
predictions of perturbative QCD to be tested over four orders of
magnitude in $Q^2$ from about $1{\rm\,GeV}^2$ to above $10^4{\rm\,GeV}^2$, 
and $x$ from below $10^{-4}$ to $0.65$. The measurements of NC and CC
$e^\pm p$ scattering cross sections provide complementary sensitivity 
to different quark distributions and the gluon distribution,
$xg(x,Q^2)$. This is used to determine the sum of 
up-type $xU$, of down-type $xD$ and of their anti-quark-type
$x\bU$ and $x\bD$ distributions, employing only H1 inclusive cross section 
data.

With the current beam energies, the HERA collider data do not give access to
the large $x$ region of deep inelastic scattering at medium $Q^2$ 
($\sim 100{\rm\,GeV}^2$).
Complementary information on quark distributions in this region is
provided by fixed-target lepton-proton data. Lepton-deuteron scattering data,
which provide further constraints on the PDFs, are not yet available
from HERA. Therefore, in the subsequent analysis, the H1 data are also
combined with the precise BCDMS muon-proton and muon-deuteron scattering
data and the results are compared for cross checks of the PDFs obtained
from the analysis of the H1 data alone.
\subsection{Ansatz}
\label{ansatz}
Traditionally, QCD analyses of inclusive deep inelastic scattering 
cross section
data have used parameterisations of the valence quark distributions and 
of a sea quark distribution, imposing additional assumptions on the flavour
decomposition of the sea~\cite{h1hiq2, zeusfit00, mrst2001, cteq6}. 
The neutral and charged current cross section
data presented here, however, are sensitive to four combinations of
up- and down-type (anti-)quark distributions which, for $Q^2$ less than the
bottom quark production threshold, are given above in eq.~\ref{ud}. 
Working in terms of these combinations weakens the influence of necessary 
assumptions on the flavour decomposition of the sea in the fit. 
The valence quark distributions are obtained from
\begin{eqnarray}\label{equvdv}
   xu_v=x\left(U-\bU \right)\,,  \quad xd_v=x\left(D-\bD \right)
\end{eqnarray}
and are not fitted directly.

In the QPM, the charged current structure function terms $\phi_{CC}^\pm$
are superpositions of the distributions given in eq.~\ref{ud} according to
\begin{eqnarray}\label{Scc}
 \phi^+_{CC} = x\bU +(1-y)^2 xD, \quad \phi^-_{CC} = xU +(1-y)^2 x\bD\,.
\end{eqnarray}
The neutral current structure function terms $\phi_{NC}^\pm$ are 
dominated by the electromagnetic structure function \Fc, which can be written 
as
\begin{eqnarray}\label{f2ud}
 F_2 = \frac{4}{9}\,x\left(U +\bU \right) + \frac{1}{9}\,x\left(D +\bD \right)\,.
\end{eqnarray}
In the high $Q^2$ neutral current data, complementary sensitivity is
obtained from the interference structure function $xF_3^{\gamma Z}=
 x [2(U-\bU) + (D-\bD)]/3$, but still higher luminosity is required
 to exploit this for a dedicated determination of the valence quarks. 

In the fit to the H1 and BCDMS data,
the isoscalar nucleon structure function $F_2^N$ is determined by
the singlet combination of parton distributions and a small
contribution from the difference of strange and charm quark
distributions,
\begin{eqnarray}\label{f2d}
 F_2^N =  \frac{5}{18}\, x\left(U +\bU  + D +\bD \right) + \frac{1}{6}\,
 x\left(c+\bc -s-\bs \right)\,.
\end{eqnarray}
The nucleon data obtained from the BCDMS muon-deuteron cross sections
are measured for $x \ge 0.07$. For these data nuclear corrections 
are applied following~\cite{nucor}.
In eq.~\ref{f2d} the charm and strange quark distributions
occur explicitly and may be constrained using experimental data 
as provided by H1 and ZEUS on the charm contribution to 
$F_2$~\cite{h1f2c,zeusf2c} and from NuTeV on the strangeness content 
of the nucleon~\cite{ccfr}.
The analysis of the H1 data, however, is rather insensitive to 
these quark distributions.
They are assumed to be fixed fractions of the up- and down-type quark
distributions respectively at the initial scale of the QCD evolution
(see section~\ref{param}).

The analysis is performed in the $\overline{MS}$ renormalisation
scheme using the DGLAP evolution equations~\cite{dglap} at
NLO~\cite{furmanski}. The structure function formulae given here are
thus replaced by integral convolutions of coefficient functions and PDFs.
An approach is used whereby all quarks are taken to be massless, including
the charm and bottom quarks, which provides an adequate
description of the parton distributions in the high $Q^2$ kinematic
range of the new data presented here. The bottom quark distribution, $xb$, 
is assumed to be zero for $Q^2<m_b^2$ where $m_b$ is the bottom quark mass.

Fits are performed to the measured cross sections calculating the
longitudinal structure functions to order $\alpha_s^2$ and assuming the
strong coupling constant to be equal to \amz$=0.1185$~\cite{pdg}. All
terms in eqs.~\ref{Snc1} and \ref{eq:cccross} are calculated, including
the weak corrections, $\Delta^{\pm,weak}_{NC, CC}$.  
The analysis uses an $x$ space program developed inside 
the H1 collaboration~\cite{qcdfit}, with cross checks performed 
using an independent program~\cite{qcdnum}.  
In the fit procedure, a $\chi^2$ function is minimised which is defined
in~\cite{h1lowq2}. The minimisation takes into account correlations of
data points caused by systematic uncertainties allowing the error
parameters (see table~\ref{qcdsys}), 
including the relative normalisation of the various data sets, 
to be determined by the fit. 
The fit to only H1 inclusive cross section
data, termed H1 PDF $2000$, uses the data sets as specified in
table~\ref{dataset}. The table additionally lists the BCDMS data used in
a further fit for comparison with the H1 PDF $2000$ fit.
\begin{table}[htb]
 \scriptsize
 \begin{center}
 \begin{tabular}{|ll|ll|rr|c|r|l|}
 \hline
 data set     & process   &  \multicolumn{2}{|c|}{$x$ range} & 
 \multicolumn{2}{|c|}{$Q^2$ range} &
$\delta^{\mathcal{L}}$ &    ref.      & \multicolumn{1}{|c|}{comment}   \\
              &       &                    &           &  (GeV$^2$)
       & (GeV$^2$)  &
  $(\%)$             &                    &           \\
 \hline\hline
 H1 minimum bias $97$ & $e^+p$ NC & $0.00008$  &  $0.02$        &  $1.5$    & 
  $12$   &  $1.7$     &~\cite{h1lowq2} &  $\sqrt{s}= 301{\rm\,GeV}$\\
 H1 low $Q^2$ $96-97$ & $e^+p$ NC & $0.000161$ &  $0.20$        &  $12$     &
  $150$  &  $1.7$     &~\cite{h1lowq2} &  $\sqrt{s}= 301{\rm\,GeV}$\\
 H1 high $Q^2$ $94-97$ & $e^+p$ NC &   $0.0032 $ &  $0.65$        &  $150$  &
  $30\,000$ &  $1.5$     &~\cite{h1hiq2}  &  $\sqrt{s}= 301{\rm\,GeV}$\\
 H1 high $Q^2$ $94-97$ & $e^+p$ CC &  $0.013  $ &  $0.40$        &  $300$   & 
  $15\,000$ &  $1.5$     &~\cite{h1hiq2}  &  $\sqrt{s}= 301{\rm\,GeV}$\\
 H1 high $Q^2$ $98-99$ & $e^-p$ NC &  $0.0032$  &  $0.65$        &  $150$   & 
  $30\,000$ &  $1.8$     &~\cite{h1elec}  &  $\sqrt{s}= 319{\rm\,GeV}$\\
 H1 high $Q^2$ $98-99$ & $e^-p$ CC &  $0.013  $ &  $0.40$        &  $300$   & 
  $15\,000$ &  $1.8$     &~\cite{h1elec}  &  $\sqrt{s}= 319{\rm\,GeV}$\\
 H1 high $Q^2$ $98-99$ & $e^-p$ NC &  $0.00131$ &  $0.0105$      &  $100$   & 
  $800$     &  $1.8$     & this rep.      &  
   $\sqrt{s}= 319{\rm\,GeV}$; high-$y$ data\\
 H1 high $Q^2$ $99-00$ & $e^+p$ NC &  $0.0032$  &  $0.65$        &  $150$   & 
  $30\,000$ &  $1.5$     & this rep.      &  
   $\sqrt{s}= 319{\rm\,GeV}$; incl. high-$y$ data\\
 H1 high $Q^2$ $99-00$ & $e^+p$ CC &  $0.013  $ &  $0.40$        &  $300$   & 
  $15\,000$ &  $1.5$     & this rep.      & $\sqrt{s}= 319{\rm\,GeV}$\\
 \hline
 BCDMS-p & $\mu p$ NC &  $0.07   $ &  $0.75$        &  $7.5$   & $230$ 
              &  $3.0$     & ~\cite{bcdms}  & require
  $y_{\mu}>0.3$ \\
 BCDMS-D & $\mu D$ NC &  $0.07   $ &  $0.75$        &  $7.5$   & $230$   
  &  $3.0$    & ~\cite{bcdms}  & require $y_{\mu}>0.3$ \\
 \hline
 \end{tabular}
 \end{center}
 \caption[RESULT]
 {\sl \label{dataset} 
 Data sets from H1 used in the H1 PDF $2000$ fit and from BCDMS
 $\mu$-proton and $\mu$-deuteron scattering used in the H1+BCDMS fit.
 As for the previous H1 QCD analysis~\cite{h1lowq2},
 the original BCDMS data are used at four different beam energies
 imposing the constraint  $y_{\mu}>0.3$. 
 The inelasticity $y_\mu$ was defined using BCDMS beam energies.
 The normalisation
 uncertainties of each data set ($\delta^\mathcal{L}$) are given as
 well as the kinematic ranges in $x$ and $Q^2$. The uncertainty 
 $\delta^\mathcal{L}$ includes a common error of $0.5\%$ for the H1
 data sets (see text). The nominal analysis and 
 high-$y$ analysis do not overlap in kinematic coverage
 (see section~\ref{ncmeas}).}
\end{table}

The correlated systematic uncertainties for the H1 cross section
measurements may be correlated across data sets as well as between 
data points, since they may arise from the same source.
They are thus not treated independently in the QCD
analysis presented here. The relationship between the error sources as
used in the fitting procedure is summarised in table~\ref{qcdsys} for
each of the eight correlated systematics considered. This leads to $18$
independent error parameters. In addition, all H1 quoted luminosity
uncertainties have a common contribution of $0.5\%$ arising from the
theoretical uncertainty on the Bethe-Heitler cross section. This common
contribution has been taken into account in the QCD analysis.
\begin{table}[htp]
 \begin{center}
 \begin{tabular}{|ll|cccccccc|}
 \hline
 data set & process & $\delta^\mathcal{L}$ &   $\delta^E$ & 
 $\delta^{\theta}$ &  $\delta^h$ &$\delta^N$&
 $\delta^B$& $\delta^V$& $\delta^S$ \\ \hline
 H1 minimum bias   $97$  &  $e^+p$ NC & $\mathcal{L}1$ & $E1$ &$\theta1$ & 
 $h1$ & $N1$ & $B1$ & $-$ & $-$ \\
 H1 low $Q^2$  $96-97$ &  $e^+p$ NC & $\mathcal{L}2$ & $E1$ &$\theta1$ & 
 $h1$ & $N1$ & $B1$ & $-$ & $-$ \\
 H1 high $Q^2$ $94-97$ &  $e^+p$ NC & $\mathcal{L}3$ & $E2$ &$\theta2$ & 
 $h2$ & $N1$ & $B2$ & $-$ & $-$ \\
 H1 high $Q^2$ $94-97$ &  $e^+p$ CC & $\mathcal{L}3$ & $-$  &   $-$    & 
 $h2$ & $N1$ & $B2$ & $V1$ & $-$ \\
 H1 high $Q^2$ $98-99$ &  $e^-p$ NC & $\mathcal{L}4$ & $E2$ &$\theta3$ & 
 $h2$ & $N1$ & $B2$ & $-$ & $S1$ \\
 H1 high $Q^2$ $98-99$ &  $e^-p$ CC & $\mathcal{L}4$ & $-$  &   $-$    & 
 $h2$ & $N1$ & $B2$ & $V2$ & $-$ \\
 H1 high $Q^2$ $99-00$ &  $e^+p$ NC & $\mathcal{L}5$ & $E2$ &$\theta3$ & 
 $h2$ & $N1$ & $B2$ & $-$ & $S1$ \\
 H1 high $Q^2$ $99-00$ &  $e^+p$ CC & $\mathcal{L}5$ & $-$  &   $-$    & 
 $h2$ & $N1$ & $B2$ & $V2$ & $-$ \\
 \hline                                                       
 \end{tabular}        
 \end{center}
 \caption 
 {\label{qcdsys} \sl Treatment of the correlated systematic error sources
    for the H1 data sets used in the fits.
    For each of the eight correlated systematic error sources, one or more 
    parameters are included in the QCD fit procedure. 
    The sources considered are due to
    the luminosity uncertainty ($\delta^{\mathcal{L}}$), the electron energy
    uncertainty ($\delta^E$), the electron polar
    angle measurement uncertainty ($\delta^{\theta}$), the hadronic energy
    uncertainty ($\delta^{h}$), the uncertainty due to noise subtraction
   ($\delta^{N}$), the photoproduction dominated background simulation error 
   ($\delta^{B}$), the uncertainty due to the cuts against photoproduction in
    the CC analysis ($\delta^{V}$) and the error on the charge symmetry
    in the high-$y$ analysis ($\delta^{S}$). 
    The table entries indicate the correlation of the
    error sources across the H1 data sets. For example, the uncertainty
    due to the noise subtraction is the same for all data sets leading
    to one common parameter in the fit ($N1$), whereas the electron
    energy uncertainty has two independently varying parameters ($E1$
    and $E2$) for the H1 NC data sets only.}
\end{table}

\subsection{Parameterisations \label{param}}
The initial parton distributions, $xP=xg,~xU,~xD,~x\bU,~x\bD$,
are parameterised at $Q^2 = Q^2_0$ in the following general form
\begin{equation}
 xP(x) = A_Px^{B_P}(1-x)^{C_P}
      [1+D_P x + E_P x^2 +F_P x^3+ G_P x^4].
 \label{eqpara}
\end{equation}
The QCD analysis requires choices to be made for the initial scale ($Q^2_0$)
and the minimum $Q^2$ of the data considered in the analysis ($Q^2_{min}$). 
Variations of both $Q^2_0$ and $Q^2_{min}$ are studied.
As in~\cite{h1lowq2} $Q^2_0$ is chosen to be
$4{\rm\,GeV}^2$ and $Q^2_{min}= 3.5{\rm\,GeV}^2$.
Reasonable variations of these choices are
considered as part of the model uncertainties on the parton
distributions (section~\ref{h1fit}).

The general ansatz, eq.~\ref{eqpara}, represents an
over-parameterisation of the data considered here.
The specific choice of these parameterisations is obtained
from saturation of the $\chi^2$: an additional parameter $D$, $E$, $F$ or 
$G$ is considered only when its introduction significantly improves the 
$\chi^2$. 
The appropriate number of parameters also depends on the data sets 
included in the fit. The H1 data requires less parameters
than the combined H1 and BCDMS data due to the precise BCDMS
proton and deuteron data in the large $x$ region, where the
cross section variations with $x$ are particularly strong.

The fit to the H1 data uses the following functional forms
\begin{eqnarray}\label{eqparh1b}
 xg(x)  &=& A_gx^{B_g}(1-x)^{C_g}   \,\,\,\,\cdot [1+D_{g}x] \nonumber\\
 xU(x)  &=& A_Ux^{B_U}(1-x)^{C_U}   \,\cdot  [1+D_{U}x+F_{U}x^3] \nonumber\\
 xD(x)  &=& A_Dx^{B_D}(1-x)^{C_D}             \cdot  [1+D_{D}x] \\
 x\bU(x)&=& A_{\bU}x^{B_{\bU}}(1-x)^{C_{\bU}} \nonumber\\
 x\bD(x) &=& A_{\bD}x^{B_{\bD}}(1-x)^{C_{\bD}}\,,\nonumber
\end{eqnarray}
in which the number of free parameters are further reduced using 
the constraints and assumptions detailed below.

The number of parameters required by the fit for the different parton 
distributions follows the expectation. A high $x$ term $E_gx^2$ is not
needed in the gluon distribution, since at large $x$
the scaling violations are due to gluon bremsstrahlung, i.e.\
are independent of the gluon distribution. The $xU$ and $xD$ distributions
require more parameters than the anti-quark distributions $x\bU$ and
$x\bD$ because the former are a superposition of valence and sea
quarks, in contrast to the latter. Due to the different electric charges,
$e_u^2=4 e_d^2$, and the $y$ dependence of the charged current cross
section, the data are much more sensitive to the up quark than to the
down quark distributions. Thus less parameters are needed for $xD$ than
for $xU$.

A number of relations between parameters can be introduced naturally
in this ansatz. At low $x$ the valence quark distributions are
expected to vanish and the sea quark and the anti-quark distributions 
can be assumed to be equal. 
Thus the low $x$ parameters $A_q$ and $B_q$ are
required to be the same for $xU,~x\bU$ and for $xD,~x\bD$.
In the absence
of deuteron data from HERA there is no distinction possible of the rise
towards low $x$ between $xU$ and $xD$.  Thus the corresponding $B$
parameters are required to be equal, i.e.\  
$B_U=B_D=B_{\overline{U}}=B_{\overline{D}}\equiv B_q$.  
Further constraints are the conventional momentum sum rule and 
the valence quark counting rules. 

The ansatz presented above allows the quark distributions
$xU,~xD,~x\bU,~x\bD$ to be determined. Further disentangling the individual
quark flavour contributions to the sea is possible only with additional
experimental information and/or assumptions.
Assuming that the strange and charm sea quark distributions $xs$ and $xc$ 
can be expressed as $x$-independent fractions $f_s$ and $f_c$ of $x\bD$ and 
$x\bU$ at the starting scale of $Q^2_0=4{\rm\,GeV}^2$ 
(see table~\ref{tabmodunc}), 
a further constraint is used in the fit: 
$A_{\bU}=A_{\bD}\cdot(1-f_s)/(1-f_c)$, which imposes that
$\bd/\bu\rightarrow 1$ as $x\rightarrow 0$.

The total number of free parameters of the five parton
distributions is thus equal to $10$ in the fit to the H1 data. The $\chi^2$ 
value is hardly improved by including any half integer power of 
$x$. The parametric form of eq.~\ref{eqparh1b} is also found 
starting from an alternative polynomial in $x^{k}$, which includes 
half integer powers up to $x^{5/2}$.
The addition of the large ${x}$ BCDMS $\mu p$ and $\mu D$ data leads to 
two additional terms, $G_U x^4$ and $F_D x^3$, in the polynomials.

\section{Results}
\label{results}

\subsection{NC and CC Cross Sections 
{\boldmath ${\rm d}\sigma/\rm{d}Q^2$}, {\boldmath ${\rm d}\sigma/\rm{d}x$} 
and {\boldmath $\sigma_{CC}^{tot}$}}
\label{integ}

\begin{figure}[htbp]
\setlength{\unitlength}{1 mm}
\begin{center}
\begin{picture}(160,180)(0,0)
\put(5,52) {\epsfig{file=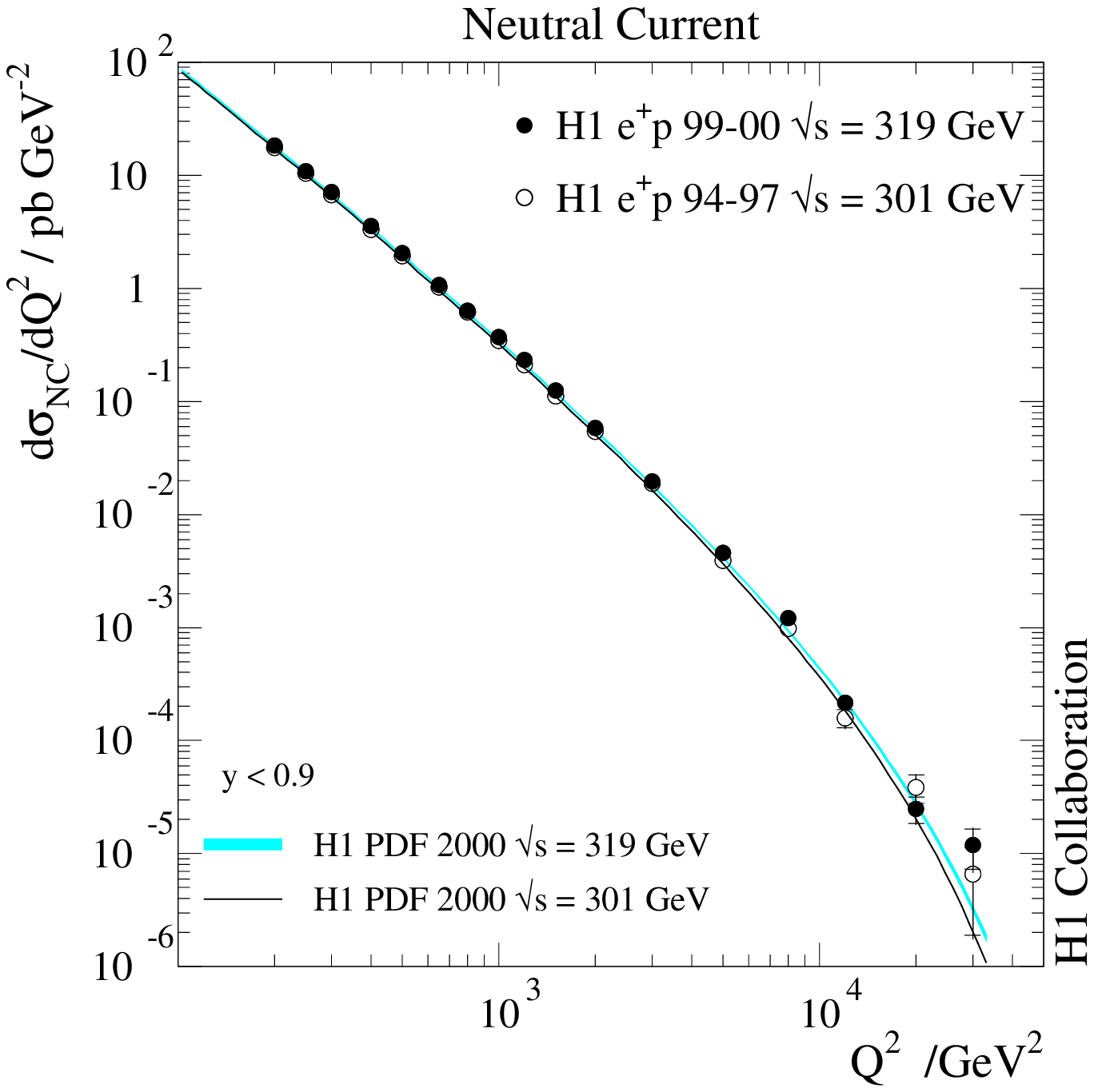,width=0.85\textwidth}}
\put(5,-12) {\epsfig{file=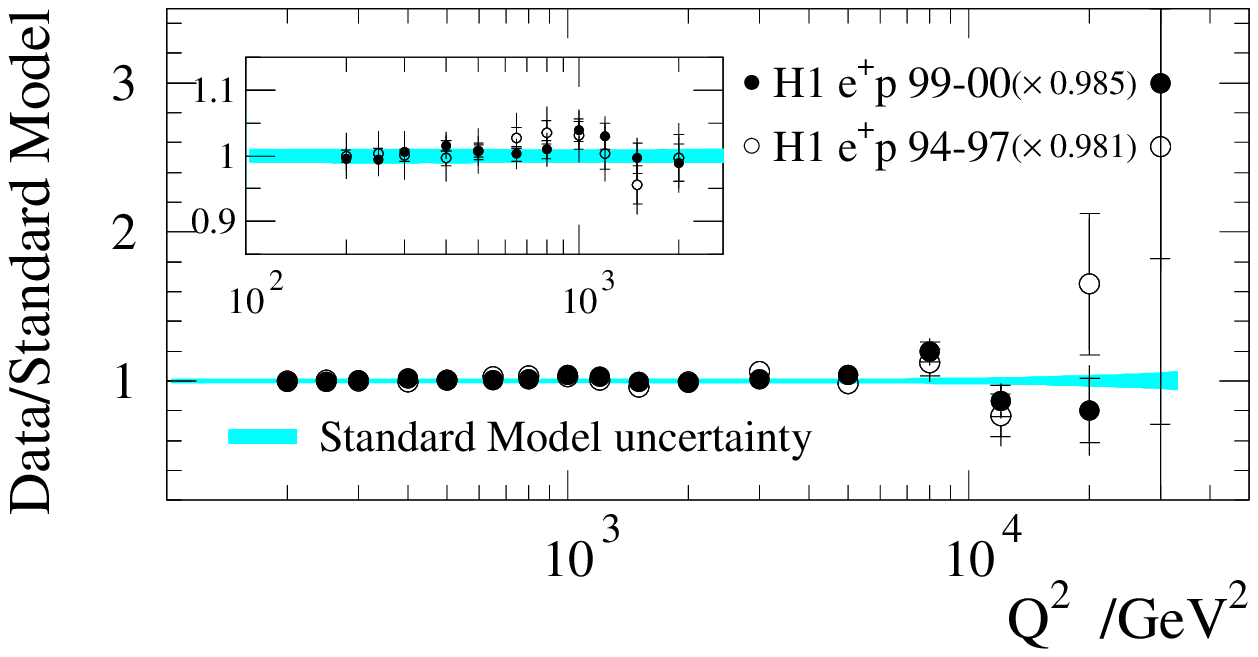,
 width=0.85\textwidth}}
\put(35,176){\bf (a)}
\put(35, 53){\bf (b)}
\end{picture}
\end{center}
\caption{\sl (a) The $Q^2$ dependence of the NC cross section
  ${\rm d}\sigma/{\rm d}Q^2$,
  shown for the new $e^+p$ (solid points) and previously published $94-97$
  $e^+p$ (open points) data. The error band and full curve represent
  the Standard Model expectations determined from 
  the H1 PDF $2000$ fit at $\sqrt{s}= 319{\rm\,GeV}$ and
  $\sqrt{s}= 301{\rm\,GeV}$, respectively.
  (b) The ratios of the $94-97$ and $99-00$ data to their corresponding
  Standard Model expectations, where the normalisation shifts as determined
  from the fit are applied to the data (see table~\ref{tab:h1fit}).
  The error band shows the Standard Model uncertainty for
  $\sqrt{s}= 319{\rm\,GeV}$ by adding in quadrature the experimental
  uncertainty as derived from the fit and the model uncertainty (see 
  section~\ref{h1fit}).
  In (a) and (b), the inner and outer error bars represent respectively
  the statistical and total errors. The luminosity uncertainty is not 
  included in the error bars.}
\label{dsdq2nc}

\end{figure}

The $e^+p$ single differential neutral current cross section
${\rm d}\sigma/\rm{ d}Q^2$ measured for $y<0.9$ is shown in
fig.~\ref{dsdq2nc}(a).
The data are compared with previous H1
$e^+p$ measurements made at $\sqrt{s}= 301{\rm\,GeV}$.  The new cross
sections are found to be higher than the measurement from $94-97$ as
expected due to the increase in centre-of-mass energy.  Both cross
sections, falling by over six orders of magnitude for the measured $Q^2$
region between $200{\rm\,GeV}^2$ and $30\,000{\rm\,GeV}^2$, are well described 
by the H1 PDF $2000$ fit. 
The error band represents the total uncertainty as derived from the QCD
analysis by adding in quadrature the experimental and model uncertainty.
The experimental uncertainty on the predicted cross sections is significantly
larger than the model uncertainty, which is discussed in section~\ref{h1fit}.
Fig.~\ref{dsdq2nc}(b) shows the ratios of the measurements to the
corresponding Standard Model expectation determined from the 
H1 PDF $2000$ fit. Note that in this lower figure the H1 data 
are scaled by the normalisation shift imposed by the QCD fit given in
table~\ref{tab:h1fit}. The new data are given in table~\ref{ncdq2}. 

\begin{figure}[htbp]
\setlength{\unitlength}{1 mm}
\begin{center}

\begin{picture}(160,180)(0,0)
\put(5,52) {\epsfig{file=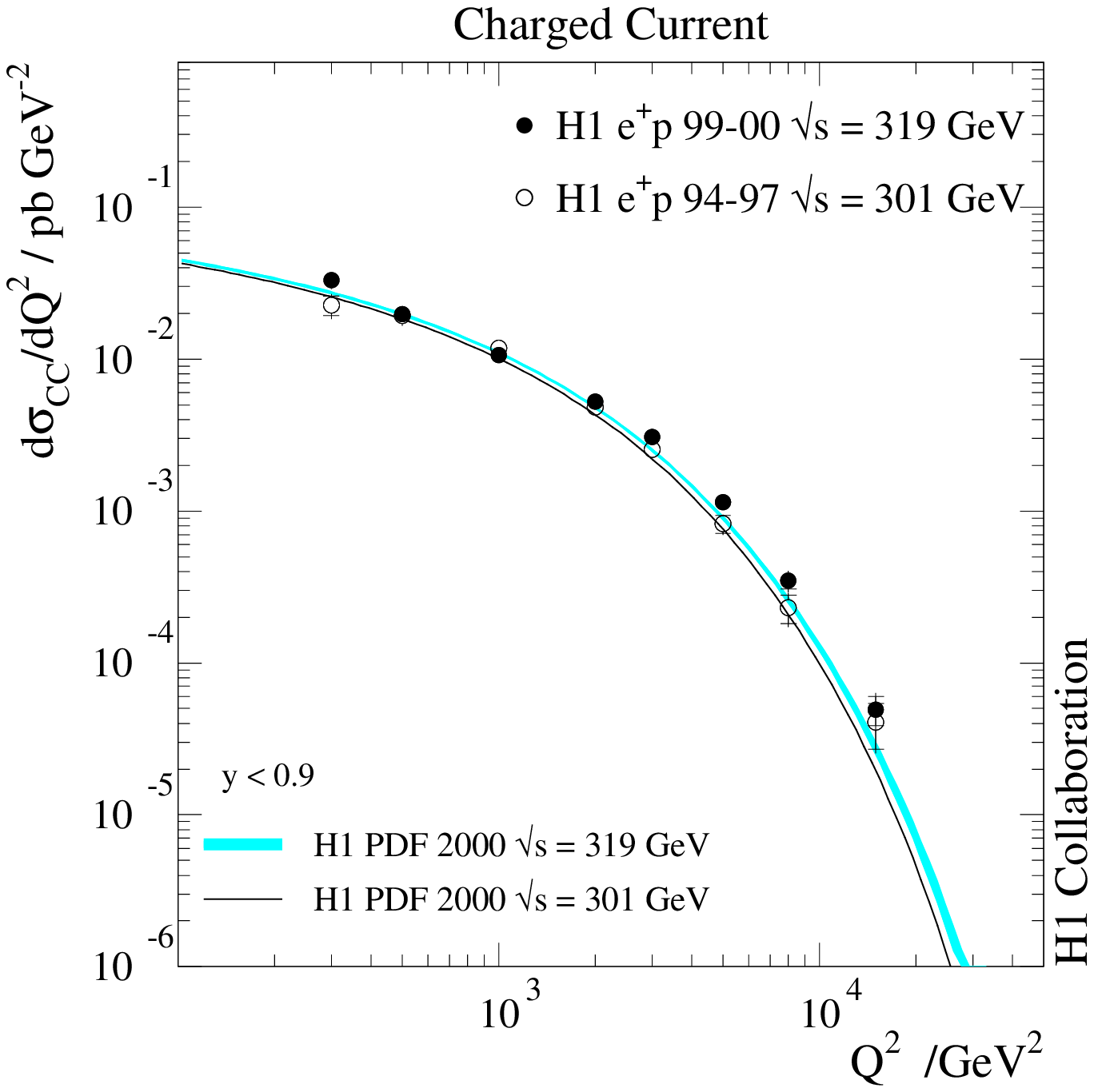,width=0.85\textwidth}}
\put(5,-12) {\epsfig{file=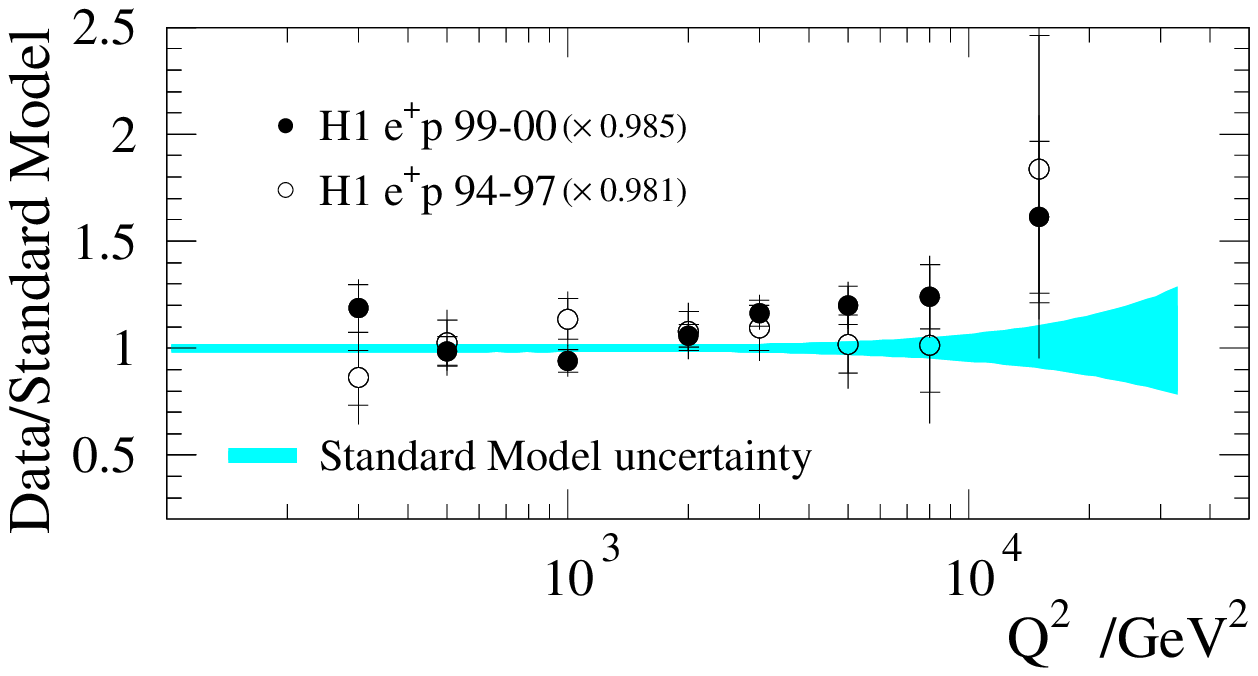,
 width=0.85\textwidth}}
\put(35,176){\bf (a)}
\put(35, 53){\bf (b)}
\end{picture}
\end{center}
\caption{\sl (a) The $Q^2$ dependence of the CC cross section 
  ${\rm d}\sigma/{\rm d}Q^2$,
  shown for the new $e^+p$ (solid points) and previously published
  $94-97$ $e^+p$ (open points) data. The error band and full curve represent
  the Standard Model expectations determined from 
  the H1 PDF $2000$ fit at $\sqrt{s}= 319{\rm\,GeV}$ and
  $\sqrt{s}= 301{\rm\,GeV}$, respectively.
  (b) The ratios of the $94-97$ and $99-00$ data to their corresponding
  Standard Model expectations, where the normalisation shifts as determined
  from the fit are applied to the data (see table~\ref{tab:h1fit}).
  The error band shows the Standard Model uncertainty for
  $\sqrt{s}= 319{\rm\,GeV}$. 
  The error bars and band are defined as for fig.~\ref{dsdq2nc}.}
\label{dsdq2cc}
\end{figure}

The $Q^2$ dependence of the charged current cross section from the
$99-00$ data is shown in fig.~\ref{dsdq2cc}(a).
For consistency with the NC cross sections, the data are presented
in the range $y<0.9$, after correction\footnote{The correction factors
  are given in table~\ref{ccdq2}.} for the kinematic cuts
$0.03<y<0.85$ and $P_{T,h}>12{\rm\,GeV}$ (section~\ref{ccmeas}).
The data are compared with the
previous measurement taken at lower centre-of-mass energy. The ratios of
data to expectations are shown in fig.~\ref{dsdq2cc}(b) together with the
Standard Model uncertainty. Again in this lower figure the
H1 data are scaled by the normalisation shift imposed by the QCD fit,
given in table~\ref{tab:h1fit}. The two data sets agree well with each other,
though the new data have a tendency to be higher than the fit result
at high $Q^2$. The data are listed in table~\ref{ccdq2}.

\begin{figure}[htbp]
\setlength{\unitlength}{1 mm}
\begin{center}
\begin{picture}(160,160)(0,0)
\put(0,0){\epsfig{file=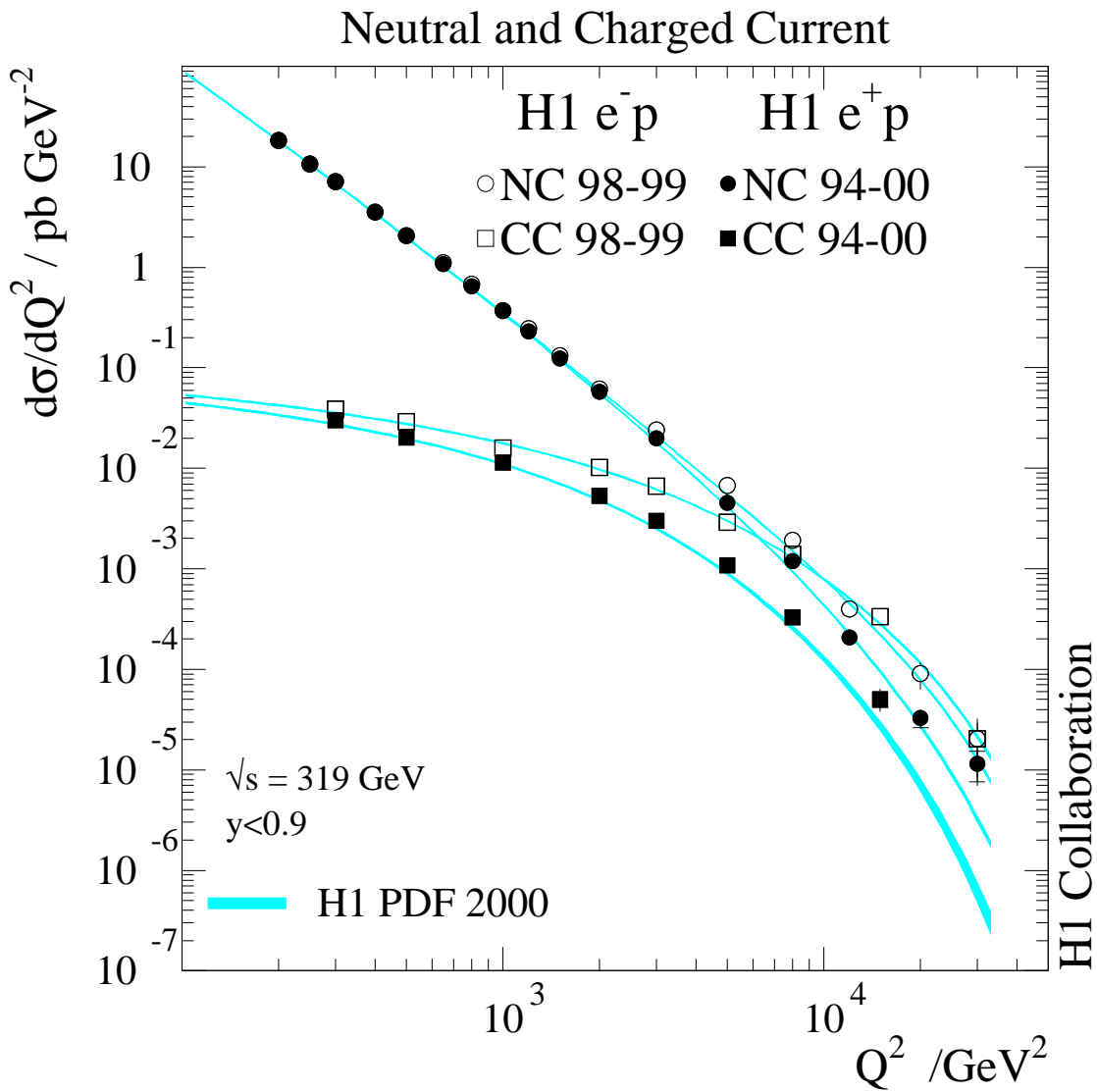,width=\textwidth}}
\end{picture}
\end{center}
\caption{\sl The $Q^2$ dependences of the NC (circles) and CC (squares)
  cross sections ${\rm d}\sigma/{\rm d}Q^2$,
  shown for the combined $94-00$ $e^+p$
  (solid points) and $98-99$ $e^-p$ (open points) data. The results are
  compared with the corresponding Standard Model expectations
  determined from the H1 PDF $2000$ fit. The error bars and bands are 
  defined as for fig.~\ref{dsdq2nc}.}
\label{fig:dsdq2nccc}
\end{figure}

Fig.~\ref{fig:dsdq2nccc} shows the $Q^2$ dependences of the NC and CC
cross sections representing the total $e^+p$ and $e^-p$ data sets
taken at HERA-I. The $e^+p$ data have been
combined after scaling the $94-97$ data to $\sqrt{s}= 319{\rm\,GeV}$, 
using the H1 PDF $2000$ fit and the procedure described in~\cite{zhang}.  
At low $Q^2$ the NC cross section exceeds the CC cross section by more than
two orders of magnitude. The sharp increase of the NC cross section 
with decreasing $Q^2$ is due to the dominating photon exchange 
cross section with the propagator term $\propto 1/Q^4$. 
In contrast the CC cross section ($\sim \left[M_W^2/(Q^2+M_W^2)\right]^2$)
approaches a constant at low $Q^2$. The CC and NC cross sections are 
of comparable size at $Q^2\sim 10^4{\rm\,GeV}^2$, where the photon and $Z^0$ 
exchange contributions to the NC cross sections are of similar size 
to those of $W^{\pm}$ exchange. 
These measurements thus illustrate unification of the electromagnetic
and the weak interactions in deep inelastic scattering. 
While the difference in NC cross sections between $e^+p$ and $e^-p$ 
scattering is due to $\gamma Z$ interference, 
the difference of CC cross sections arises from the difference between
the up and down quark distributions and the less favourable helicity factor 
in the $e^+p$ cross section (see eq.~\ref{Scc}).

\begin{figure}[htbp]
\setlength{\unitlength}{1 mm}
\begin{center}
\begin{picture}(90,72)(0,0)
\put(-40,-10){\epsfig{file=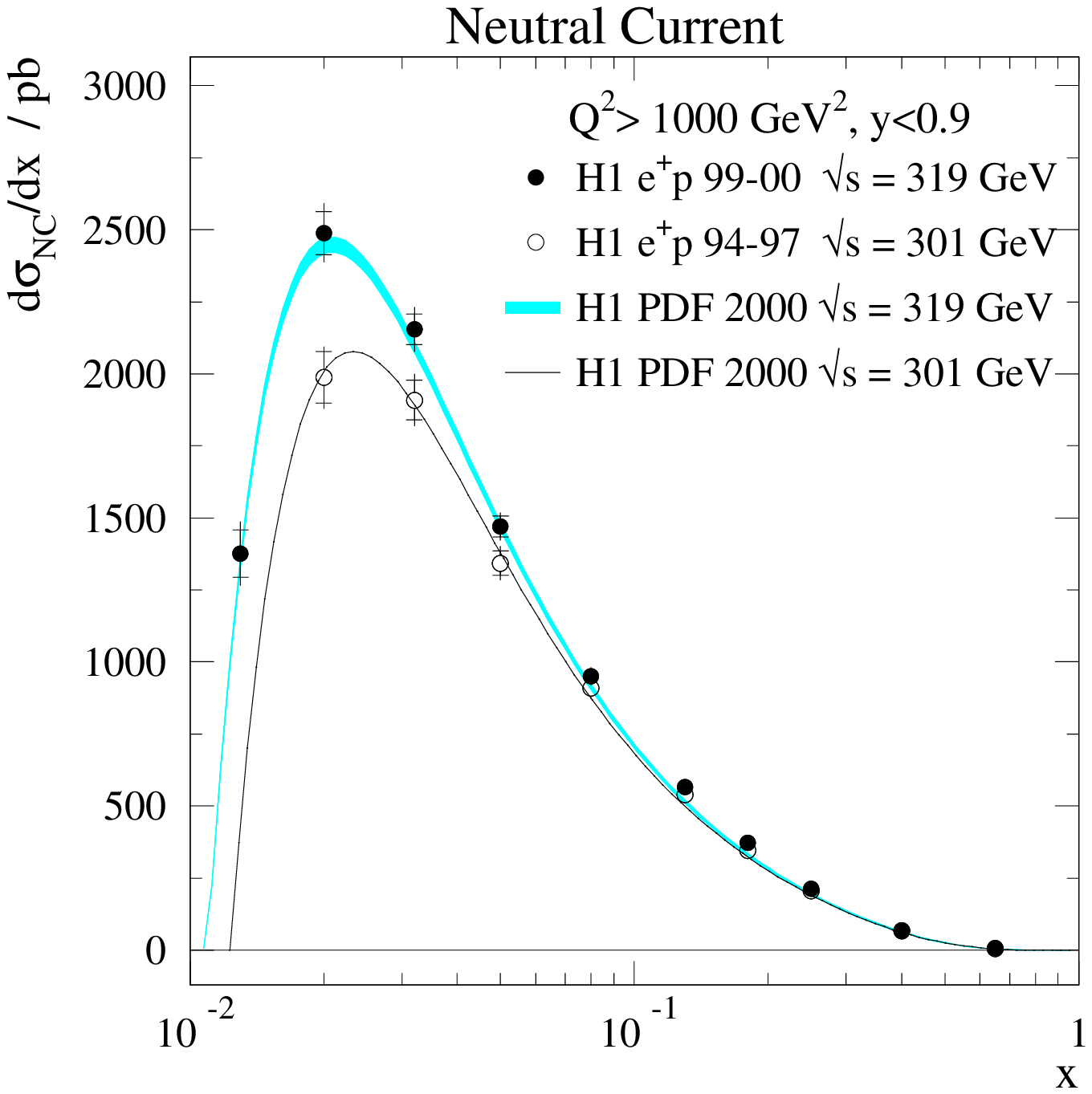,width=0.55\textwidth}}
\put( 40,-10){\epsfig{file=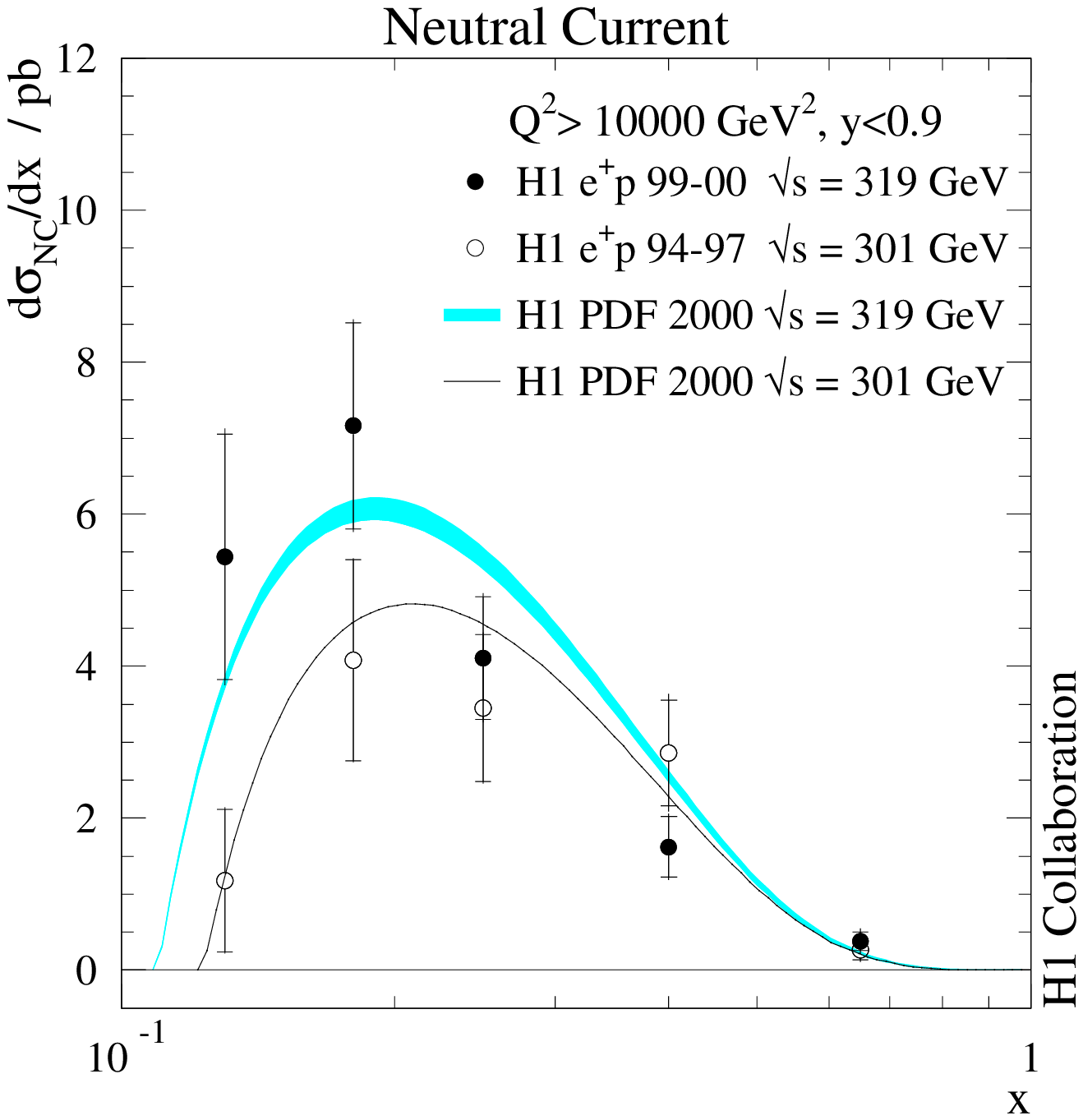,width=0.55\textwidth}}
\put(-20,62){\bf (a)}
\put( 60,62){\bf (b)}
\end{picture}
\end{center}
\caption{\sl The $x$ dependence of the NC cross section 
  ${\rm d}\sigma/{\rm d}x$ 
  for (a) $Q^2>1\,000{\rm\,GeV}^2$ and (b) $Q^2>10\,000{\rm\,GeV}^2$,
  shown for the new $e^+p$ (solid points) and previously published $94-97$
  $e^+p$ (open points) data. The error bands and full curves represent
  the corresponding Standard Model expectations determined from 
  the H1 PDF $2000$ fit at $\sqrt{s}= 319{\rm\,GeV}$ and 
  $\sqrt{s}= 301{\rm\,GeV}$, respectively. 
  The error bars and bands are defined as for fig.~\ref{dsdq2nc}. }
\label{fig:ncdsdx}
\end{figure}

\begin{figure}[htbp]
\setlength{\unitlength}{1 mm}
\begin{center}
\begin{picture}(90,72)(0,0)
\put(0,-10){\epsfig{file=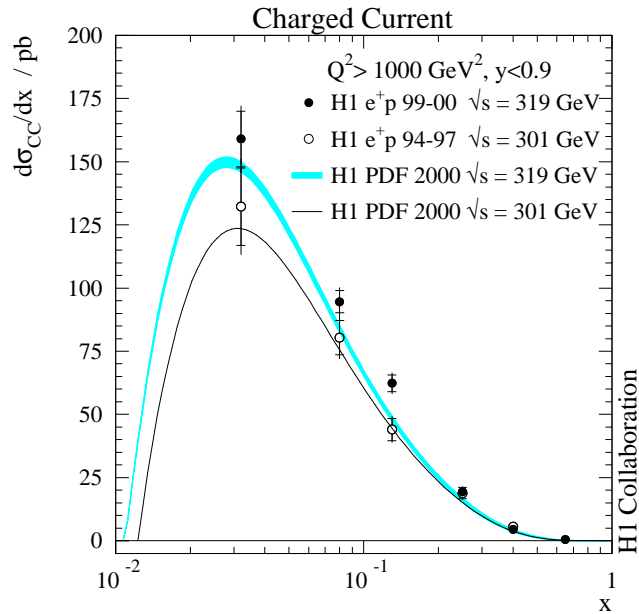,width=0.55\textwidth}}
\end{picture}
\end{center}
\caption{\sl The $x$ dependence of the CC cross section 
  ${\rm d}\sigma/{\rm d}x$ for $Q^2>1\,000{\rm\,GeV}^2$,
  shown for the new $e^+p$ (solid points) and
  previously published $94-97$ $e^+p$ (open points) data. 
  The error band and full curve represent the corresponding Standard Model 
  expectations determined from the H1 PDF $2000$ fit at 
  $\sqrt{s}= 319{\rm\,GeV}$ and $\sqrt{s}= 301{\rm\,GeV}$, respectively. 
  The error bars and band are defined as for fig.~\ref{dsdq2nc}.}
\label{fig:ccdsdx}
\end{figure}

The single differential cross sections ${\rm d}\sigma/{\rm d}x$ are
measured for $Q^2>1\,000{\rm\,GeV}^2$ for both NC and CC 
and also for $Q^2>10\,000{\rm\,GeV}^2$ in the NC case. 
The NC data are compared in fig.~\ref{fig:ncdsdx} with
the previous H1 $e^+p$ measurement at $\sqrt{s}=
301{\rm\,GeV}$ and the corresponding expectations from the fit. 
A similar comparison for the CC data is shown in fig.~\ref{fig:ccdsdx}.
Increases with $\sqrt{s}$ are observed in both the NC and the CC 
cross sections, in agreement with the expectations. 
The fall in the cross sections at low $x$ is due to the restriction $y<0.9$. 
The measurements are summarised in tables~\ref{ncdx1}-\ref{ccdx}.

The total CC cross section has been measured in the region
\mbox{$Q^2>1\,000{\rm\,GeV}^2$} and $y<0.9$ after applying 
a small correction factor of $1.03$ for the $y$ and $P_{T,h}$ cuts, 
determined from the H1 PDF $2000$ fit.
The result is
$$\sigma^{tot}_{CC}(e^+p;\sqrt{s}=319\,{\rm GeV}) = 19.19 \pm 
0.61(\rm stat.)  \pm 0.82(\rm syst.) \rm ~pb\,,$$
where the $1.5\%$ normalisation
uncertainty is included in the systematic error. This is to be compared
with the value from the H1 PDF $2000$ fit
$\sigma^{tot}_{CC}(e^+p)=16.76\pm 0.32{\rm\,pb}$.
The difference between the measurement and the fit is $2.3$ 
standard deviations assuming the correlation of uncertainties 
between measurement and fit is negligible. 
An unbiased theoretical expectation for
$\sigma^{tot}_{CC}(e^+p)$ may be obtained by repeating the H1 PDF $2000$
fit but excluding the new $99-00$ CC data, which leads to 
$16.66\pm 0.54{\rm\,pb}$.

Additionally, the analysis has been performed on the $94-97$ data set at 
the lower centre-of-mass energy, yielding
$$\sigma^{tot}_{CC}(e^+p;\sqrt{s}=301\,{\rm GeV}) = 16.41 \pm 
0.80(\rm stat.) \pm 0.90(\rm syst.) \rm ~pb\,.$$ 

This is to be compared with the cross section obtained from 
the H1 PDF $2000$ fit $\sigma^{tot}_{CC}(e^+p) = 14.76 \pm 0.30$\,pb. 
Assuming that the systematic uncertainties are fully correlated and 
part of the luminosity uncertainties are common (section~\ref{ansatz}),
the $94-97$ and $99-00$ measurements are combined~\cite{zhang} 
yielding a value of 
$$\sigma^{tot}_{CC}(e^+p;\sqrt{s}=319\,{\rm GeV}) = 18.99 \pm 
0.52(\rm stat.) \pm 0.81(\rm syst.) \rm ~pb\,.$$ 

\subsection{NC and CC Double Differential Cross Sections}

\begin{figure}[htbp]
\center
\epsfig{file=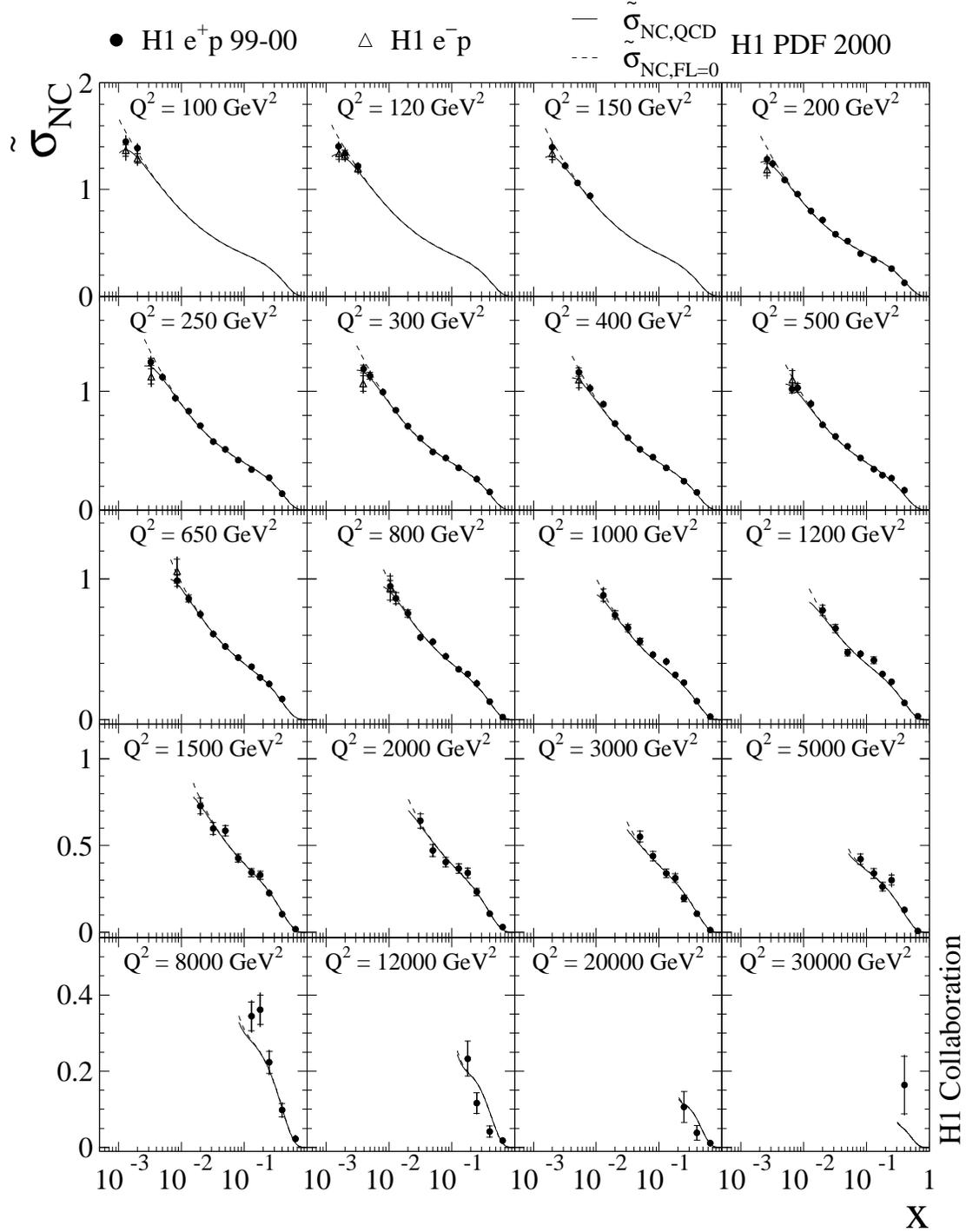,bbllx=20,bblly=45,bburx=570,
 bbury=805,width=14.5cm}
\caption{\sl The NC reduced cross section $\tilde{\sigma}_{NC}(x,Q^2)$,
  shown for the new $e^+p$ (solid points) and high-$y$ analysis 
  of the $98-99$ $e^-p$ (open triangles) data.
  The full (dashed) curves, labelled as $\tilde{\sigma}_{NC, QCD}$
  ($\tilde{\sigma}_{NC, FL=0}$), show the Standard Model expectations
  determined from the H1 PDF $2000$ fit by including (excluding) the $F_L$ 
  contribution in the reduced cross section.
  The error bars are defined as for fig.~\ref{dsdq2nc}.}
\label{nc_stamp} 
\end{figure}

The double differential NC reduced cross section, $\tilde{\sigma}_{NC}$
(defined in eq.~\ref{Rnc}), is shown in fig.~\ref{nc_stamp}
for both the nominal and high-$y$ $99-00$ $e^+p$ data.
In addition the new high-$y$ $98-99$ $e^-p$ data are presented.  
The data agree well with the
expectations of the H1 PDF $2000$ fit, which are also shown\footnote{The
  normalisation factors as determined by the QCD fit (table~\ref{tab:h1fit})
  are not applied to the data shown in the figure.}. The rise of the
Standard Model DIS cross section towards low $x$ (high $y$) 
departs from the monotonic behaviour of $F_2$ 
due to the contribution of the longitudinal structure function $F_L$. 
This allows $F_L$ to be determined in the high $y$ region
(section~\ref{sec:fl}).

\begin{figure}[htbp]
\setlength{\unitlength}{1 mm}
\begin{center}
\epsfig{file=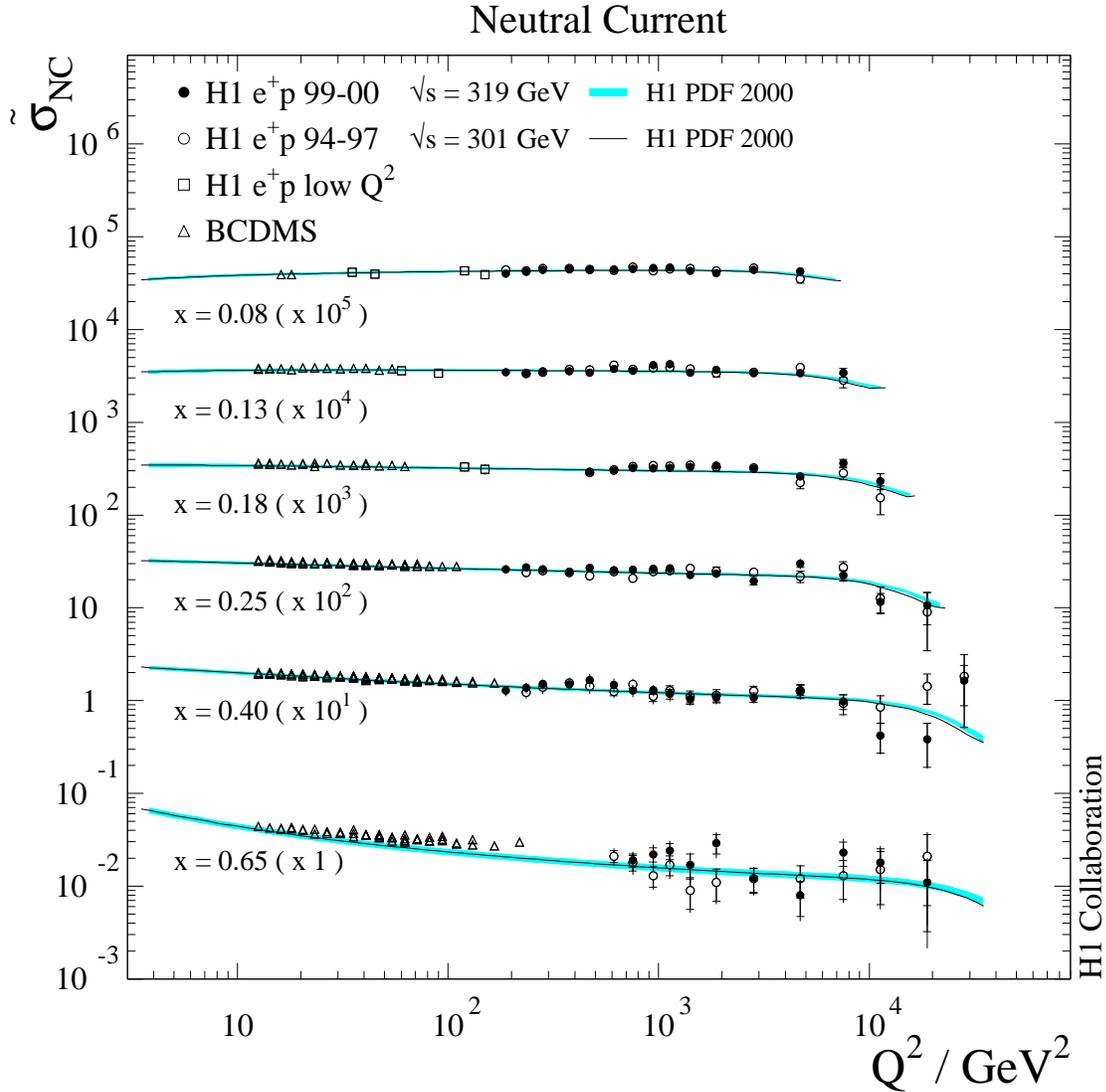,width=15cm}
\end{center}
\caption{\sl The NC reduced cross section $\tilde{\sigma}_{NC}(x,Q^2)$,
  shown for the new $e^+p$ (solid points) and previously published $94-97$
  $e^+p$ (open points) data. The results are compared with 
  the corresponding Standard Model
  expectations determined from the H1 PDF $2000$ fit at
  $\sqrt{s}=319{\rm\,GeV}$ (error bands) and $\sqrt{s}=301{\rm\,GeV}$ 
  (full curves), respectively.
  Also shown are data from H1 measured at lower $Q^2$ (open squares), 
  as well as from the fixed-target experiment BCDMS (open triangles).
  The BCDMS data are not used in the fit. The error bars and bands are 
  defined as for fig.~\ref{dsdq2nc}.}
\label{fig:nc_hixc} 
\end{figure}

In fig.~\ref{fig:nc_hixc} the $e^+p$ NC large $x$ cross section data at
$\sqrt{s}= 319{\rm\,GeV}$ are compared with the data obtained
previously~\cite{h1hiq2} at $\sqrt{s}= 301{\rm\,GeV}$.  The two data sets
are found to be in agreement with each other and with the H1 PDF $2000$
fit. Fig.~\ref{fig:nc_hixc} also shows the data from the recent H1
measurement at lower $Q^2$~\cite{h1lowq2} and the fixed-target data from
BCDMS~\cite{bcdms}. 
The fit description of the BCDMS data, which are not used in the fit,
is remarkably good except at very large $x=0.65$.
A similar observation has already been reported in~\cite{h1hiq2, h1lowq2}.
At the highest $Q^2$ a decrease of the cross section is expected 
due to the negative $\gamma Z$ interference in $e^+p$ scattering.

\begin{figure}[htbp]
\center 
\epsfig{file=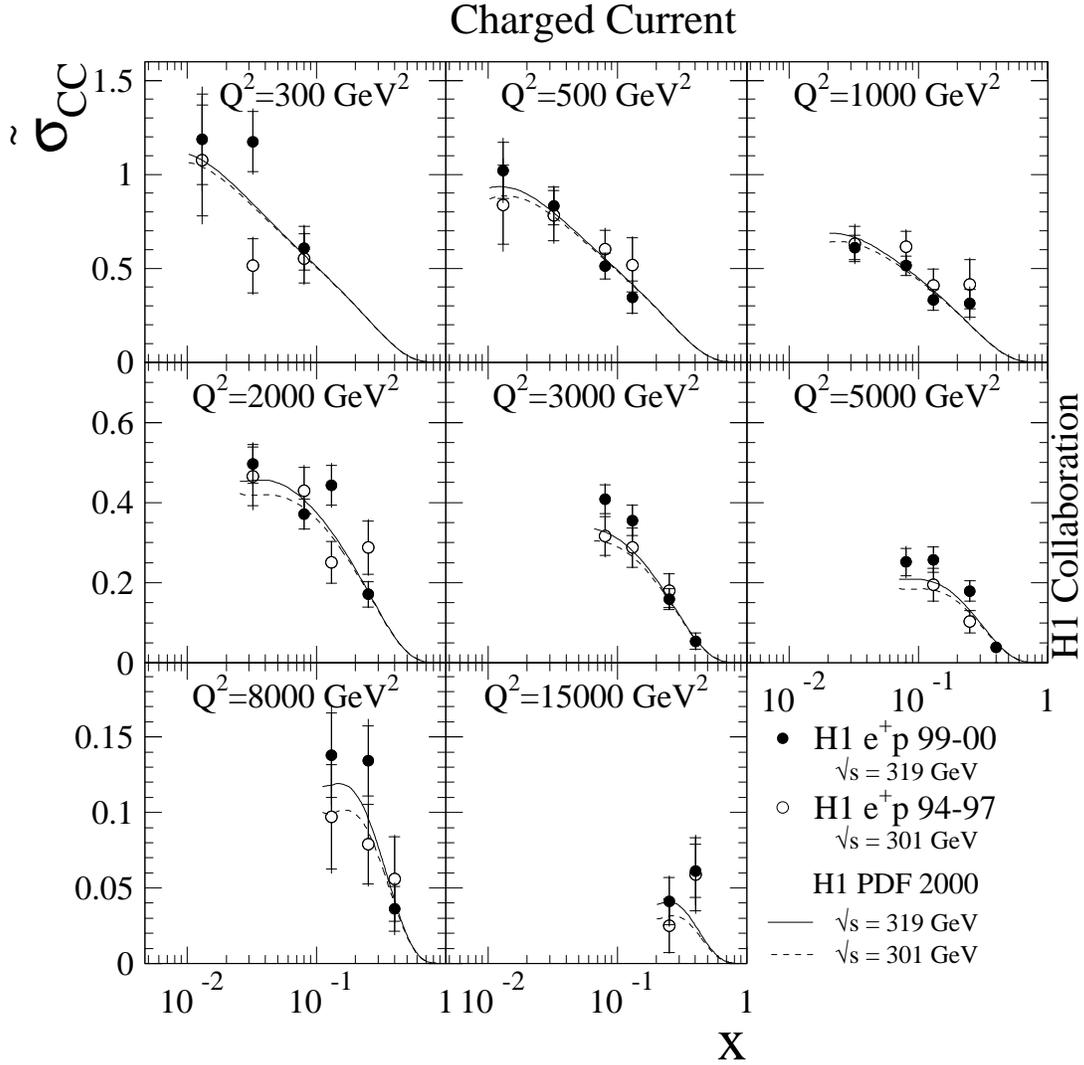,width=15cm}
\caption{\sl The CC reduced cross section $\tilde{\sigma}_{CC}(x,Q^2)$,
  shown for the new $e^+p$ (solid points) and previously published $94-97$
  $e^+p$ (open points) data. The results are compared with 
  the corresponding Standard Model
  expectations determined from the H1 PDF $2000$ fit at
  $\sqrt{s} = 319{\rm\,GeV}$ (full curves) and $\sqrt{s}=301{\rm\,GeV}$ 
  (dashed curves), respectively.
  The error bars are defined as for fig.~\ref{dsdq2nc}.}
\label{cc_stamp}
\end{figure}

In fig.~\ref{cc_stamp} the reduced CC cross section,
$\tilde{\sigma}_{CC}$ (defined in eq.~\ref{Rnc}), is shown for the new
data and the data taken at lower energy between $1994$ and $1997$. 
These data are found to be compatible with each other
considering the weak energy dependence of the reduced CC cross section. 
An extension of the $x$ range for $Q^2=3\,000{\rm\,GeV}^2$ and 
$5\,000{\rm\,GeV}^2$ is achieved 
due to the improved trigger efficiency.
The combined $94-00$ result is compared in fig.~\ref{cc_stampcomb}
with the expectation from the H1 PDF $2000$ fit. 
Also shown is the expected contribution 
of the $xD$ distribution, which dominates the cross section at large $x$.
The HERA $e^+p$ CC data can thus be used to constrain the $d$ quark 
distribution in the valence region.

All double differential measurements together with the contributions 
of each of the major systematic uncertainties are listed in 
tables~\ref{ncdxdq2}-\ref{ncextra}.

\subsection{Fit Results \label{h1fit}}
%
In this section, the results of the QCD analysis are presented.
The $\chi^2$ value for each data set is given in table~\ref{tab:h1fit}
as well as the optimised relative normalisation as determined from the fit.
The total $\chi^2$ value\footnote{In the calculation of the $\chi^2$, 
  the assumption is made that the uncorrelated errors among different data 
  points within one data set stay uncorrelated with the corresponding 
  data points from an independent data set.} 
per degree of freedom ($\chi^2$/ndf) is $540/(621-10)=0.88$. 
The NLO QCD fit requires the lowest $Q^2$ data (H1 minimum bias $97$, 
$Q^2\leq 12{\rm\,GeV}^2$) to be raised by $3.7\%$, corresponding to $2.3$
standard deviations in terms of the normalisation uncertainty excluding
the common error of $0.5\%$ (see section~\ref{ansatz}). 
In contrast all data for 
$Q^2\gtrsim 100{\rm\,GeV}^2$ are lowered, by at most $1.9\%$.
It can not yet be decided whether this behaviour is due to inadequacies 
in the theory (e.g.\ 
the missing higher order terms in $\ln Q^2$) or experimental effects.
\begin{table}[htb]
 \begin{center}
 \begin{tabular}{|ll|r|r|c|c|}
 \hline
 data set &  process & data points   & $\chi^2$ (unc.\ err.) & $\chi^2$ (corr.\ err.) & 
 normalisation \\
\hline
 H1 minimum bias    $97$    & $e^+p$ NC &  $ 45\hspace{6mm}$ & 
 $ 37.5\hspace{6mm}$ & $5.9$ & $1.037$ \\
 H1 low $Q^2$   $96-97$ & $e^+p$ NC &  $ 80\hspace{6mm}$ & 
 $ 71.2\hspace{6mm}$ & $1.3$ & $1.008$ \\
 H1 high $Q^2$  $94-97$ & $e^+p$ NC &  $130\hspace{6mm}$ & 
 $ 89.7\hspace{6mm}$ & $2.1$ & $0.981$ \\
 H1 high $Q^2$  $94-97$ & $e^+p$ CC &  $ 25\hspace{6mm}$ & 
 $ 18.0\hspace{6mm}$ & $0.4$ & $0.981$ \\
 H1 high $Q^2$  $98-99$ & $e^-p$ NC &  $139\hspace{6mm}$ & 
 $114.7\hspace{6mm}$ & $1.0$ & $0.991$ \\
 H1 high $Q^2$  $98-99$ & $e^-p$ CC &  $ 27\hspace{6mm}$ & 
 $ 19.5\hspace{6mm}$ & $0.7$ & $0.991$ \\
 H1 high $Q^2$  $99-00$ & $e^+p$ NC &  $147\hspace{6mm}$ & 
 $142.6\hspace{6mm}$ & $2.6$ & $0.985$ \\
 H1 high $Q^2$  $99-00$ & $e^+p$ CC &  $ 28\hspace{6mm}$ & 
 $ 32.4\hspace{6mm}$ & $0.9$ & $0.985$ \\
 \hline
 \multicolumn{2}{|c|}{Total}                   &  $621\hspace{6mm}$ & 
 \multicolumn{2}{|c|}{$540$} & $-$     \\
 \hline
 \end{tabular}
 \end{center}
 \caption[RESULT]
{\sl \label{tab:h1fit} For each data set used in the H1 PDF $2000$ fit, 
 the number of data points is shown, along with the $\chi^2$ 
 contribution determined using the uncorrelated errors (unc.\ err.).
 Each of the correlated error sources (see table~\ref{qcdsys})
 leads to an additional contribution~\cite{h1lowq2}, which is
 listed as $\chi^2$ (corr.\ err.).
 Also shown is the optimised normalisation of the data set as 
 determined by the fit. The H1 NC $98-99$ $e^-p$ and H1 NC $99-00$ $e^+p$ 
 data include the high-$y$ analyses.}
\end{table}

\begin{figure}[p] 
\setlength{\unitlength}{1 mm}
\begin{center}
\begin{picture}(160,150)(0,0)
\put(0,0) {\epsfig{file=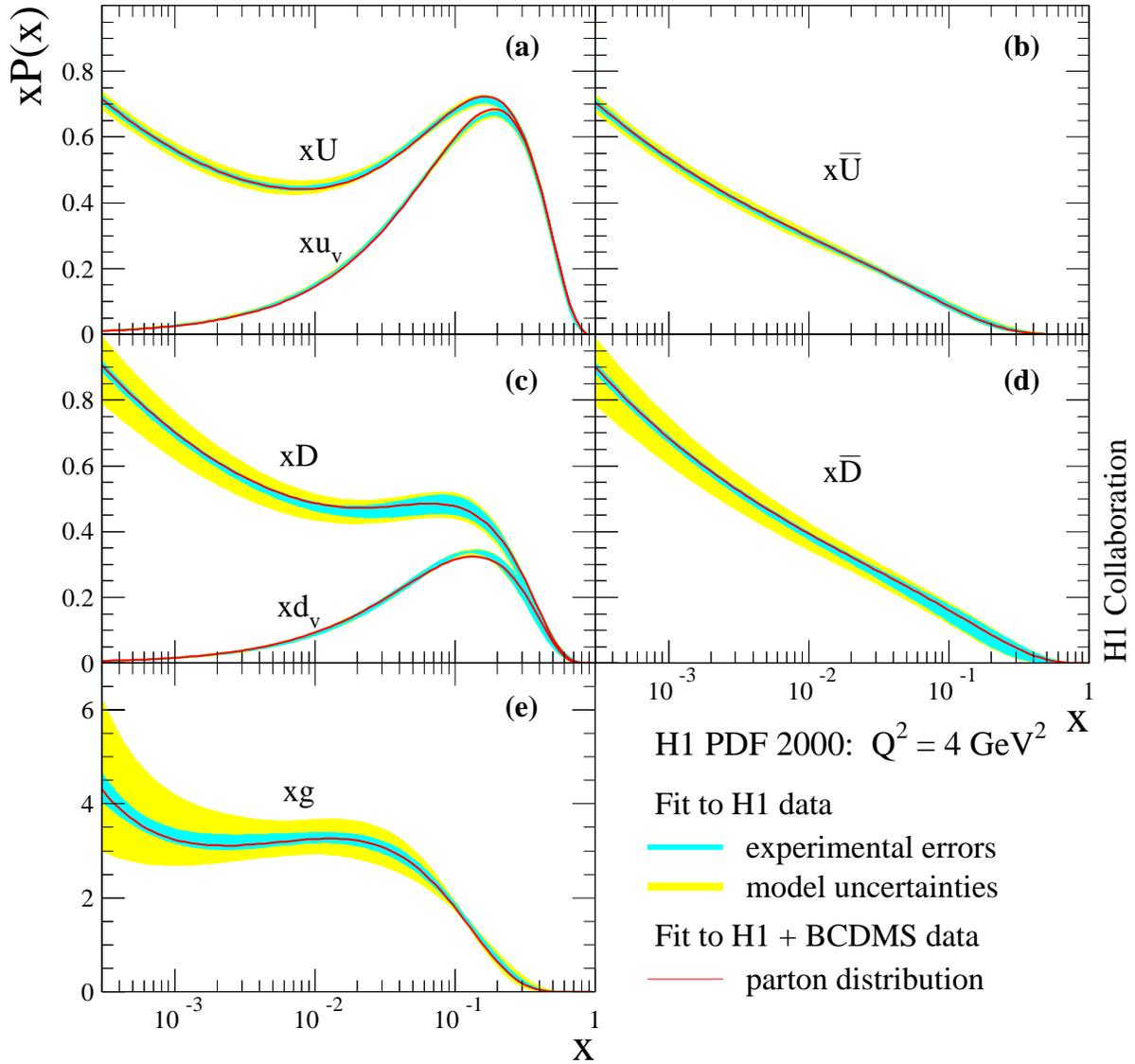,width=16cm}}
\put( 70,145){\bf (a)}
\put(140,145){\bf (b)}
\put( 70, 98){\bf (c)}
\put(140, 98){\bf (d)}
\put( 70, 52){\bf (e)}
\end{picture}
\end{center}
\caption{Parton distributions (a) $xU$, (b) $x\bU$, (c) $xD$, (d)
  $x\bD$ and (e) $xg$ as determined from the H1 PDF $2000$ fit to H1
  data only. The distributions are shown at the initial scale 
  $Q^2_0=4{\rm\,GeV}^2$. The inner error band represents the
  experimental uncertainty as determined from the fit. The outer error
  band shows the total uncertainty by adding in quadrature the experimental
  and model uncertainties (see text). The valence quark distributions
  $xu_v$ (a) and $xd_v$ (c) are also shown. For comparison, the parton
  distributions from the fit to H1 and BCDMS data are shown as the full
  curves.}
\label{figpdfhb}
\end{figure}

The parameters of the initial parton distributions are given in
table~\ref{tabpdfpar} (see also \cite{pdfsinfo}) and 
the distributions are shown in fig.~\ref{figpdfhb}. 
The inner error band describes the experimental
uncertainty, while the outer band represents the experimental and
model uncertainties added in quadrature. 

The experimental accuracy of the initial distributions is typically 
a few percent in the bulk of the phase space of the H1 data. This accuracy 
has negligible dependence on $Q^2$ but a strong dependence on $x$.
The best precision is achieved for the $xU$ quark distribution, which amounts
to about $1\%$ for $x= 0.01$ and reaches $3\%$ and 
$7\%$ at $x=0.4$ and $0.65$, respectively. The $xD$ quark distribution
is only determined with moderate precision as it is predominantly 
constrained by the CC $e^+p$ cross sections, which are still subject to
limited precision.
The corresponding uncertainties on $xD$ at the three quoted $x$ values are 
respectively $\sim 2\%$, $\sim 10\%$ and $\sim 30\%$.

These uncertainties reflect the kinematic dependence and size of 
the measurement errors. However the error size also depends significantly on 
the fit assumptions.
If, for example, the constraint between $A_{\overline{U}}$ and 
$A_{\overline{D}}$ on the low $x$ behaviour of the anti-quark
distributions is relaxed, the small uncertainty at low $x=0.01$ is much
increased to $\sim 6\%$ and $\sim 20\%$ respectively for $xU$ and $xD$.
The measurement of the low $x$ behaviour of up and down quarks and
their possible distinction requires electron-deuteron data to be taken 
at HERA.

The model parameter uncertainties on the PDFs are determined in a similar 
manner to~\cite{h1lowq2} and the sources of uncertainty are specified in
table~\ref{tabmodunc}. The model uncertainties are relatively small with 
respect to those from experimental sources except at small $x$ and 
low $Q^2$, where they reach $\sim 2\%$ and $\sim 6\%$ respectively 
for $xU$ and $xD$ at $x=0.01$ and $Q^2=10{\rm\,GeV}^2$.

Within the functional form considered (see eq.~\ref{eqpara}),
the parameterisation given in eq.~\ref{eqparh1b} is
found uniquely. Possible variations within the $\Delta\chi^2 \simeq 1$
region of the parameter space do not lead to noticeably different
distributions. Thus in this analysis no account is made of uncertainties
due to the choice of parameterisations.
A completely different ansatz, however,
may well lead to different initial distributions, as seen, for example,
in the complicated shape of $xg$ chosen in~\cite{cteq6}. 
The gluon distribution determined in this analysis is consistent
with the distribution obtained previously by H1~\cite{h1lowq2}
if the effects of the
different heavy flavour treatments are taken into account.
\begin{table}[htb]
 \begin{center}
 \begin{tabular}{|c|c|c|c|c|c|}
 \hline
 $P$    & $A$      & $B$      & $C$    & $D$     &  $F$   \\
 \hline
 $xg$   & $0.0183$ & $-0.872$ & $8.97$ & $3450.$ &        \\
 $xU$   & $0.112$  & $-0.227$ & $5.08$ & $48.0$  & $373.$ \\
 $xD$   & $0.142$  & $-0.227$ & $4.93$ & $23.5$  &        \\
 $x\bU$ & $0.112$  & $-0.227$ & $7.28$ &         &        \\
 $x\bD$ & $0.142$  & $-0.227$ & $4.36$ &         &        \\
 \hline
 \end{tabular}
 \end{center}
 \caption[RESULT]
 {\sl \label{tabpdfpar} Parameters of the H1 PDF $2000$ fit to the H1 data 
 alone for the initial distributions at $Q_0^2=4{\rm\,GeV}^2$. 
 Equal parameter values reflect the constraints imposed by the fit 
 (see section~\ref{param}). The uncertainties and their correlations are 
 available in~\cite{pdfsinfo}.}
\end{table}
\begin{table}[htb]
 \begin{center}
  \begin{tabular}{|ll|c|c|} \hline
   \multicolumn{2}{|c|}{source of  uncertainty} &  central value & variation 
 \\ \hline
   $ Q^2_{min}$       & (\gv)  & $ 3.5   $  &  $2.0\:-\:5.0$       \\
   $ Q^2_{0} $        & (\gv)  & $ 4.0   $  &  $2.0\:-\:6.0$       \\
   $\alpha_s(M_Z^2)$  &        & $ 0.1185$  &  $0.1165\:-\:0.1205$ \\
   \multicolumn{2}{|l|}{$f_s$, strange fraction of $x\bD$} & $ 0.33  $  &  
    $0.25\:-\:0.40$     \\ 
   \multicolumn{2}{|l|}{$f_c$, charm  fraction of $x\bU$}  & $ 0.15  $  &  
    $0.10\:-\:0.20$     \\ 
   $ m_c$             & (GeV) & $ 1.4   $  &  $1.2\:-\:1.6$       \\ 
   $ m_b$             & (GeV) & $ 4.5   $  &  $4.0\:-\:5.0$       \\
   \hline
    \end{tabular}
    \caption{ \sl Model uncertainties considered 
     in the QCD analysis.}
    \label{tabmodunc}
  \end{center}
\end{table}

The full curve in fig.~\ref{figpdfhb} is the result~\cite{hbfit00} 
of the fit to H1 and BCDMS data, which gives a
$\chi^2$/ndf$=883/(1014-12)=0.88$.
Excellent agreement of the PDFs between the two fits is observed.
For large $x$, the high $Q^2$ data of H1 allow 
distinction between up and down flavours yielding results compatible 
with those from BCDMS proton and deuteron data. 
At low $x$ only HERA data are available and thus
the two fits are forced to be in agreement, apart from possible small
effects due to sum rules.

\begin{figure}[p] 
\setlength{\unitlength}{1 mm}
\begin{center}
\begin{picture}(160,150)(0,0)
\put(0,0) {\epsfig{file=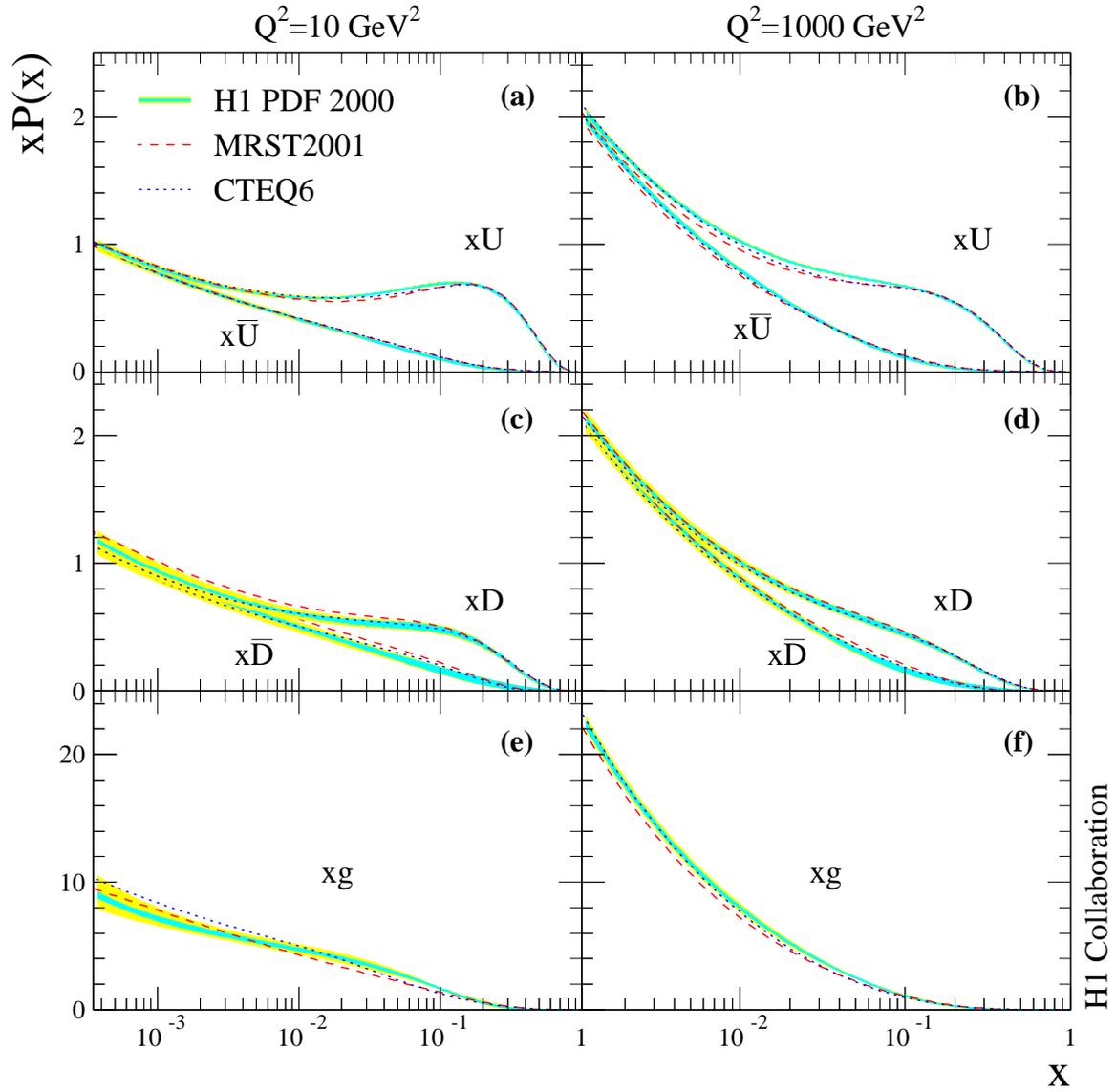,width=16cm}}
\put( 70,142){\bf (a)}
\put(140,142){\bf (b)}
\put( 70, 97){\bf (c)}
\put(140, 97){\bf (d)}
\put( 70, 52){\bf (e)}
\put(140, 52){\bf (f)}
\end{picture}
\end{center}
\caption{Parton distributions (a, b) $xU$ and $x\bU$, (c, d) $xD$
  and $x\bD$, and (e, f) $xg$ as determined from
  the H1 PDF $2000$ fit to H1 data only. The distributions are shown at
  $Q^2=10{\rm\,GeV}^2$ (a, c, e) and at $Q^2=1\,000{\rm\,GeV}^2$ 
  (b, d, f).
  The error bands are defined as for fig.~\ref{figpdfhb}. 
  For comparison, recent results from the MRST and CTEQ 
  groups are also shown.}
\label{figpdfglob}
\end{figure}

The PDFs from the H1 PDF $2000$ fit are further compared in 
fig.~\ref{figpdfglob} with recent results from the MRST~\cite{mrst2001} 
and CTEQ~\cite{cteq6} groups for two values of $Q^2$ at $10{\rm\,GeV}^2$ and
$1\,000{\rm\,GeV}^2$. The H1 PDF $2000$ fit is in remarkable agreement with
the MRST and in particular the CTEQ analyses, given the many differences 
in terms of the data sets used, the assumptions made and the treatment of 
heavy flavours. 

\subsection{\boldmath Extraction of the Proton Structure Function $F_2$}
\label{sec:f2}
The NC structure function term $\phi_{NC}$ is obtained from the
measured NC double differential cross section according to eq.~\ref{Snc1}.
It is dominated by the structure function $F_2$ in most of the kinematic range.
The structure function $F_2$ is extracted using
\begin{equation}
 \label{f2corr}
 F_2=\frac{\phi_{NC}}{Y_+}(1+\Delta_{F_2}+\Delta_{F_3}+\Delta_{F_L})^{-1}\,.
\end{equation}
Here the correction terms $\Delta_{F_2}$ and $\Delta_{F_3}$ account for
the effects of $Z^0$ exchange on $\Ftwo$ and $x\Fz$ 
(eqs.~\ref{eq:phinc}-\ref{f3p}) and
$\Delta_{F_L}$ originates from the longitudinal structure function $\FL$.
These corrections, shown in table~\ref{ncdxdq2}, are
determined using the H1 PDF $2000$ fit (see section~\ref{qcdana}).  
At high $y$ and $Q^2< 1\,500{\rm\,GeV}^2$, $\Delta_{F_L}$ is sizeable and 
the extraction of $F_2$ in this $Q^2$ region is thus restricted to 
the kinematic range $y<0.6$. 
It is extended to higher $y$ at larger $Q^2$ ($\geq 2\,000{\rm\,GeV}^2$)
where the predicted contribution of $\tilde{F_L}$ is small.

\begin{figure}[htbp]
\center 
\epsfig{file=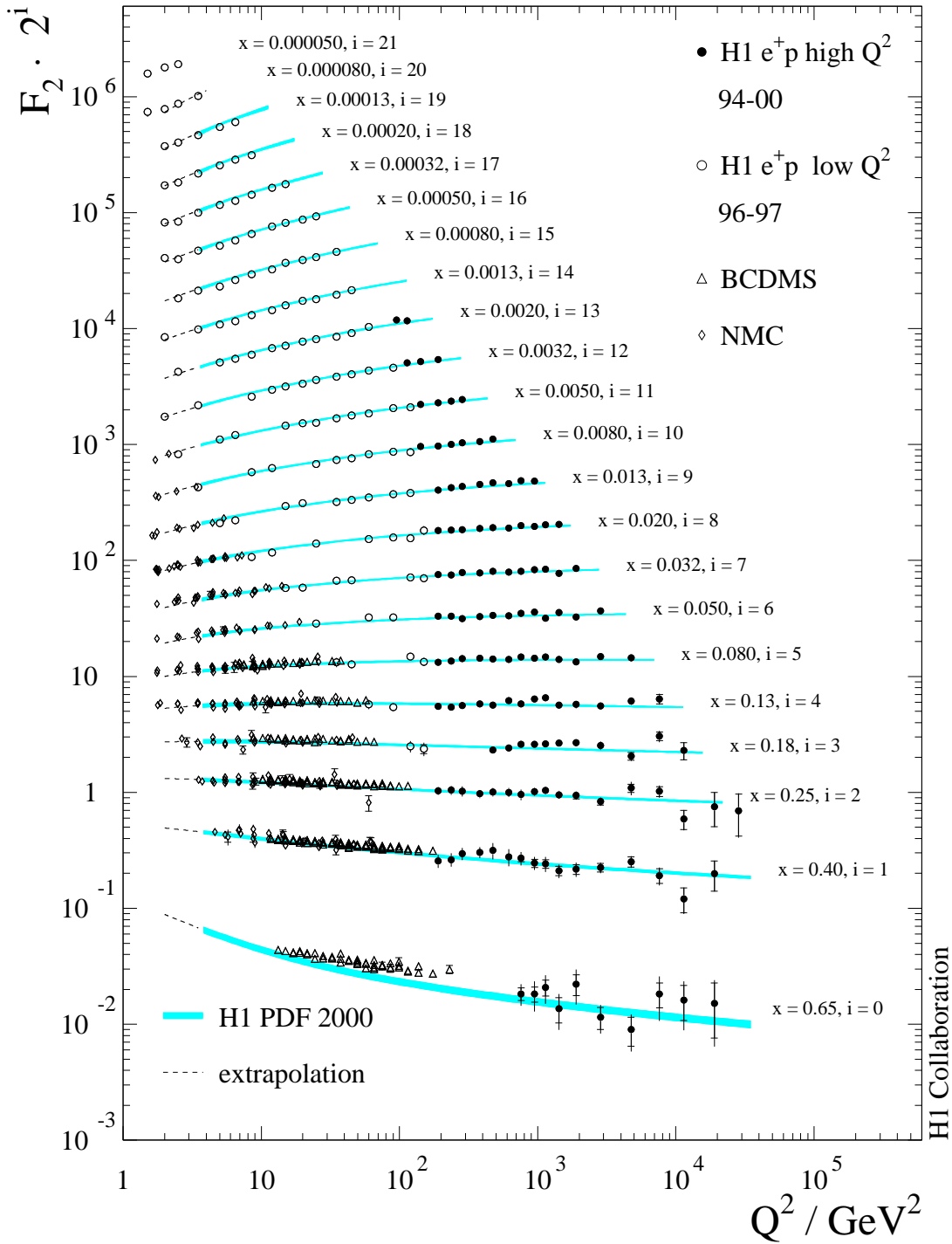,width=0.925\textwidth}
\caption{\sl The proton structure function $F_2$ shown for 
  the combined $94-00$ $e^+p$ (solid points) and previously published
  low $Q^2$ (open circles) data. The results are compared with 
  the corresponding Standard Model
  expectation determined from the H1 PDF $2000$ fit (error bands).
  The dashed curves show the backward extrapolation of the fit to 
  $Q^2<Q^2_{min}$. 
  Also shown are the $F_2$ data from BCDMS and NMC,
  which are not used in the fit. The error bars and bands are defined as for 
  fig.~\ref{dsdq2nc}.}
\label{f2_plot}
\end{figure}

The extracted $F_2$ using the $99-00$ data is presented in 
table~\ref{ncdxdq2}. Fig.~\ref{f2_plot} shows the $F_2$ data using 
the combined $94-97$ and $99-00$ high $Q^2$ $e^+p$ data sets.
Also shown in the figure are the recent H1 $F_2$ data at lower 
$Q^2$~\cite{h1lowq2} and structure function data from
BCDMS~\cite{bcdms} and NMC~\cite{nmc}. 
The full H1 data cover a range of four orders of magnitude in $x$ and $Q^2$. 
The H1 PDF $2000$ fit provides a good description of the data over 
the whole region except for the BCDMS data at $x=0.65$, as
seen in fig.~\ref{fig:nc_hixc}.
The fit also gives a good description of the scaling violations observed 
in the measurements. 

\subsection{\boldmath Determination of the Longitudinal Structure Function 
 $F_L$}
\label{sec:fl}
The structure function term $\phi_{NC}$ is used to determine $F_L$ 
at $y>0.63$ and $Q^2 < 890{\rm\,GeV}^2$.
For statistical reasons, the measured cross sections in two neighbouring
$Q^2$ bins are combined, assuming that the systematic uncertainties
are fully correlated.
The longitudinal structure function is then determined using the formula
\begin{equation}
 \tilde{F}_L = \frac{1}{y^2}\left[ Y_+\Ftwo \mp Y_-x\tilde{F}_3
      -\phi^\pm_{NC}\right],
 \label{eqfl:ew}
\end{equation}
for $e^\pm p$ scattering which, neglecting the small electroweak
contributions in the region of this extraction, reduces to the expression
\begin{equation}
 F_L = \frac{1}{y^2} \left[Y_+ F_2 - \phi_{NC}^\pm \right].
 \label{eqfl}
\end{equation}
The extraction of $F_L$ relies upon the extrapolation of the fit result
for $F_2$ into the high $y$ region, that is, to larger
$Q^2$ for given $x$.
In order to avoid a possible influence of the high $y$ data region on this
calculation, a dedicated NLO QCD fit (H1 Low $y$ fit) is performed to 
H1 data with $y<0.35$ only and the results
are extrapolated using the DGLAP evolution equations. This method was
introduced in~\cite{flpaper}. 

Apart from the $y$ range restriction, the H1 Low $y$ fit follows 
the same procedure as described in section~\ref{qcdana}. It results in a
$\chi^2/$ndf$=417/(455-10)=0.94$ and agrees very well with 
the H1 PDF $2000$ fit over the full $y$ range. The normalisation shifts
of the data sets used are found to be within $1\%$ of those from the H1
PDF $2000$ fit.

In the extraction of the longitudinal structure function, the
experimental cross sections are slightly modified 
using the results of the H1 Low $y$ fit for the renormalisation and 
small shifts from the correlated uncertainties common to the low $y$
and the high $y$ region.
The combined HERA-I measurements of the structure function term $\phi_{NC}$ 
and the extracted values of $F_L$ are listed in table~\ref{fltab}.
The statistical precision is due directly to the cross section measurement
at high $y$. The systematic uncertainties arise from the measurement
errors at high $y$ and the model uncertainties related to the extrapolation 
of $F_2$ from the low $y$ to the high $y$ region.
The correlations in the systematic uncertainties between low and high $y$
are taken into account.

\begin{figure}[htbp]
\center 
\epsfig{file=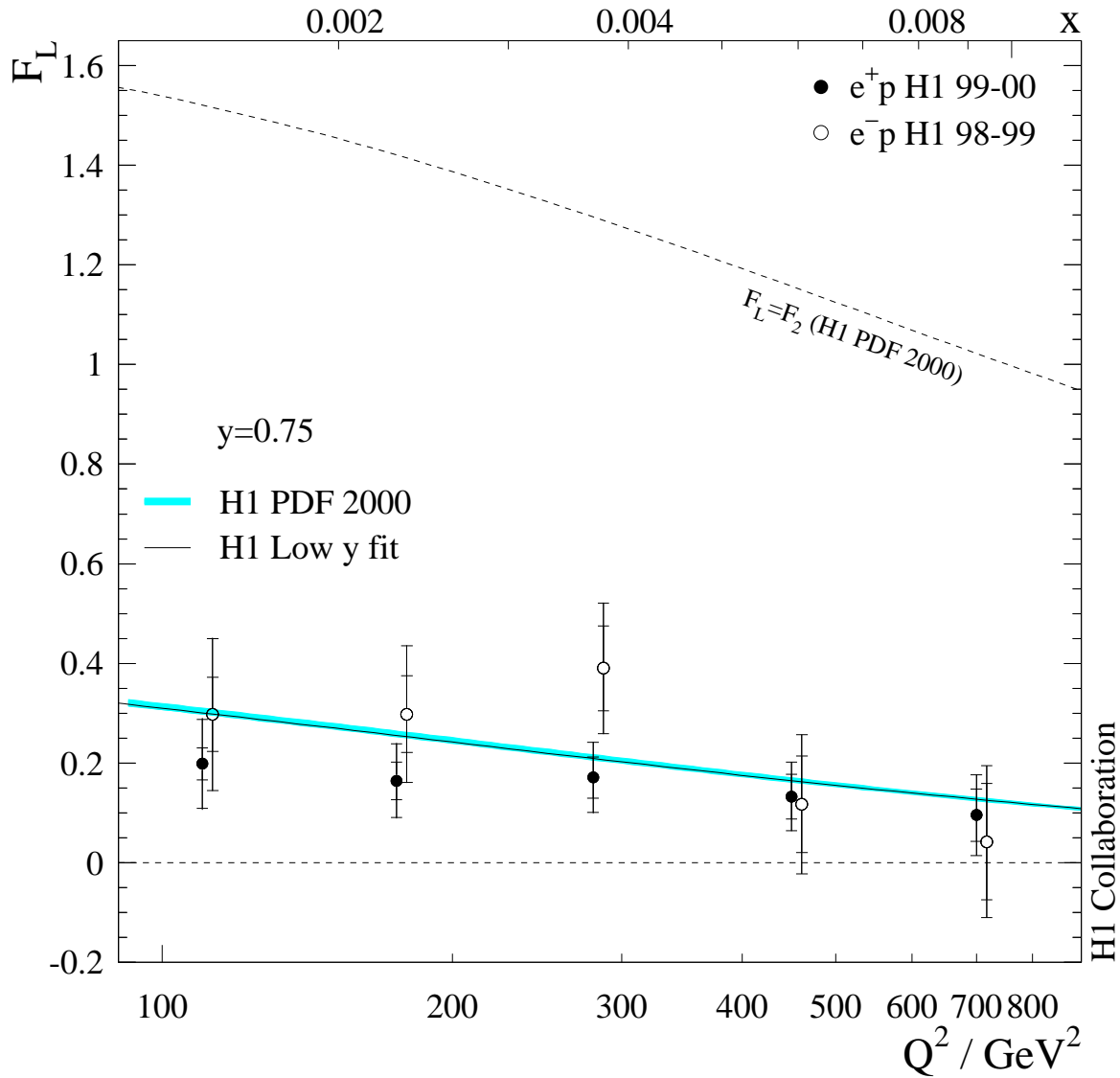,width=\textwidth}
\caption{\sl Determination of $F_L$ shown for the high $y$ $e^-p$ and $e^+p$ 
  data at fixed $y=0.75$ as a function of $Q^2$ (lower scale), or
  equivalently $x$ (upper scale). The inner error bar represents the
  statistical error, the intermediate error bar shows the statistical
  and systematic errors added in quadrature and the outer error bar also
  includes the uncertainty arising from the extrapolation of $F_2$. 
  The error band, defined as for fig.~\ref{dsdq2nc}, shows the 
  expectation for $F_L$ and its uncertainty, determined from 
  the H1 PDF $2000$ fit.
  The full curve shows the expectation for $F_L$ determined from 
  the H1 Low-$y$ fit. The upper and lower dashed curves are  
  the maximum and minimum allowed values for $F_L$.}
\label{fl_plot}
\end{figure}

In fig.~\ref{fl_plot} the determinations of $F_L$ at high $Q^2$
are shown for both the $e^+p$ and the $e^-p$ data sets. 
The results from both data sets are mutually consistent and are 
in agreement with the H1 Low~$y$ fit prediction for $F_L$, based on
the gluon distribution derived from the scaling
violations of $F_2$ at larger $x$. The extreme values allowed for $F_L$ 
($F_L=0$ and $F_L=F_2$) are clearly excluded by the data.
A model independent measurement of $F_L$ and the $x$ dependence can, however, 
only be achieved with reduced beam energies at HERA.

\subsection{\boldmath Measurement of the Generalised Structure Function
$x\tilde{F}_3$}
\label{sec:xf3}
At high $Q^2$, the NC cross section in $e^+p$ scattering has been observed
to be significantly smaller than that in $e^-p$ scattering~\cite{h1elec},
confirming the Standard Model expectation of $\gamma Z$ interference.
These earlier H1 data were used to obtain a first measurement of 
the generalised structure function $x\tilde{F}_3$ in the kinematic range 
$0.02\leq x\leq 0.65$ and 
$1\,500{\rm\,GeV}^2\leq Q^2\leq 12\,000{\rm\,GeV}^2$~\cite{h1elec}.
A similar measurement has been reported recently by 
ZEUS~\cite{zeusxf3}.

\begin{figure}[htbp]
\begin{center}
\begin{picture}(50,195)
\put(-40,90){\epsfig{file=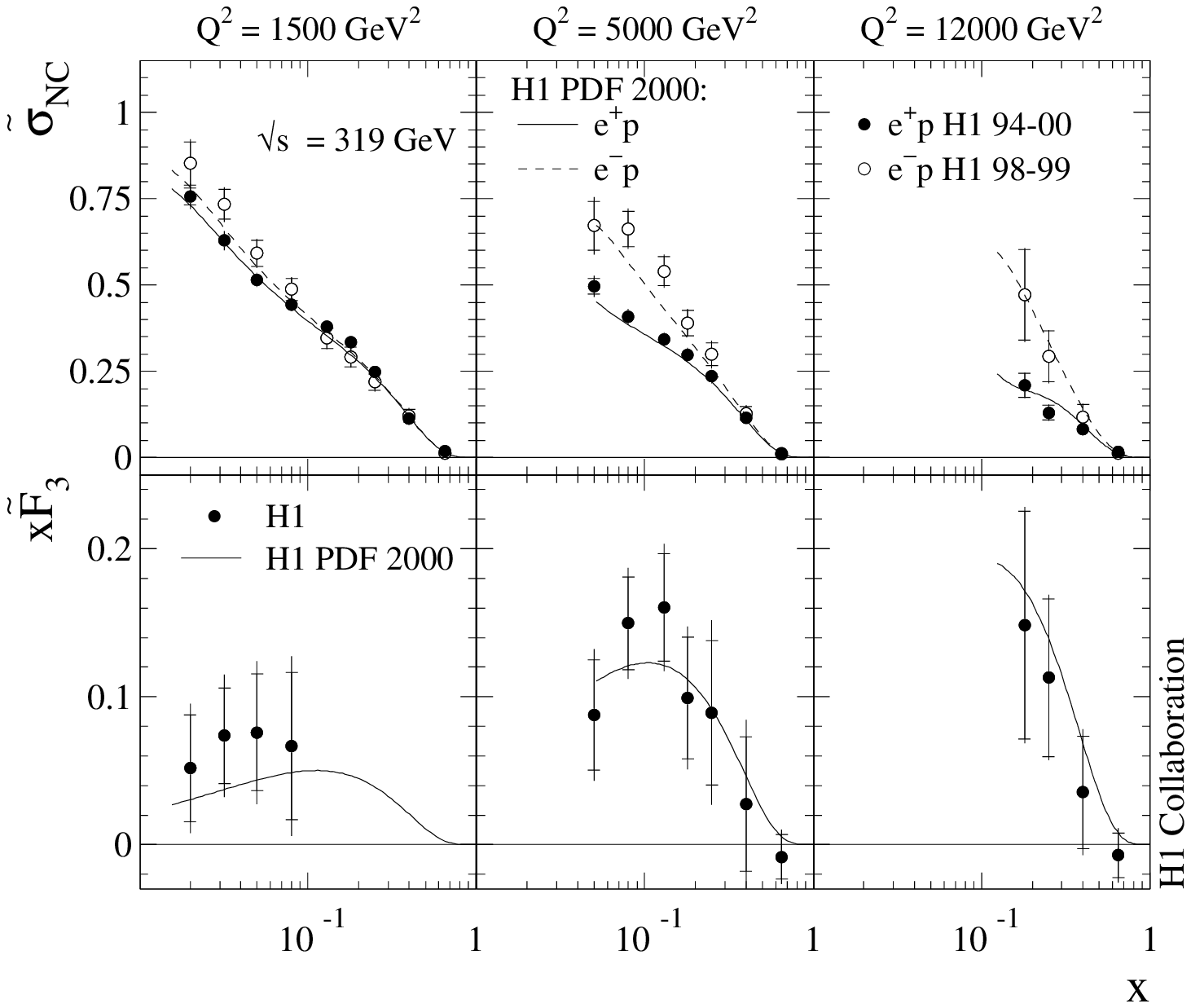,
 width=0.78\textwidth}}
\put(-40,-10){\epsfig{file=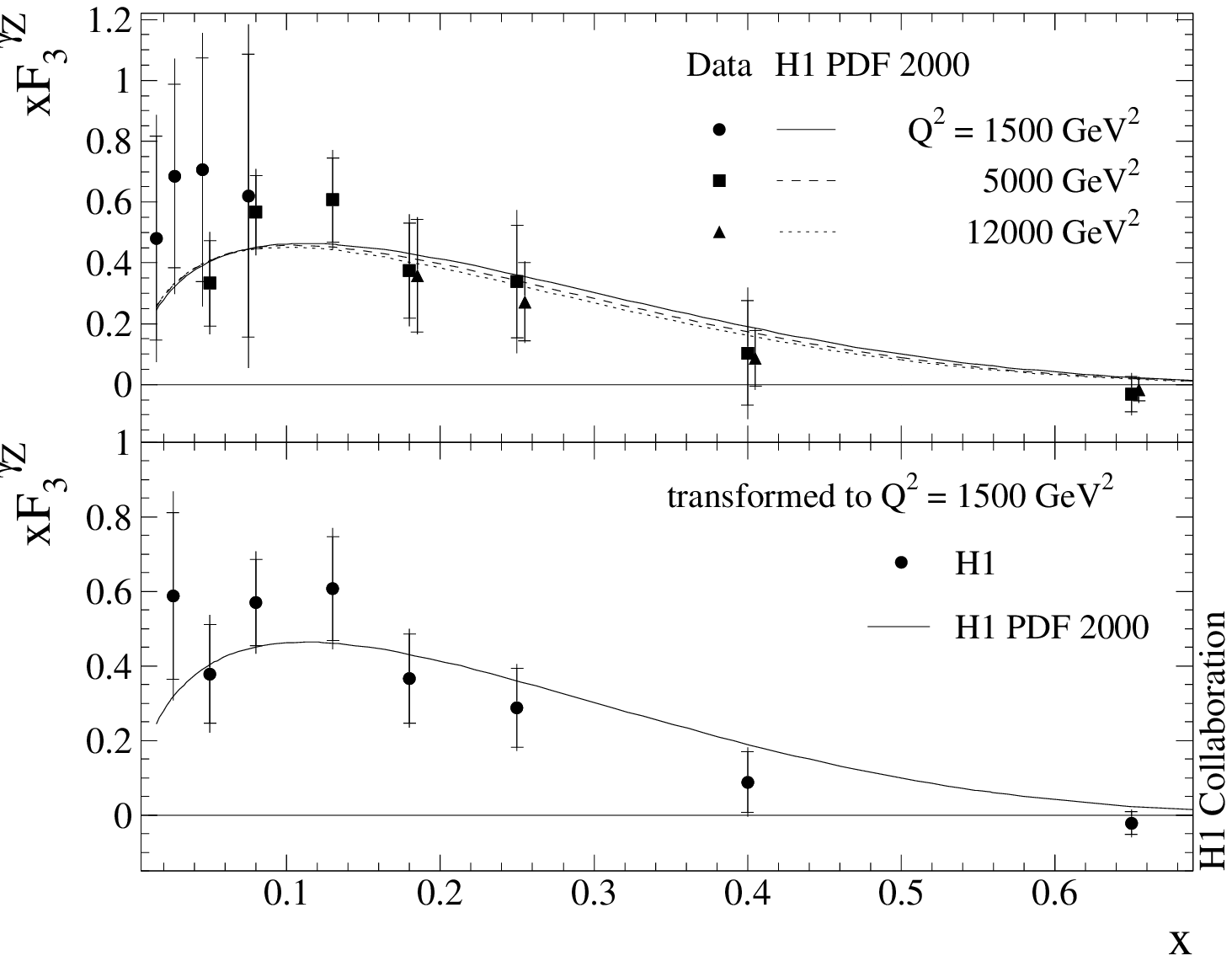,
 width=0.78\textwidth}}
\put(50,152){\bf (a)}
\put(50,135){\bf (b)}
\put(0, 80){\bf (c)}
\put(0, 36){\bf (d)}
\end{picture}
\end{center}
\caption{\sl The measured NC reduced cross sections 
 $\tilde{\sigma}^\pm_{NC}(x,Q^2)$ (a),
 structure functions $x\tilde{F}_3$ (b) and 
 $xF_3^{\gamma Z}$ (c), shown for three different $Q^2$ values.
 The results are compared with the 
 corresponding Standard Model expectations determined from the 
 H1 PDF $2000$ fit.
 In (d), the averaged structure function $xF_3^{\gamma Z}$ for a $Q^2$ 
 value of $1\,500{\rm\,GeV}^2$ is compared with the expectation 
 determined from the same fit.
 The error bars are defined as for fig.~\ref{dsdq2nc}. The normalisation
 uncertainties of the $e^-p$ and $e^+p$ data sets are included in the 
 systematic errors.}
\label{fig:xf3}
\end{figure}

Profiting from the enlarged statistics and the reduced systematic 
uncertainties, the previous measurement of $x\tilde{F}_3$~\cite{h1elec} 
is updated here by using the same published $e^-p$ and the full $e^+p$ data 
obtained by H1 at HERA-I.
Fig.~\ref{fig:xf3}(a) shows the comparison of the $e^-p$ and $e^+p$
data for three different $Q^2$ values at $1\,500{\rm\,GeV}^2$, 
$5\,000{\rm\,GeV}^2$ and $12\,000{\rm\,GeV}^2$, together with 
the expectations determined from the H1 PDF $2000$ fit. 
The generalised structure function $x\tilde{F}_3$,
given in table~\ref{tab:xf3}, is obtained from
\begin{equation}
 x\tilde{F}_3=\frac{1}{2Y_-}
 \left[ \phi^-_{NC}-\phi^+_{NC}\right]
\end{equation}
and is compared in fig.~\ref{fig:xf3}(b) with the calculation. 
Since at high $x$ and low $Q^2$ the expected sensitivity to
$x\tilde{F}_3$ is smaller than the luminosity uncertainty, 
the measurement is not performed in this region.
The dominant contribution to $x\tilde{F}_3$ arises from $\gamma Z$ 
interference, which allows
$xF^{\gamma Z}_3$ to be extracted (eq.~\ref{f3p}) according to 
$xF^{\gamma Z}_3\simeq -x\tilde{F}_3(Q^2+M^2_Z)/(a_e\kappa Q^2)$ by neglecting
the pure $Z$ exchange contribution, which is suppressed by the small vector
coupling $v_e$. 
This structure function is non-singlet and has little dependence on $Q^2$.
This is illustrated in fig.~\ref{fig:xf3}(c).
The measured $xF_3^{\gamma Z}$ at these $Q^2$ values can thus be averaged
taking into account the small $Q^2$ dependence. The two lowest $x$ bins 
at $x=0.020$ and $0.032$ are averaged as well. The averaged $xF_3^{\gamma Z}$,
determined for a $Q^2$ value of $1\,500{\rm\,GeV}^2$, is shown 
in table~\ref{tab:xf3} and fig.~\ref{fig:xf3}(d) in comparison
with the QCD fit result.
The structure function $xF^{\gamma Z}_3$ determines both the shape and
magnitude of the valence quark distributions independent of
the sea quark distributions. The calculation from the QCD fit, 
in which the parton densities in the valence region are principally 
constrained by the NC and CC cross sections rather than the difference
between the $e^\pm$ NC cross sections, gives a good description of
the measurement.
The averaged structure function is integrated~\cite{rizvi00} over 
the measured $x$ range, yielding
$$\int_{0.026}^{0.650} F_3^{\gamma Z}(x,Q^2=1\,500{\rm\,GeV}^2){\rm d}x
=1.28\pm 0.17{\rm (stat.)}\pm 0.11{\rm (syst.)}\,,$$
which is in agreement with $1.06\pm 0.02$, as predicted from
the H1 PDF $2000$ fit.

\subsection{The Quark Distributions {\boldmath $xu$} and {\boldmath
    $xd$} at Large {\boldmath $x$}}
\label{quarks}

\begin{figure}[htbp]
\center 
\epsfig{file=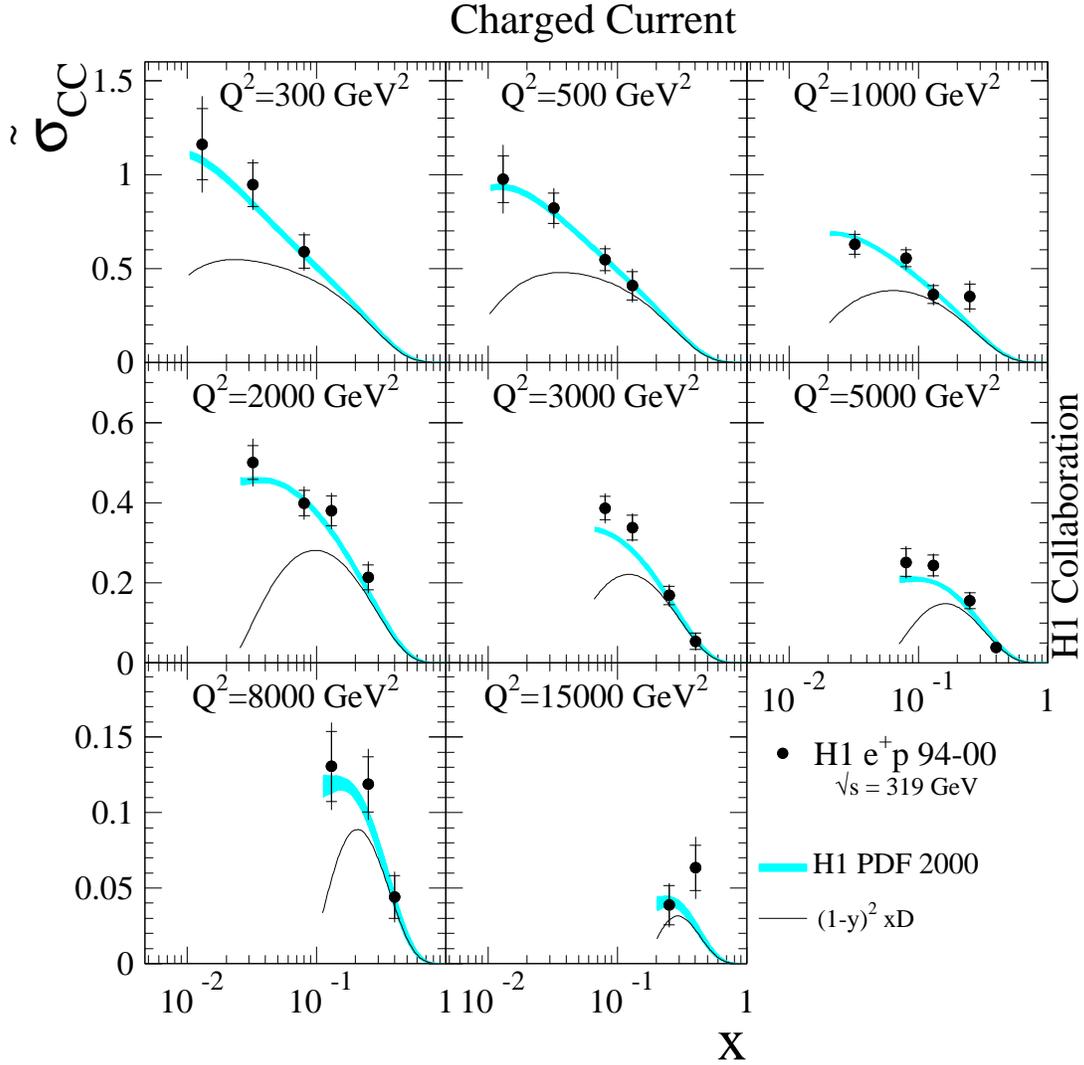,width=15cm}
\caption{\sl The CC reduced cross section $\tilde{\sigma}_{CC}(x,Q^2)$,
  shown for the combined $94-00$ $e^+p$ data (solid points).
  The results are compared with the
  corresponding Standard Model expectation (error bands) determined from
  the H1 PDF $2000$ fit at $\sqrt{s}=319{\rm\,GeV}$.
  The full curves indicate the expected $xD$ contributions.
  The error bars and bands are defined as for fig.~\ref{dsdq2nc}.}
\label{cc_stampcomb}
\end{figure}

\begin{figure}[htbp]
\begin{center}
\begin{picture}(50,160)
\put(-60,0){\epsfig{file=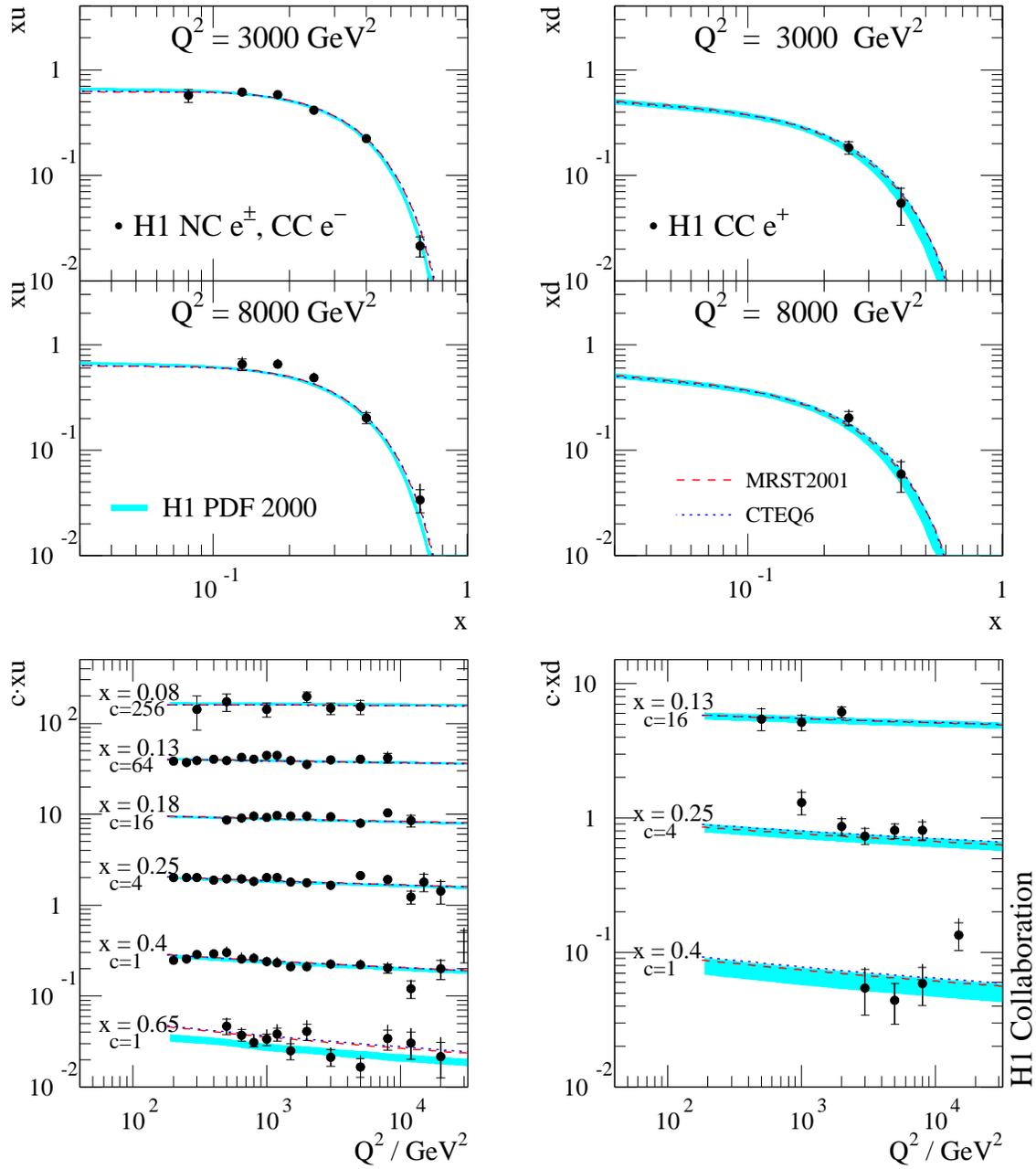,
width=\textwidth}}
\end{picture}
\end{center}
\caption{\sl The quark distributions $xu$ and $xd$ obtained
 from the local extraction method (solid points) in comparison
 with the expectations from the H1 PDF $2000$ fit
 (error bands), MRST (dashed curves) and CTEQ (dotted curves).}
\label{fig:xuxd}
\end{figure}

The flavour composition of the proton at high $x$ may be disentangled by
exploiting the NC and CC cross section measurements. The $e^+p$ CC cross
section at large $x$ is dominated by the $d$ quark contribution as may
be inferred from fig.~\ref{cc_stampcomb}. Similarly the $u$ distribution
dominates the $e^-p$ CC and $e^\pm p$ NC cross sections at large $x$.
Using data points for which the $xu$ or $xd$ contribution
provides at least $70\%$ of the cross section, as determined from the H1
PDF $2000$ fit, the up and down quark distributions are determined
locally, using the method described in~\cite{h1hiq2,zhang}. 
The extraction relies on weighting the
differential cross section measurement with the calculated local flavour
contribution and is illustrated in fig.~\ref{fig:xuxd}, where $xu$ is the
combined result from three independent extractions from the NC $e^\pm p$ and 
CC $e^-p$ data and $xd$ is determined from the CC $e^+p$ data only. 
This method is complementary to performing a QCD fit, 
since it is based on the {\em local} cross section measurements and
is less sensitive to the parameterisations and dynamical assumptions used 
in the fit.

The extracted $xu$ and $xd$ distributions are further compared in 
fig.~\ref{fig:xuxd}
with the results of the H1 PDF $2000$ fit by subtracting $xc$ and $xs$
from the fitted $xU$ and $xD$. The two determinations are in good
agreement. They also compare well with the recent parameterisations
from the MRST~\cite{mrst2001} and CTEQ~\cite{cteq6} groups 
except for $xu$ at large $x=0.65$,
where the results of MRST and CTEQ, being constrained mainly by 
the BCDMS data, yield a larger up quark distribution.

\section{Summary}
\label{summary}

New measurements are presented of inclusive deep inelastic neutral and 
charged current scattering cross sections at high momentum transfers 
$Q^2 \ge 100{\rm\,GeV}^2$ from recent $e^+p$ data recorded in $1999$ 
and $2000$ by the H1 experiment at HERA.
This analysis, together with previous analyses of the $1994-1997$ $e^+p$ 
and $1998-1999$ $e^-p$ data,
completes the H1 measurements of the inclusive cross sections at high $Q^2$ 
from the first phase of HERA operation.

The accuracy of the neutral current (NC) measurements presented here 
has reached the level of a few percent in the medium $Q^2$ range of 
$Q^2<3\,000{\rm\,GeV}^2$. 
The very high $Q^2$ NC and charged current (CC) data
are still limited by their statistical precision,
which is expected to improve in the high luminosity phase of HERA.

For both $e^+p$ and $e^-p$ scattering data, the region of very
large inelasticity is explored, which allows a determination of the
longitudinal structure function $F_L(x,Q^2)$ for the first time in the
large momentum transfer range, $110{\rm\,GeV}^2\leq Q^2\leq700{\rm\,GeV}^2$.
The observed interference of the photon and $Z$ exchange, differing
between $e^+ p$ and $e^-p$ NC scattering at high $Q^2$, is used to measure 
the structure function $x\tilde{F}_3$,
superseding the earlier measurement.
 
The NC and CC cross sections in $e^\pm p$ scattering are sensitive to
the sums of up- and down-type quark and anti-quark distributions,
$xU$, $xD$, $x\bU$ and $x\bD$. 
Based on these quark distribution decompositions,
a novel NLO QCD analysis is performed,
resulting in a first determination of the partonic nucleon structure from
inclusive DIS scattering data from H1 alone. The experimental precision 
achieved in this analysis is about $3\%$ and $10\%$ respectively for 
$xU$ and $xD$ at $x=0.4$.
The resulting parton distributions
are found to be in agreement with those obtained in an analysis 
also including the BCDMS muon-nucleon data at large $x$.
The QCD analysis leads to a good description of all the fitted NC and CC 
cross section data and of the derived structure functions over more than 
four orders of magnitude in $x$ and $Q^2$.

\section*{Acknowledgments}
We are grateful to the HERA machine group whose outstanding
efforts have made this experiment possible. 
We thank
the engineers and technicians for their work in constructing and 
maintaining the H1 detector, our funding agencies for 
financial support, the
DESY technical staff for continual assistance, 
and the DESY directorate for support and for the 
hospitality which they extend to the non-DESY 
members of the collaboration.
We would like to thank D.~Y.~Bardin, T.~Riemann and H.~Spiesberger for 
helpful discussions.


\newpage



%
\clearpage
 \begin{table}[htb]
 \begin{center}
 \begin{tabular}{|r|l|l||r|r|r|r||r|r|r|r|r|}
 \hline
 $Q^2$  &
 ${\rm d}\sigma_{NC}/{\rm d}Q^2$ &
 $k_{\rm cor}$ &
 $\delta_{stat}$ & $\delta_{unc}$ & $\delta_{cor}$ &
 $\delta_{tot}$ &
 $\delta_{cor}^{E^+}$ &
 $\delta_{cor}^{\theta^+}$&
 $\delta_{cor}^{h^+}$&
 $\delta_{cor}^{N^+}$&
 $\delta_{cor}^{B^+}$\\
 $(\rm GeV^2)$ &$(\rm pb / \rm GeV^2)$ &
      & 
 $(\%)$ &$(\%)$ &$(\%)$ &$(\%)$ &$(\%)$ &
 $(\%)$ &$(\%)$ &$(\%)$ &$(\%)$ \\
 &$y<0.9$&&&&&&&&&&\\
 \hline
$  200$&$1.835 \cdot 10^{ 1}$&$1.012$&$  0.5$&$  1.4$&$  0.6$&$  1.7$&
 $ -0.4$&$ -0.5$&$  0.0$&$  0.2$&$ -0.1$\\
$  250$&$1.080 \cdot 10^{ 1}$&$1.009$&$  0.6$&$  1.9$&$  0.9$&$  2.2$&
 $  0.7$&$ -0.6$&$  0.0$&$  0.2$&$ -0.1$\\
$  300$&$7.076 \cdot 10^{ 0}$&$1.006$&$  0.6$&$  2.0$&$  1.0$&$  2.3$&
 $  0.9$&$ -0.5$&$  0.1$&$  0.2$&$ -0.1$\\
$  400$&$3.565 \cdot 10^{ 0}$&$1.000$&$  0.8$&$  1.6$&$  0.8$&$  2.0$&
 $  0.8$&$ -0.3$&$  0.1$&$  0.2$&$ -0.1$\\
$  500$&$2.049 \cdot 10^{ 0}$&$1.000$&$  0.9$&$  1.8$&$  1.0$&$  2.2$&
 $  1.0$&$ -0.3$&$  0.0$&$  0.2$&$ -0.1$\\
$  650$&$1.065 \cdot 10^{ 0}$&$1.000$&$  1.1$&$  1.7$&$  1.1$&$  2.3$&
 $  0.9$&$ -0.7$&$  0.0$&$  0.1$&$  0.0$\\
$  800$&$0.638 \cdot 10^{ 0}$&$1.000$&$  1.4$&$  1.8$&$  1.0$&$  2.5$&
 $  0.9$&$ -0.6$&$  0.1$&$  0.1$&$  0.0$\\
$ 1000$&$0.373 \cdot 10^{ 0}$&$1.000$&$  1.6$&$  1.9$&$  1.1$&$  2.7$&
 $  0.9$&$ -0.6$&$  0.0$&$  0.1$&$ -0.2$\\
$ 1200$&$0.231 \cdot 10^{ 0}$&$1.000$&$  1.9$&$  1.9$&$  1.3$&$  3.0$&
 $  0.8$&$ -0.9$&$  0.1$&$  0.1$&$ -0.2$\\
$ 1500$&$0.125 \cdot 10^{ 0}$&$1.000$&$  2.3$&$  1.9$&$  0.9$&$  3.1$&
 $  0.7$&$ -0.6$&$  0.0$&$  0.0$&$ -0.2$\\
$ 2000$&$0.580 \cdot 10^{-1}$&$1.000$&$  2.8$&$  2.4$&$  1.3$&$  3.9$&
 $  1.1$&$ -0.7$&$  0.1$&$  0.0$&$ -0.2$\\
$ 3000$&$0.195 \cdot 10^{-1}$&$1.000$&$  2.7$&$  2.6$&$  1.2$&$  4.0$&
 $  1.1$&$ -0.5$&$  0.0$&$  0.0$&$ -0.2$\\
$ 5000$&$0.460 \cdot 10^{-2}$&$1.000$&$  3.7$&$  3.5$&$  0.8$&$  5.1$&
 $  0.7$&$ -0.4$&$  0.1$&$  0.1$&$ -0.2$\\
$ 8000$&$0.122 \cdot 10^{-2}$&$1.000$&$  5.7$&$  5.2$&$  1.5$&$  7.8$&
 $  1.4$&$ -0.3$&$  0.2$&$  0.1$&$ -0.2$\\
$12000$&$0.215 \cdot 10^{-3}$&$1.000$&$ 12.0$&$  6.6$&$  0.9$&$ 13.7$&
 $  0.9$&$ -0.1$&$  0.1$&$  0.0$&$ -0.3$\\
$20000$&$0.250 \cdot 10^{-4}$&$1.000$&$ 26.3$&$ 10.6$&$  1.8$&$ 28.4$&
 $  1.8$&$  0.1$&$  0.1$&$  0.0$&$ -0.6$\\
$30000$&$0.119 \cdot 10^{-4}$&$1.000$&$ 38.7$&$ 18.7$&$  4.3$&$ 43.1$&
 $  2.6$&$  3.4$&$  0.1$&$  0.0$&$ -0.7$\\
 \hline
 \end{tabular}
 \end{center}
 \caption[RESULT]
 {\sl \label{ncdq2} The NC $e^+p$ cross section
 ${\rm d}\sigma_{NC}/{\rm d}Q^2$ for $y<0.9$ after
 correction ($k_{\rm cor}$) according to the
 Standard Model expectation determined from the H1 PDF $2000$
 fit for the kinematic cut $E_e^{\prime}>6{\rm\,GeV}$
 for $Q^2<890{\rm\,GeV}^2$. The
 statistical ($\delta_{stat}$),
 uncorrelated systematic ($\delta_{unc}$),
 correlated systematic ($\delta_{cor}$)
 and total ($\delta_{tot}$) errors are provided.
 In addition the correlated systematic 
 error contributions from a
 positive variation of one 
 standard deviation of the
 electron energy error ($\delta_{cor}^{E^+}$), of
 the polar electron angle error
 ($\delta_{cor}^{\theta^+}$), of the hadronic
 energy error ($\delta_{cor}^{h^+}$), of the error
 due to noise subtraction ($\delta_{cor}^{N^+}$) and
 of the error due to background subtraction
 ($\delta_{cor}^{B^+}$) are given.
 The normalisation uncertainty of $1.5\%$ is
 not included in the errors.}
 \end{table}
 \begin{table}[htb]
 \begin{center}
 \begin{tabular}{|r|l|l||r|r|r|r||r|r|r|r|r|}
 \hline
 $Q^2$ & 
 ${\rm d}\sigma_{CC}/{\rm d}Q^2$ &
 $k_{\rm cor}$ &
 $\delta_{stat}$ & $\delta_{unc}$ & $\delta_{cor}$ &
 $\delta_{tot}$ &
 $\delta_{cor}^{V^+}$ &
 $\delta_{cor}^{h^+}$&
 $\delta_{cor}^{N^+}$&
 $\delta_{cor}^{B^+}$& $\Delta^{QED}_{CC}$ \\
 $(\rm GeV^2)$ &$(\rm pb / \rm GeV^2)$ &
  & 
 $(\%)$ &$(\%)$ &$(\%)$ &$(\%)$ &$(\%)$&
 $(\%)$ &$(\%)$ &$(\%)$ &$(\%)$ \\
  & $ y < 0.9$ & & & & & & & & & & \\
 \hline
$  300$&$0.330 \cdot 10^{-1}$&$  1.40$&$  9.4$&$  6.8$&$  4.5$&$ 12.4$&
 $  3.8$&$ -2.0$&$  1.2$&$ -0.1$&$  1.4$\\
$  500$&$0.198 \cdot 10^{-1}$&$  1.18$&$  6.9$&$  4.7$&$  2.5$&$  8.7$&
 $  2.2$&$ -1.1$&$  0.8$&$  0.0$&$ -2.2$\\
$ 1000$&$0.106 \cdot 10^{-1}$&$  1.05$&$  5.6$&$  4.2$&$  2.1$&$  7.3$&
 $  1.8$&$ -0.9$&$  0.4$&$  0.0$&$ -2.4$\\
$ 2000$&$0.527 \cdot 10^{-2}$&$  1.03$&$  5.0$&$  3.8$&$  1.3$&$  6.4$&
 $  1.1$&$ -0.3$&$  0.6$&$  0.0$&$ -5.0$\\
$ 3000$&$0.307 \cdot 10^{-2}$&$  1.03$&$  5.3$&$  3.7$&$  1.3$&$  6.6$&
 $  0.8$&$  1.0$&$  0.3$&$  0.0$&$ -7.1$\\
$ 5000$&$0.114 \cdot 10^{-2}$&$  1.04$&$  7.2$&$  5.1$&$  3.1$&$  9.4$&
 $  0.6$&$  2.7$&$  1.4$&$  0.0$&$-12.0$\\
$ 8000$&$0.347 \cdot 10^{-3}$&$  1.04$&$ 11.9$&$  9.4$&$  5.4$&$ 16.1$&
 $  0.3$&$  5.1$&$  1.8$&$  0.0$&$-11.1$\\
$15000$&$0.492 \cdot 10^{-4}$&$  1.06$&$ 21.7$&$ 16.9$&$  6.5$&$ 28.2$&
 $  0.2$&$  6.4$&$  0.9$&$  0.0$&$-15.7$\\
 \hline
 \end{tabular}
 \end{center}
 \caption[RESULT]
 {\sl \label{ccdq2} The CC $e^+p$ cross section
 ${\rm d}\sigma_{CC}/{\rm d}Q^2$
 for $y<0.9$ after correction ($k_{cor}$)
 according to the Standard Model expectation
 determined from the H1 PDF $2000$ fit for the kinematic
 cuts $0.03<y<0.85$ and $P_{T,h}>12{\rm\,GeV}$.
 The statistical ($\delta_{stat}$),
 uncorrelated systematic ($\delta_{unc}$),
 correlated systematic ($\delta_{cor}$)
 and total ($\delta_{tot}$) errors are also given.
 In addition the correlated systematic 
 error contributions from a
 positive variation of one standard deviation of the
 error due to the cuts against photoproduction
 ($\delta_{cor}^{V^+}$), of the hadronic
 energy error ($\delta_{cor}^{h^+}$), of the error
 due to noise subtraction ($\delta_{cor}^{N^+}$)
 and of the error due to background subtraction
 ($\delta_{cor}^{B^+}$) are given.
 The normalisation uncertainty of $1.5\%$ is
 not included in the errors.
 The last column gives the correction for QED
 radiative effects ($\Delta^{QED}_{CC}$).}
 \end{table}
 \begin{table}[htb]
 \begin{center}
 \small
 \begin{tabular}{|c|c||r|r|r|r||r|r|r|r|r|}
 \hline
  $x$ & ${\rm d}\sigma_{NC}/{\rm d}x~(\rm pb )$ & $\delta_{stat}$ & 
   $\delta_{unc}$ & $\delta_{cor}$ & $\delta_{tot}$ & $\delta_{cor}^{E^+}$ &
    $\delta_{cor}^{\theta^+}$ & $\delta_{cor}^{h^+}$ & $\delta_{cor}^{N^+}$ &
     $\delta_{cor}^{B^+}$ \\
   & $Q^2>1\,000{\rm\,GeV}^2$ & $(\%)$ & $(\%)$ & $(\%)$ & $(\%)$ & $(\%)$ & 
    $(\%)$ & $(\%)$ & $(\%)$ & $(\%)$ \\
  & $y<0.9$ & & & & & & & & & \\
 \hline
$0.013$&$0.138 \cdot 10^{ 4}$&$  5.9$&$  4.2$&$  3.7$&$  8.1$&$ -0.8$&
 $ -1.0$&$ -1.5$&$ -0.8$&$ -3.1$\\
$0.020$&$0.249 \cdot 10^{ 4}$&$  3.0$&$  2.3$&$  1.4$&$  4.1$&$  0.3$&
 $ -1.2$&$  0.4$&$  0.3$&$ -0.6$\\
$0.032$&$0.215 \cdot 10^{ 4}$&$  2.6$&$  2.3$&$  1.2$&$  3.6$&$  0.6$&
 $ -0.7$&$  0.7$&$  0.4$&$ -0.1$\\
$0.050$&$0.147 \cdot 10^{ 4}$&$  2.5$&$  2.0$&$  1.0$&$  3.3$&$  0.3$&
 $ -0.7$&$  0.5$&$  0.4$&$ -0.1$\\
$0.080$&$0.951 \cdot 10^{ 3}$&$  2.4$&$  2.1$&$  1.1$&$  3.4$&$  0.5$&
 $ -0.6$&$  0.5$&$  0.6$&$  0.0$\\
$0.130$&$0.566 \cdot 10^{ 3}$&$  2.8$&$  2.2$&$  1.1$&$  3.7$&$  0.9$&
 $ -0.5$&$  0.3$&$  0.4$&$  0.0$\\
$0.180$&$0.372 \cdot 10^{ 3}$&$  3.0$&$  2.6$&$  1.6$&$  4.3$&$  1.3$&
 $ -0.8$&$ -0.3$&$  0.5$&$  0.0$\\
$0.250$&$0.212 \cdot 10^{ 3}$&$  3.3$&$  3.4$&$  2.1$&$  5.2$&$  2.0$&
 $ -0.2$&$ -0.6$&$ -0.3$&$  0.0$\\
$0.400$&$0.646 \cdot 10^{ 2}$&$  4.7$&$  5.7$&$  5.3$&$  9.1$&$  3.7$&
 $ -0.4$&$ -1.9$&$ -3.5$&$  0.0$\\
$0.650$&$0.650 \cdot 10^{ 1}$&$  9.8$&$ 13.3$&$ 11.6$&$ 20.2$&$  7.4$&
 $  0.8$&$ -4.0$&$ -8.1$&$  0.0$\\
 \hline
 \end{tabular}
 \end{center}
 \caption[RESULT]
 {\sl \label{ncdx1}
 The NC $e^+p$ cross section
 ${\rm d}\sigma_{NC}/{\rm d}x$ measured for $y<0.9$
 and $Q^2>1\,000{\rm\,GeV}^2$.
 The statistical ($\delta_{stat}$),
 uncorrelated systematic ($\delta_{unc}$),
 correlated systematic ($\delta_{cor}$)
 and total ($\delta_{tot}$) errors are provided.
 In addition the correlated systematic 
 error contributions from a
 positive variation of one 
 standard deviation of the
 electron energy error ($\delta_{cor}^{E^+}$), of
 the polar electron angle error
 ($\delta_{cor}^{\theta^+}$), of the hadronic
 energy error ($\delta_{cor}^{h^+}$), of the error
 due to noise subtraction ($\delta_{cor}^{N^+}$) and
 of the error due to background subtraction
 ($\delta_{cor}^{B^+}$) are given.
 The normalisation uncertainty of $1.5\%$ is
 not included in the errors.}
 \end{table}
 \begin{table}[htb]
 \begin{center}
 \small
 \begin{tabular}{|c|c||r|r|r|r||r|r|r|r|r|}
 \hline
  $x$  &
 ${\rm d}\sigma_{NC}/{\rm d}x~(\rm pb )$ &
 $\delta_{stat}$ & $\delta_{unc}$ & $\delta_{cor}$ &
 $\delta_{tot}$ &
 $\delta_{cor}^{E^+}$ &
 $\delta_{cor}^{\theta^+}$&
 $\delta_{cor}^{h^+}$&
 $\delta_{cor}^{N^+}$&
 $\delta_{cor}^{B^+}$\\
  &$Q^2>10\,000{\rm\,GeV}^2$&
 $(\%)$ &$(\%)$ &$(\%)$ &$(\%)$ &
 $(\%)$ &$(\%)$ &$(\%)$ &$(\%)$ &$(\%)$ \\
 &$y<0.9$&&&&&&&&&\\
 \hline
$0.130$&$0.544 \cdot 10^{ 1}$&$ 29.7$&$  8.9$&$  1.9$&$ 31.0$&$  0.1$&
        $ -1.2$&$ -1.1$&$ -0.6$&$ -0.8$\\
$0.180$&$0.716 \cdot 10^{ 1}$&$ 18.9$&$  4.2$&$  2.2$&$ 19.5$&$ -1.1$&
        $ -1.9$&$  0.4$&$  0.1$&$ -0.5$\\
$0.250$&$0.411 \cdot 10^{ 1}$&$ 19.7$&$  6.3$&$  1.4$&$ 20.7$&$  1.1$&
        $ -0.5$&$  0.9$&$  0.3$&$ -0.1$\\
$0.400$&$0.162 \cdot 10^{ 1}$&$ 24.5$&$ 11.5$&$  2.0$&$ 27.1$&$  1.9$&
        $  0.6$&$  0.5$&$  0.2$&$ -0.3$\\
$0.650$&$0.038 \cdot 10^{ 1}$&$ 31.6$&$ 30.5$&$  8.4$&$ 44.8$&$  6.3$&
        $  5.1$&$ -2.2$&$ -0.8$&$  0.0$\\
 \hline
 \end{tabular}
 \end{center}
 \caption[RESULT]
 {\sl \label{ncdx2}
 The NC $e^+p$ cross section
 ${\rm d}\sigma_{NC}/{\rm d}x$ measured for $y<0.9$
 and $Q^2>10\,000{\rm\,GeV}^2$.
 The statistical ($\delta_{stat}$),
 uncorrelated systematic ($\delta_{unc}$),
 correlated systematic ($\delta_{cor}$)
 and total ($\delta_{tot}$) errors are provided.
 In addition the correlated systematic 
 error contributions from a
 positive variation of one 
 standard deviation of the
 electron energy error ($\delta_{cor}^{E^+}$), of
 the polar electron angle error
 ($\delta_{cor}^{\theta^+}$), of the hadronic
 energy error ($\delta_{cor}^{h^+}$), of the error
 due to noise subtraction ($\delta_{cor}^{N^+}$) and
 of the error due to background subtraction
 ($\delta_{cor}^{B^+}$) are given.
 The normalisation uncertainty of $1.5\%$ is
 not included in the errors.}
 \end{table}
 \begin{table}[htb]
 \begin{center}
 \begin{tabular}{|r|c|c||r|r|r|r||r|r|r|r|}
 \hline
  $x$  &
 ${\rm d}\sigma_{CC}/{\rm d}x$ $(\rm pb )$& $k_{\rm cor}$ &
 $\delta_{stat}$ & $\delta_{unc}$ & $\delta_{cor}$ &
 $\delta_{tot}$ &
 $\delta_{cor}^{V^+}$ &
 $\delta_{cor}^{h^+}$&
 $\delta_{cor}^{N^+}$&
 $\delta_{cor}^{B^+}$\\
  & $Q^2>1\,000{\rm\,GeV}^2$ & &
 $(\%)$ &$(\%)$ &$(\%)$ &$(\%)$ &
 $(\%)$ &$(\%)$ &$(\%)$ &$(\%)$ \\
  & $y<0.9$& & & & & & & & & \\
 \hline
$0.032$&$0.159 \cdot 10^{ 3}$&$1.05$&$  6.9$&$  4.0$&$  2.3$&$  8.3$&
        $  2.1$&$ -0.5$&$  0.6$&$ -0.4$\\
$0.080$&$0.946 \cdot 10^{ 2}$&$1.02$&$  4.6$&$  3.4$&$  1.2$&$  5.9$&
        $  0.9$&$  0.5$&$  0.7$&$  0.0$\\
$0.130$&$0.623 \cdot 10^{ 2}$&$1.01$&$  5.2$&$  3.7$&$  2.0$&$  6.7$&
        $  0.3$&$  1.2$&$  1.5$&$  0.0$\\
$0.250$&$0.194 \cdot 10^{ 2}$&$1.00$&$  7.2$&$  5.8$&$  2.2$&$  9.5$&
        $  0.1$&$  2.2$&$  0.1$&$  0.0$\\
$0.400$&$0.451 \cdot 10^{ 1}$&$1.06$&$ 16.6$&$ 11.3$&$  6.0$&$ 21.0$&
        $  0.0$&$  4.9$&$ -3.6$&$  0.0$\\
$0.650$&$0.469 \cdot 10^{ 0}$&$1.26$&$ 70.6$&$ 24.5$&$ 21.3$&$ 77.7$&
        $  0.0$&$  9.2$&$-19.2$&$  0.0$\\
 \hline
 \end{tabular}
 \end{center}
 \caption[RESULT]
 {\sl \label{ccdx}
 The CC $e^+p$ cross section
 ${\rm d}\sigma_{CC}/{\rm d}x$ for $y<0.9$ and $Q^2>1\,000{\rm\,GeV}^2$
 after correction ($k_{cor}$) according to the Standard Model
 expectation determined from the H1 PDF $2000$ fit for the
 kinematic cuts  $0.03<y<0.85$ and $P_{T,h}>12{\rm\,GeV}$.
 The statistical ($\delta_{stat}$),
 uncorrelated systematic ($\delta_{unc}$),
 correlated systematic ($\delta_{cor}$)
 and total ($\delta_{tot}$) errors are also given.
 In addition the correlated systematic 
 error contributions from a
 positive variation of one standard deviation of the
 error due to the cuts against photoproduction
 ($\delta_{cor}^{V^+}$), of the hadronic
 energy error ($\delta_{cor}^{h^+}$), of the error
 due to noise subtraction ($\delta_{cor}^{N^+}$)
 and of the error due to background subtraction
 ($\delta_{cor}^{B^+}$) are given.
 The normalisation uncertainty of $1.5\%$ is
 not included in the errors.}
 \end{table}
\newcommand{\tnca}{
 \begin{tabular}{|r|c|c||c|r|r|r||r|r|r||r|r|r|r|r|r|r||c|r|r|r|r|}
 \hline
 $Q^2$  &$x$ &$y$ & $\tilde{\sigma}_{NC}$ &
 $\delta_{stat}$ & $\delta_{sys}$ & $\delta_{tot}$ &
 $\delta_{unc}$ &
 $\delta_{unc}^{E}$ &
 $\delta_{unc}^{h}$&
 $\delta_{cor}$ &
 $\delta_{cor}^{E^+}$ &
 $\delta_{cor}^{\theta^+}$&
 $\delta_{cor}^{h^+}$&
 $\delta_{cor}^{N^+}$&
 $\delta_{cor}^{B^+}$&
 $\delta_{cor}^{S^+}$ &
 $\phi_{NC}/Y_+$ & $F_2$&
 $\Delta_{F_2}$& $\Delta_{F_3}$& $\Delta_{F_L}$ \\
 $(\rm GeV^2)$ & & & &
 $(\%)$ &$(\%)$ &$(\%)$ & 
 $(\%)$ &$(\%)$ &$(\%)$ &$(\%)$ &$(\%)$ &
 $(\%)$ &$(\%)$ &$(\%)$ &$(\%)$ &$(\%)$ & &
        &$(\%)$ &$(\%)$ &$(\%)$ \\ \hline \hline
$  100$&$0.00131$&$0.750$&$1.450$&$  1.6$&$  3.6$&$  4.0$&
$  3.5$&$  0.6$&$  0.0$&$  1.2$&$  0.4$&$  0.5$&$ -0.1$&$  0.1$&$    -$&$1.1$&
$1.450$ & $  -  $ & $  -  $ & $  -  $ & $  -  $ \\
$  100$&$0.00200$&$0.492$&$1.388$&$  1.4$&$  2.7$&$  3.1$&
$  2.1$&$  0.8$&$  0.3$&$  1.8$&$ -0.3$&$ -0.6$&$  1.3$&$  0.1$&$ -0.9$&$-$&
$1.388$ & $1.442$ & $  0.1$ & $  0.0$ & $ -3.8$ \\ \hline
$  120$&$0.00158$&$0.750$&$1.404$&$  1.8$&$  2.7$&$  3.3$&
$  2.7$&$  0.3$&$  0.0$&$  1.1$&$ -0.4$&$ -0.3$&$  0.2$&$  0.1$&$    -$&$0.9$&
$1.404$ & $  -  $ & $  -  $ & $  -  $ & $  -  $ \\
$  120$&$0.00200$&$0.591$&$1.338$&$  1.5$&$  2.5$&$  2.9$&
$  2.0$&$  0.6$&$  0.3$&$  1.5$&$ -0.4$&$ -0.3$&$  0.8$&$  0.2$&$ -1.2$&$-$&
$1.338$ & $1.418$ & $  0.2$ & $  0.0$ & $ -5.8$ \\
$  120$&$0.00320$&$0.369$&$1.220$&$  1.3$&$  2.2$&$  2.6$&
$  1.8$&$  0.4$&$  0.3$&$  1.4$&$ -0.3$&$ -1.1$&$  0.6$&$  0.1$&$ -0.5$&$-$&
$1.220$ & $1.241$ & $  0.2$ & $  0.0$ & $ -1.8$ \\ \hline
$  150$&$0.00197$&$0.750$&$1.398$&$  2.0$&$  2.8$&$  3.5$&
$  2.7$&$  0.7$&$  0.1$&$  1.2$&$ -0.7$&$  0.3$&$  0.3$&$  0.1$&$    -$&$0.9$&
$1.398$ & $  -  $ & $  -  $ & $  -  $ & $  -  $ \\
$  150$&$0.00320$&$0.462$&$1.225$&$  1.2$&$  2.6$&$  2.9$&
$  2.1$&$  0.4$&$  0.8$&$  1.6$&$  0.2$&$ -0.7$&$  1.0$&$  0.9$&$ -0.6$&$-$&
$1.225$ & $1.262$ & $  0.2$ & $ -0.1$ & $ -3.0$ \\
$  150$&$0.00500$&$0.295$&$1.061$&$  1.2$&$  2.2$&$  2.6$&
$  2.0$&$  0.5$&$  0.4$&$  1.0$&$ -0.1$&$ -0.6$&$  0.4$&$  0.7$&$ -0.1$&$-$&
$1.061$ & $1.071$ & $  0.2$ & $  0.0$ & $ -1.0$ \\
$  150$&$0.00800$&$0.185$&$0.940$&$  1.6$&$  2.9$&$  3.4$&
$  2.6$&$  1.3$&$  0.7$&$  1.4$&$ -0.6$&$ -0.6$&$ -0.9$&$ -0.8$&$  0.0$&$-$&
$0.940$ & $0.942$ & $  0.1$ & $  0.0$ & $ -0.3$ \\ \hline
$  200$&$0.00263$&$0.750$&$1.284$&$  2.2$&$  2.7$&$  3.5$&
$  2.6$&$  0.2$&$  0.1$&$  1.2$&$  0.2$&$ -0.9$&$  0.1$&$  0.1$&$    -$&$0.7$&
$1.285$ & $  -  $ & $  -  $ & $  -  $ & $  -  $ \\
$  200$&$0.00320$&$0.615$&$1.242$&$  1.9$&$  2.5$&$  3.1$&
$  2.2$&$  0.2$&$  0.2$&$  1.2$&$ -0.2$&$  0.0$&$  0.3$&$  0.5$&$ -1.1$&$-$&
$1.242$ & $  -  $ & $  -  $ & $  -  $ & $  -  $ \\
$  200$&$0.00500$&$0.394$&$1.091$&$  1.3$&$  2.5$&$  2.8$&
$  2.1$&$  0.2$&$  0.8$&$  1.4$&$  0.0$&$ -0.6$&$  0.9$&$  0.9$&$ -0.2$&$-$&
$1.091$ & $1.111$ & $  0.2$ & $ -0.1$ & $ -1.9$ \\
$  200$&$0.00800$&$0.246$&$0.956$&$  1.3$&$  2.3$&$  2.7$&
$  2.2$&$  0.9$&$  0.5$&$  0.9$&$  0.2$&$ -0.6$&$  0.5$&$  0.6$&$  0.0$&$-$&
$0.956$ & $0.961$ & $  0.2$ & $ -0.1$ & $ -0.6$ \\
$  200$&$0.01300$&$0.151$&$0.801$&$  1.4$&$  2.2$&$  2.6$&
$  2.0$&$  0.5$&$  0.2$&$  0.9$&$  0.3$&$ -0.7$&$  0.0$&$  0.5$&$  0.0$&$-$&
$0.801$ & $0.802$ & $  0.2$ & $ -0.1$ & $ -0.2$ \\
$  200$&$0.02000$&$0.098$&$0.713$&$  1.6$&$  2.2$&$  2.7$&
$  2.1$&$  0.5$&$  0.2$&$  0.7$&$ -0.4$&$ -0.4$&$ -0.4$&$  0.2$&$  0.0$&$-$&
$0.713$ & $0.713$ & $  0.2$ & $ -0.1$ & $ -0.1$ \\
$  200$&$0.03200$&$0.062$&$0.583$&$  1.8$&$  2.8$&$  3.3$&
$  2.5$&$  1.4$&$  0.4$&$  1.2$&$ -0.7$&$ -0.6$&$ -0.8$&$  0.1$&$  0.0$&$-$&
$0.583$ & $0.583$ & $  0.2$ & $ -0.1$ & $  0.0$ \\
$  200$&$0.05000$&$0.039$&$0.516$&$  2.0$&$  4.3$&$  4.7$&
$  3.4$&$  2.6$&$  0.1$&$  2.6$&$ -1.6$&$ -0.7$&$ -0.6$&$  1.9$&$  0.0$&$-$&
$0.516$ & $0.516$ & $  0.2$ & $ -0.1$ & $  0.0$ \\
$  200$&$0.08000$&$0.025$&$0.401$&$  2.3$&$  4.1$&$  4.7$&
$  3.4$&$  2.5$&$  0.6$&$  2.2$&$ -1.7$&$ -0.4$&$ -0.6$&$  1.3$&$  0.0$&$-$&
$0.401$ & $0.401$ & $  0.1$ & $  0.0$ & $  0.0$ \\
$  200$&$0.13000$&$0.015$&$0.345$&$  2.6$&$  4.5$&$  5.2$&
$  3.2$&$  1.5$&$  1.4$&$  3.3$&$ -0.9$&$ -0.7$&$ -0.9$&$ -3.0$&$  0.0$&$-$&
$0.345$ & $0.345$ & $  0.1$ & $  0.0$ & $  0.0$ \\
$  200$&$0.25000$&$0.008$&$0.259$&$  3.7$&$  7.5$&$  8.4$&
$  4.2$&$  0.9$&$  2.8$&$  6.3$&$  0.2$&$ -0.7$&$ -2.0$&$ -5.9$&$  0.0$&$-$&
$0.259$ & $0.259$ & $  0.1$ & $  0.0$ & $  0.0$ \\
$  200$&$0.40000$&$0.005$&$0.129$&$  4.8$&$ 12.3$&$ 13.2$&
$  4.3$&$  1.8$&$  1.7$&$ 11.5$&$ -1.2$&$ -0.8$&$ -0.8$&$-11.4$&$  0.0$&$-$&
$0.129$ & $0.128$ & $  0.1$ & $  0.0$ & $  0.0$ \\ \hline
$  250$&$0.00328$&$0.750$&$1.250$&$  2.5$&$  2.9$&$  3.8$&
$  2.7$&$  0.3$&$  0.1$&$  1.4$&$ -0.3$&$ -0.9$&$  0.6$&$  0.0$&$    -$&$0.8$&
$1.251$ & $  -  $ & $  -  $ & $  -  $ & $  -  $ \\
$  250$&$0.00500$&$0.492$&$1.120$&$  1.6$&$  2.4$&$  2.9$&
$  2.1$&$  0.1$&$  0.6$&$  1.2$&$ -0.1$&$ -0.3$&$  0.8$&$  0.8$&$ -0.4$&$-$&
$1.120$ & $1.155$ & $  0.3$ & $ -0.1$ & $ -3.1$ \\
$  250$&$0.00800$&$0.308$&$0.943$&$  1.5$&$  2.5$&$  2.9$&
$  2.2$&$  0.7$&$  0.8$&$  1.2$&$  0.4$&$ -0.5$&$  0.6$&$  0.9$&$  0.0$&$-$&
$0.943$ & $0.951$ & $  0.3$ & $ -0.1$ & $ -1.0$ \\
$  250$&$0.01300$&$0.189$&$0.833$&$  1.5$&$  2.5$&$  2.9$&
$  2.3$&$  1.1$&$  0.3$&$  0.9$&$  0.5$&$ -0.4$&$ -0.4$&$  0.6$&$  0.0$&$-$&
$0.833$ & $0.835$ & $  0.2$ & $ -0.1$ & $ -0.3$ \\
$  250$&$0.02000$&$0.123$&$0.710$&$  1.5$&$  2.6$&$  3.0$&
$  2.4$&$  1.3$&$  0.3$&$  1.0$&$  0.6$&$ -0.7$&$  0.2$&$  0.6$&$  0.0$&$-$&
$0.710$ & $0.710$ & $  0.3$ & $ -0.1$ & $ -0.1$ \\
$  250$&$0.03200$&$0.077$&$0.577$&$  1.7$&$  3.2$&$  3.6$&
$  2.8$&$  2.0$&$  0.0$&$  1.5$&$  1.0$&$ -0.8$&$ -0.4$&$  0.6$&$  0.0$&$-$&
$0.577$ & $0.577$ & $  0.2$ & $ -0.1$ & $  0.0$ \\
$  250$&$0.05000$&$0.049$&$0.514$&$  1.8$&$  3.2$&$  3.6$&
$  2.6$&$  1.6$&$  0.2$&$  1.8$&$  0.5$&$ -0.6$&$ -0.4$&$  1.6$&$  0.0$&$-$&
$0.513$ & $0.513$ & $  0.2$ & $ -0.1$ & $  0.0$ \\
$  250$&$0.08000$&$0.031$&$0.420$&$  1.9$&$  3.5$&$  4.0$&
$  2.5$&$  1.5$&$  0.2$&$  2.4$&$  1.0$&$ -1.0$&$ -0.5$&$  1.9$&$  0.0$&$-$&
$0.420$ & $0.420$ & $  0.2$ & $ -0.1$ & $  0.0$ \\
$  250$&$0.13000$&$0.019$&$0.342$&$  2.0$&$  4.0$&$  4.5$&
$  3.5$&$  2.7$&$  1.0$&$  1.9$&$  1.2$&$ -0.8$&$ -0.8$&$ -1.1$&$  0.0$&$-$&
$0.342$ & $0.342$ & $  0.2$ & $ -0.1$ & $  0.0$ \\
$  250$&$0.25000$&$0.010$&$0.273$&$  2.7$&$  8.2$&$  8.6$&
$  5.7$&$  4.6$&$  2.3$&$  5.9$&$  2.5$&$ -1.1$&$ -1.5$&$ -5.0$&$  0.0$&$-$&
$0.273$ & $0.273$ & $  0.1$ & $  0.0$ & $  0.0$ \\
$  250$&$0.40000$&$0.006$&$0.137$&$  3.7$&$ 12.2$&$ 12.8$&
$  5.5$&$  4.0$&$  2.4$&$ 11.0$&$  2.0$&$ -1.0$&$ -1.5$&$-10.7$&$  0.0$&$-$&
$0.137$ & $0.137$ & $  0.1$ & $  0.0$ & $  0.0$ \\ \hline
$  300$&$0.00394$&$0.750$&$1.189$&$  2.7$&$  2.8$&$  3.9$&
$  2.7$&$  0.2$&$  0.2$&$  1.0$&$ -0.2$&$ -0.5$&$  0.6$&$  0.1$&$    -$&$0.7$&
$1.190$ & $  -  $ & $  -  $ & $  -  $ & $  -  $ \\
$  300$&$0.00500$&$0.591$&$1.132$&$  2.5$&$  2.9$&$  3.8$&
$  2.6$&$  0.9$&$  0.2$&$  1.4$&$ -0.9$&$ -0.7$&$  0.3$&$  0.3$&$ -0.7$&$-$&
$1.133$ & $1.186$ & $  0.4$ & $ -0.2$ & $ -4.7$ \\
$  300$&$0.00800$&$0.369$&$0.995$&$  1.7$&$  2.5$&$  3.1$&
$  2.2$&$  0.4$&$  0.8$&$  1.2$&$  0.5$&$  0.4$&$  0.8$&$  0.8$&$ -0.1$&$-$&
$0.995$ & $1.008$ & $  0.3$ & $ -0.2$ & $ -1.4$ \\
$  300$&$0.01300$&$0.227$&$0.842$&$  1.7$&$  2.4$&$  3.0$&
$  2.2$&$  0.2$&$  0.8$&$  1.0$&$  0.1$&$ -0.1$&$  0.7$&$  0.8$&$  0.0$&$-$&
$0.842$ & $0.845$ & $  0.3$ & $ -0.2$ & $ -0.4$ \\
$  300$&$0.02000$&$0.148$&$0.708$&$  1.8$&$  2.5$&$  3.1$&
$  2.3$&$  1.0$&$  0.2$&$  1.0$&$  0.8$&$ -0.6$&$  0.1$&$  0.2$&$  0.0$&$-$&
$0.708$ & $0.708$ & $  0.3$ & $ -0.1$ & $ -0.2$ \\
$  300$&$0.03200$&$0.092$&$0.607$&$  1.9$&$  2.9$&$  3.5$&
$  2.6$&$  1.6$&$  0.1$&$  1.3$&$  0.9$&$ -0.2$&$ -0.5$&$  0.8$&$  0.0$&$-$&
$0.607$ & $0.606$ & $  0.3$ & $ -0.1$ & $ -0.1$ \\
$  300$&$0.05000$&$0.059$&$0.491$&$  2.1$&$  3.3$&$  3.9$&
$  2.9$&$  2.0$&$  0.0$&$  1.5$&$  1.0$&$ -0.8$&$  0.0$&$  0.9$&$  0.0$&$-$&
$0.491$ & $0.490$ & $  0.3$ & $ -0.1$ & $  0.0$ \\
$  300$&$0.08000$&$0.037$&$0.440$&$  2.1$&$  3.8$&$  4.3$&
$  3.0$&$  2.1$&$  0.2$&$  2.3$&$  0.9$&$ -0.8$&$ -0.3$&$  2.0$&$  0.0$&$-$&
$0.440$ & $0.440$ & $  0.2$ & $ -0.1$ & $  0.0$ \\
$  300$&$0.13000$&$0.023$&$0.355$&$  2.2$&$  4.1$&$  4.7$&
$  3.7$&$  2.9$&$  0.9$&$  1.8$&$  1.6$&$ -0.6$&$ -0.6$&$ -0.2$&$  0.0$&$-$&
$0.355$ & $0.355$ & $  0.2$ & $ -0.1$ & $  0.0$ \\
$  300$&$0.25000$&$0.012$&$0.260$&$  3.0$&$  9.8$&$ 10.2$&
$  6.4$&$  5.2$&$  2.6$&$  7.5$&$  3.4$&$ -1.0$&$ -2.2$&$ -6.3$&$  0.0$&$-$&
$0.260$ & $0.260$ & $  0.2$ & $  0.0$ & $  0.0$ \\
$  300$&$0.40000$&$0.007$&$0.152$&$  4.1$&$ 11.3$&$ 12.1$&
$  6.8$&$  5.8$&$  1.6$&$  9.1$&$  3.9$&$ -1.6$&$ -0.9$&$ -8.0$&$  0.0$&$-$&
$0.152$ & $0.152$ & $  0.1$ & $  0.0$ & $  0.0$ \\ \hline
 \multicolumn{22}{c}{\begin{minipage}{220mm}
\caption{\sl \label{ncdxdq2} The NC $e^+p$ reduced
  cross section $\tilde{\sigma}_{NC}(x,Q^2)$,
  shown with statistical
  ($\delta_{stat}$), systematic ($\delta_{sys}$) and
  total ($\delta_{tot}$) errors. Also shown are
  the total uncorrelated systematic ($\delta_{unc}$)
  errors and two of its contributions: the
  electron energy error ($\delta_{unc}^{E}$) 
  and the hadronic energy error 
  ($\delta_{unc}^{h}$). 
  The effect of the other uncorrelated
  systematic errors is included in ($\delta_{unc}$). 
  The table also provides the correlated systematic
  error ($\delta_{cor}$) and its contributions from a
  positive variation of one 
  standard deviation of the error on the
  electron energy ($\delta_{cor}^{E^+}$) and polar angle
  ($\delta_{cor}^{\theta^+}$), of the hadronic
  energy error ($\delta_{cor}^{h^+}$), of the error
  due to noise subtraction ($\delta_{cor}^{N^+}$) and
  background subtraction ($\delta_{cor}^{B^+}$) and
  of the error due to charge symmetry background subtraction 
  in the high-$y$ analysis ($\delta_{cor}^{S^+}$).
  The normalisation uncertainty of $1.5\%$ is
  not included in the errors.
  The NC structure function term scaled by the helicity 
  factor $Y_+$ $\phi_{NC}/Y_+$ is given
  as well as the electromagnetic structure function $F_2$
  with the corrections 
  $\Delta_{F_2}$, $\Delta_{F_3}$ and $\Delta_{F_L}$
  as defined in eq.~\ref{f2corr}.
  For $Q^2<2\,000{\rm\,GeV}^2$, the extraction of $F_2$ 
  is restricted to the region of $y<0.6$.
  The table continues on the next $2$ pages.}
\end{minipage}}
 \end{tabular}}

 \begin{table}[htbp]
 \begin{center}
  \tiny
  \rotatebox{90}{\tnca}
 \end{center}
 \end{table}

\newcommand{\tncb}{
 \begin{tabular}{|r|c|c||c|r|r|r||r|r|r|r|r|r|r|r|r|r||c|r|r|r|r|}
 \hline
 $Q^2$  &$x$ &$y$ & $\tilde{\sigma}_{NC}$ &
 $\delta_{stat}$ & $\delta_{sys}$ & $\delta_{tot}$ &
 $\delta_{unc}$ &
 $\delta_{unc}^{E}$ &
 $\delta_{unc}^{h}$&
 $\delta_{cor}$ &
 $\delta_{cor}^{E^+}$ &
 $\delta_{cor}^{\theta^+}$&
 $\delta_{cor}^{h^+}$&
 $\delta_{cor}^{N^+}$&
 $\delta_{cor}^{B^+}$&
 $\delta_{cor}^{S^+}$ &
 $\phi_{NC}/Y_+$ & $F_2$&
 $\Delta_{F_2}$& $\Delta_{F_3}$& $\Delta_{F_L}$ \\
 $(\rm GeV^2)$ & & & &
 $(\%)$ &$(\%)$ &$(\%)$ & 
 $(\%)$ &$(\%)$ &$(\%)$ &$(\%)$ &$(\%)$ &
 $(\%)$ &$(\%)$ &$(\%)$ &$(\%)$ &$(\%)$ & &
      &$(\%)$ &$(\%)$ &$(\%)$ \\ \hline \hline
$  400$&$0.00525$&$0.750$&$1.162$&$  2.9$&$  3.0$&$  4.2$&
$  2.9$&$  0.0$&$  0.2$&$  1.1$&$  0.0$&$ -0.8$&$  0.4$&$  0.0$&$    -$&$0.6$&
$1.163$ & $  -  $ & $  -  $ & $  -  $ & $  -  $ \\
$  400$&$0.00800$&$0.492$&$1.026$&$  2.2$&$  2.8$&$  3.5$&
$  2.4$&$  0.5$&$  0.7$&$  1.5$&$  0.5$&$ -1.0$&$  0.8$&$  0.6$&$ -0.4$&$-$&
$1.027$ & $1.053$ & $  0.5$ & $ -0.3$ & $ -2.7$ \\
$  400$&$0.01300$&$0.303$&$0.893$&$  2.0$&$  2.5$&$  3.2$&
$  2.3$&$  0.0$&$  0.7$&$  1.0$&$  0.1$&$  0.1$&$  0.7$&$  0.8$&$  0.0$&$-$&
$0.894$ & $0.899$ & $  0.5$ & $ -0.3$ & $ -0.8$ \\
$  400$&$0.02000$&$0.197$&$0.731$&$  2.1$&$  2.8$&$  3.5$&
$  2.5$&$  1.1$&$  0.5$&$  1.4$&$  1.0$&$  0.3$&$  0.4$&$  0.8$&$  0.0$&$-$&
$0.732$ & $0.732$ & $  0.4$ & $ -0.2$ & $ -0.3$ \\
$  400$&$0.03200$&$0.123$&$0.608$&$  2.1$&$  2.4$&$  3.2$&
$  2.3$&$  0.6$&$  0.4$&$  0.9$&$  0.6$&$ -0.4$&$  0.2$&$  0.5$&$  0.0$&$-$&
$0.608$ & $0.608$ & $  0.4$ & $ -0.2$ & $ -0.1$ \\
$  400$&$0.05000$&$0.079$&$0.513$&$  2.3$&$  3.1$&$  3.9$&
$  2.7$&$  1.5$&$  0.6$&$  1.6$&$  1.4$&$ -0.2$&$ -0.7$&$  0.5$&$  0.0$&$-$&
$0.513$ & $0.512$ & $  0.4$ & $ -0.2$ & $  0.0$ \\
$  400$&$0.08000$&$0.049$&$0.445$&$  2.4$&$  3.2$&$  4.0$&
$  2.4$&$  0.4$&$  0.4$&$  2.1$&$  0.3$&$ -0.2$&$ -0.1$&$  2.1$&$  0.0$&$-$&
$0.445$ & $0.444$ & $  0.4$ & $ -0.2$ & $  0.0$ \\
$  400$&$0.13000$&$0.030$&$0.356$&$  2.4$&$  3.2$&$  4.0$&
$  2.8$&$  1.5$&$  0.8$&$  1.5$&$  1.3$&$ -0.5$&$ -0.5$&$  0.4$&$  0.0$&$-$&
$0.356$ & $0.356$ & $  0.3$ & $ -0.1$ & $  0.0$ \\
$  400$&$0.25000$&$0.016$&$0.244$&$  3.3$&$  7.5$&$  8.2$&
$  4.3$&$  2.8$&$  1.9$&$  6.1$&$  2.6$&$ -0.6$&$ -1.7$&$ -5.3$&$  0.0$&$-$&
$0.244$ & $0.244$ & $  0.2$ & $ -0.1$ & $  0.0$ \\
$  400$&$0.40000$&$0.010$&$0.150$&$  4.7$&$ 10.3$&$ 11.3$&
$  4.8$&$  2.0$&$  2.5$&$  9.1$&$  1.9$&$ -0.1$&$ -1.4$&$ -8.8$&$  0.0$&$-$&
$0.150$ & $0.150$ & $  0.2$ & $ -0.1$ & $  0.0$ \\ \hline
$  500$&$0.00656$&$0.750$&$1.021$&$  3.3$&$  3.1$&$  4.5$&
$  2.9$&$  0.1$&$  0.3$&$  1.1$&$  0.1$&$ -0.9$&$  0.5$&$  0.1$&$    -$&$0.3$&
$1.022$ & $  -  $ & $  -  $ & $  -  $ & $  -  $ \\
$  500$&$0.00800$&$0.615$&$1.032$&$  3.5$&$  3.1$&$  4.7$&
$  3.0$&$  0.7$&$  0.3$&$  1.0$&$ -0.8$&$ -0.1$&$  0.3$&$  0.2$&$ -0.5$&$-$&
$1.033$ & $  -  $ & $  -  $ & $  -  $ & $  -  $ \\
$  500$&$0.01300$&$0.379$&$0.896$&$  2.4$&$  3.2$&$  4.0$&
$  2.7$&$  1.1$&$  0.7$&$  1.8$&$  1.2$&$ -1.1$&$  0.8$&$  0.6$&$ -0.1$&$-$&
$0.897$ & $0.906$ & $  0.6$ & $ -0.4$ & $ -1.3$ \\
$  500$&$0.02000$&$0.246$&$0.720$&$  2.5$&$  2.9$&$  3.8$&
$  2.5$&$  0.4$&$  0.8$&$  1.4$&$ -0.4$&$ -1.0$&$  0.5$&$  0.9$&$  0.0$&$-$&
$0.721$ & $0.722$ & $  0.7$ & $ -0.4$ & $ -0.4$ \\
$  500$&$0.03200$&$0.154$&$0.619$&$  2.5$&$  2.6$&$  3.6$&
$  2.5$&$  0.8$&$  0.2$&$  1.0$&$  0.9$&$ -0.2$&$  0.2$&$  0.4$&$  0.0$&$-$&
$0.619$ & $0.619$ & $  0.6$ & $ -0.3$ & $ -0.1$ \\
$  500$&$0.05000$&$0.098$&$0.536$&$  2.6$&$  3.3$&$  4.2$&
$  2.8$&$  1.6$&$  0.1$&$  1.7$&$  1.6$&$  0.1$&$  0.0$&$  0.8$&$  0.0$&$-$&
$0.536$ & $0.536$ & $  0.5$ & $ -0.3$ & $  0.0$ \\
$  500$&$0.08000$&$0.062$&$0.440$&$  2.8$&$  3.0$&$  4.1$&
$  2.6$&$  1.0$&$  0.1$&$  1.4$&$  1.1$&$  0.4$&$ -0.6$&$  0.8$&$  0.0$&$-$&
$0.440$ & $0.439$ & $  0.6$ & $ -0.3$ & $  0.0$ \\
$  500$&$0.13000$&$0.038$&$0.344$&$  3.4$&$  3.5$&$  4.9$&
$  2.7$&$  0.7$&$  0.1$&$  2.3$&$  0.7$&$  0.7$&$ -0.4$&$  2.1$&$  0.0$&$-$&
$0.344$ & $0.344$ & $  0.4$ & $ -0.2$ & $  0.0$ \\
$  500$&$0.18000$&$0.027$&$0.294$&$  3.7$&$  4.5$&$  5.9$&
$  3.7$&$  2.0$&$  1.4$&$  2.6$&$  2.0$&$  0.6$&$ -0.8$&$ -1.3$&$  0.0$&$-$&
$0.294$ & $0.294$ & $  0.4$ & $ -0.2$ & $  0.0$ \\
$  500$&$0.25000$&$0.020$&$0.269$&$  4.6$&$  7.1$&$  8.5$&
$  4.6$&$  2.6$&$  1.8$&$  5.5$&$  2.7$&$  0.2$&$ -1.1$&$ -4.7$&$  0.0$&$-$&
$0.269$ & $0.268$ & $  0.4$ & $ -0.1$ & $  0.0$ \\
$  500$&$0.40000$&$0.012$&$0.166$&$  6.6$&$ 14.7$&$ 16.1$&
$  8.0$&$  5.3$&$  3.4$&$ 12.3$&$  5.3$&$  1.1$&$ -3.3$&$-10.6$&$  0.0$&$-$&
$0.166$ & $0.166$ & $  0.3$ & $ -0.1$ & $  0.0$ \\ \hline
$  650$&$0.00853$&$0.750$&$0.989$&$  3.8$&$  3.8$&$  5.4$&
$  3.7$&$  0.5$&$  0.1$&$  0.9$&$ -0.6$&$ -0.5$&$  0.5$&$  0.1$&$    -$&$0.1$&
$0.989$ & $  -  $ & $  -  $ & $  -  $ & $  -  $ \\
$  650$&$0.01300$&$0.492$&$0.862$&$  3.0$&$  2.9$&$  4.2$&
$  2.7$&$  0.2$&$  0.7$&$  1.1$&$ -0.2$&$ -0.4$&$  0.8$&$  0.6$&$ -0.2$&$-$&
$0.863$ & $0.882$ & $  0.9$ & $ -0.7$ & $ -2.3$ \\
$  650$&$0.02000$&$0.320$&$0.750$&$  3.0$&$  2.9$&$  4.2$&
$  2.8$&$  0.2$&$  1.0$&$  1.0$&$  0.2$&$  0.2$&$  0.8$&$  0.6$&$  0.0$&$-$&
$0.751$ & $0.755$ & $  0.8$ & $ -0.7$ & $ -0.7$ \\
$  650$&$0.03200$&$0.200$&$0.609$&$  2.9$&$  3.1$&$  4.2$&
$  2.7$&$  0.8$&$  0.7$&$  1.4$&$  0.8$&$ -0.7$&$  0.5$&$  0.8$&$  0.0$&$-$&
$0.610$ & $0.609$ & $  0.9$ & $ -0.6$ & $ -0.2$ \\
$  650$&$0.05000$&$0.128$&$0.519$&$  3.0$&$  3.0$&$  4.2$&
$  2.7$&$  0.9$&$  0.1$&$  1.2$&$  1.0$&$ -0.5$&$  0.1$&$  0.6$&$  0.0$&$-$&
$0.520$ & $0.518$ & $  0.8$ & $ -0.5$ & $ -0.1$ \\
$  650$&$0.08000$&$0.080$&$0.442$&$  3.2$&$  3.3$&$  4.6$&
$  2.8$&$  1.1$&$  0.1$&$  1.7$&$  1.1$&$ -0.9$&$ -0.6$&$  0.9$&$  0.0$&$-$&
$0.442$ & $0.441$ & $  0.7$ & $ -0.4$ & $  0.0$ \\
$  650$&$0.13000$&$0.049$&$0.375$&$  3.7$&$  3.5$&$  5.1$&
$  3.0$&$  0.7$&$  0.2$&$  1.9$&$  0.8$&$ -0.9$&$ -0.1$&$  1.5$&$  0.0$&$-$&
$0.375$ & $0.374$ & $  0.7$ & $ -0.3$ & $  0.0$ \\
$  650$&$0.18000$&$0.036$&$0.299$&$  4.3$&$  5.7$&$  7.1$&
$  4.5$&$  2.9$&$  1.6$&$  3.4$&$  2.9$&$ -1.1$&$ -1.1$&$ -1.0$&$  0.0$&$-$&
$0.299$ & $0.298$ & $  0.6$ & $ -0.3$ & $  0.0$ \\
$  650$&$0.25000$&$0.026$&$0.253$&$  5.2$&$  7.8$&$  9.4$&
$  5.6$&$  3.1$&$  2.8$&$  5.5$&$  3.1$&$ -1.0$&$ -2.1$&$ -3.9$&$  0.0$&$-$&
$0.253$ & $0.252$ & $  0.6$ & $ -0.2$ & $  0.0$ \\
$  650$&$0.40000$&$0.016$&$0.147$&$  8.1$&$ 14.7$&$ 16.8$&
$  7.9$&$  4.5$&$  3.2$&$ 12.5$&$  4.5$&$ -1.8$&$ -2.7$&$-11.2$&$  0.0$&$-$&
$0.147$ & $0.147$ & $  0.5$ & $ -0.1$ & $  0.0$ \\ \hline
$  800$&$0.01050$&$0.750$&$0.948$&$  4.5$&$  4.4$&$  6.3$&
$  4.2$&$  1.2$&$  0.1$&$  1.5$&$  0.9$&$ -1.2$&$  0.2$&$  0.0$&$    -$&$0.2$&
$0.949$ & $  -  $ & $  -  $ & $  -  $ & $  -  $ \\
$  800$&$0.01300$&$0.606$&$0.864$&$  4.6$&$  3.8$&$  6.0$&
$  3.5$&$  0.2$&$  0.2$&$  1.5$&$  0.3$&$ -1.5$&$  0.2$&$  0.3$&$ -0.3$&$-$&
$0.865$ & $  -  $ & $  -  $ & $  -  $ & $  -  $ \\
$  800$&$0.02000$&$0.394$&$0.755$&$  3.4$&$  3.3$&$  4.7$&
$  3.0$&$  0.1$&$  1.1$&$  1.3$&$  0.0$&$ -0.5$&$  1.0$&$  0.8$&$ -0.1$&$-$&
$0.756$ & $0.764$ & $  1.1$ & $ -1.0$ & $ -1.1$ \\
$  800$&$0.03200$&$0.246$&$0.584$&$  3.6$&$  3.0$&$  4.7$&
$  2.9$&$  0.7$&$  0.3$&$  0.8$&$ -0.8$&$  0.0$&$  0.2$&$  0.3$&$ -0.1$&$-$&
$0.585$ & $0.586$ & $  1.1$ & $ -0.9$ & $ -0.3$ \\
$  800$&$0.05000$&$0.158$&$0.554$&$  3.5$&$  3.0$&$  4.6$&
$  2.9$&$  0.5$&$  0.5$&$  0.9$&$  0.6$&$  0.5$&$  0.5$&$  0.3$&$  0.0$&$-$&
$0.555$ & $0.553$ & $  1.1$ & $ -0.8$ & $ -0.1$ \\
$  800$&$0.08000$&$0.098$&$0.451$&$  3.8$&$  3.2$&$  5.0$&
$  3.0$&$  0.7$&$  0.2$&$  1.2$&$  0.7$&$ -0.5$&$ -0.4$&$  0.8$&$  0.0$&$-$&
$0.451$ & $0.450$ & $  1.0$ & $ -0.6$ & $  0.0$ \\
$  800$&$0.13000$&$0.061$&$0.358$&$  4.5$&$  4.1$&$  6.1$&
$  3.5$&$  1.1$&$  0.9$&$  2.1$&$  1.1$&$ -1.4$&$  0.0$&$  1.3$&$  0.0$&$-$&
$0.358$ & $0.357$ & $  1.0$ & $ -0.5$ & $  0.0$ \\
$  800$&$0.18000$&$0.044$&$0.324$&$  4.9$&$  4.7$&$  6.7$&
$  4.1$&$  1.8$&$  1.2$&$  2.2$&$  1.8$&$ -0.3$&$ -0.4$&$  1.1$&$  0.0$&$-$&
$0.324$ & $0.323$ & $  0.9$ & $ -0.4$ & $  0.0$ \\
$  800$&$0.25000$&$0.032$&$0.257$&$  5.8$&$  9.0$&$ 10.7$&
$  6.2$&$  4.0$&$  2.5$&$  6.6$&$  4.0$&$ -1.3$&$ -2.0$&$ -4.7$&$  0.0$&$-$&
$0.257$ & $0.256$ & $  0.8$ & $ -0.3$ & $  0.0$ \\
$  800$&$0.40000$&$0.020$&$0.128$&$  9.5$&$ 12.1$&$ 15.4$&
$  8.5$&$  5.0$&$  3.4$&$  8.6$&$  5.2$&$ -1.8$&$ -1.8$&$ -6.5$&$  0.0$&$-$&
$0.128$ & $0.127$ & $  0.7$ & $ -0.2$ & $  0.0$ \\
$  800$&$0.65000$&$0.012$&$0.0186$&$ 16.4$&$ 16.5$&$ 23.3$&
$ 10.4$&$  2.8$&$  2.2$&$ 12.9$&$  2.9$&$  0.7$&$ -2.2$&$-12.4$&$  0.0$&$-$&
$0.0186$ & $0.0182$ & $  2.5$ & $ -0.1$ & $  0.0$ \\ \hline
$ 1000$&$0.01300$&$0.757$&$0.886$&$  5.0$&$  4.0$&$  6.4$&
$  3.7$&$  0.1$&$  0.3$&$  1.5$&$ -0.2$&$ -0.6$&$  0.6$&$  0.1$&$ -1.2$&$-$&
$0.888$ & $  -  $ & $  -  $ & $  -  $ & $  -  $ \\
$ 1000$&$0.02000$&$0.492$&$0.745$&$  3.9$&$  3.4$&$  5.2$&
$  3.0$&$  0.0$&$  1.2$&$  1.5$&$  0.1$&$ -0.7$&$  1.1$&$  0.8$&$ -0.2$&$-$&
$0.746$ & $0.761$ & $  1.6$ & $ -1.6$ & $ -1.9$ \\
$ 1000$&$0.03200$&$0.308$&$0.653$&$  4.0$&$  4.0$&$  5.6$&
$  3.3$&$  1.5$&$  1.0$&$  2.2$&$  1.5$&$ -1.5$&$  0.7$&$  0.6$&$  0.0$&$-$&
$0.653$ & $0.656$ & $  1.5$ & $ -1.4$ & $ -0.5$ \\
$ 1000$&$0.05000$&$0.197$&$0.556$&$  4.1$&$  3.6$&$  5.5$&
$  3.1$&$  1.4$&$  0.0$&$  1.7$&$  1.3$&$ -1.1$&$  0.0$&$  0.2$&$  0.0$&$-$&
$0.557$ & $0.557$ & $  1.4$ & $ -1.2$ & $ -0.2$ \\
$ 1000$&$0.08000$&$0.123$&$0.461$&$  4.3$&$  3.1$&$  5.3$&
$  3.0$&$  0.9$&$  0.0$&$  0.9$&$  0.9$&$  0.0$&$  0.0$&$  0.4$&$  0.0$&$-$&
$0.462$ & $0.460$ & $  1.4$ & $ -1.0$ & $  0.0$ \\
$ 1000$&$0.13000$&$0.076$&$0.412$&$  5.2$&$  4.0$&$  6.5$&
$  3.4$&$  0.7$&$  0.0$&$  2.1$&$  0.8$&$  0.1$&$  0.3$&$  1.9$&$  0.0$&$-$&
$0.413$ & $0.410$ & $  1.3$ & $ -0.8$ & $  0.0$ \\
$ 1000$&$0.18000$&$0.055$&$0.319$&$  5.7$&$  3.8$&$  6.8$&
$  3.6$&$  1.1$&$  0.0$&$  1.2$&$  1.1$&$ -0.6$&$ -0.3$&$ -0.1$&$  0.0$&$-$&
$0.319$ & $0.317$ & $  1.2$ & $ -0.6$ & $  0.0$ \\
$ 1000$&$0.25000$&$0.039$&$0.263$&$  6.3$&$  5.8$&$  8.6$&
$  4.8$&$  2.2$&$  2.0$&$  3.3$&$  2.3$&$  0.8$&$ -1.4$&$ -1.8$&$  0.0$&$-$&
$0.263$ & $0.261$ & $  1.2$ & $ -0.5$ & $  0.0$ \\
$ 1000$&$0.40000$&$0.025$&$0.130$&$ 10.1$&$  9.6$&$ 14.0$&
$  7.3$&$  3.6$&$  2.9$&$  6.2$&$  3.6$&$  0.3$&$ -1.3$&$ -4.9$&$  0.0$&$-$&
$0.130$ & $0.129$ & $  1.0$ & $ -0.3$ & $  0.0$ \\
$ 1000$&$0.65000$&$0.015$&$0.0221$&$ 17.7$&$ 21.6$&$ 27.9$&
$ 12.3$&$  5.4$&$  4.8$&$ 17.7$&$  5.4$&$  0.4$&$ -4.6$&$-16.3$&$  0.0$&$-$&
$0.0221$ & $0.0214$ & $  3.5$ & $ -0.2$ & $  0.0$ \\ \hline
$ 1200$&$0.02000$&$0.591$&$0.777$&$  4.8$&$  3.4$&$  5.9$&
$  3.0$&$  0.4$&$  0.9$&$  1.5$&$  0.3$&$ -0.9$&$  1.1$&$  0.7$&$ -0.1$&$-$&
$0.779$ & $0.803$ & $  2.1$ & $ -2.3$ & $ -2.9$ \\
$ 1200$&$0.03200$&$0.369$&$0.648$&$  4.5$&$  3.0$&$  5.4$&
$  2.8$&$  0.2$&$  1.2$&$  1.2$&$  0.2$&$  0.2$&$  1.0$&$  0.7$&$  0.0$&$-$&
$0.649$ & $0.655$ & $  2.0$ & $ -2.0$ & $ -0.8$ \\
$ 1200$&$0.05000$&$0.236$&$0.476$&$  4.9$&$  3.2$&$  5.8$&
$  2.6$&$  0.9$&$  0.3$&$  1.8$&$  0.9$&$ -1.6$&$ -0.3$&$  0.4$&$  0.0$&$-$&
$0.477$ & $0.477$ & $  1.9$ & $ -1.8$ & $ -0.2$ \\
$ 1200$&$0.08000$&$0.148$&$0.468$&$  4.8$&$  3.1$&$  5.7$&
$  2.7$&$  0.9$&$  0.3$&$  1.5$&$  0.9$&$ -1.2$&$  0.2$&$  0.4$&$  0.0$&$-$&
$0.469$ & $0.467$ & $  1.9$ & $ -1.4$ & $ -0.1$ \\
$ 1200$&$0.13000$&$0.091$&$0.422$&$  5.7$&$  3.8$&$  6.8$&
$  3.3$&$  1.7$&$  0.4$&$  1.8$&$  1.6$&$ -0.7$&$ -0.3$&$  0.5$&$  0.0$&$-$&
$0.423$ & $0.420$ & $  1.7$ & $ -1.1$ & $  0.0$ \\
$ 1200$&$0.18000$&$0.066$&$0.324$&$  6.5$&$  3.3$&$  7.3$&
$  3.1$&$  0.7$&$  0.1$&$  1.2$&$  0.7$&$ -0.7$&$ -0.1$&$  0.8$&$  0.0$&$-$&
$0.324$ & $0.322$ & $  1.6$ & $ -0.9$ & $  0.0$ \\
$ 1200$&$0.25000$&$0.047$&$0.268$&$  7.0$&$  4.4$&$  8.3$&
$  3.8$&$  1.6$&$  1.3$&$  2.1$&$  1.6$&$ -0.8$&$ -1.0$&$ -0.3$&$  0.0$&$-$&
$0.268$ & $0.266$ & $  1.6$ & $ -0.7$ & $  0.0$ \\
$ 1200$&$0.40000$&$0.030$&$0.118$&$ 11.6$&$ 12.4$&$ 16.9$&
$  7.8$&$  5.2$&$  3.8$&$  9.6$&$  5.2$&$ -1.3$&$ -3.2$&$ -7.3$&$  0.0$&$-$&
$0.118$ & $0.117$ & $  1.4$ & $ -0.4$ & $  0.0$ \\
$ 1200$&$0.65000$&$0.018$&$0.0238$&$ 19.6$&$ 20.3$&$ 28.2$&
$ 13.6$&$  9.1$&$  5.4$&$ 15.0$&$  9.0$&$ -2.8$&$ -4.0$&$-11.0$&$  0.0$&$-$&
$0.0239$ & $0.0230$ & $  3.9$ & $ -0.3$ & $  0.0$ \\ \hline
 \end{tabular}}
 
 \begin{table}[htbp]
 \begin{center}
  \tiny
  \rotatebox{90}{\tncb}
 \end{center}
 \end{table}

\newcommand{\tncc}{
 \begin{tabular}{|r|c|c||c|r|r|r||r|r|r|r|r|r|r|r|r|r||c|r|r|r|r|}
 \hline
 $Q^2$  &$x$ &$y$ & $\tilde{\sigma}_{NC}$ &
 $\delta_{stat}$ & $\delta_{sys}$ & $\delta_{tot}$ &
 $\delta_{unc}$ &
 $\delta_{unc}^{E}$ &
 $\delta_{unc}^{h}$&
 $\delta_{cor}$ &
 $\delta_{cor}^{E^+}$ &
 $\delta_{cor}^{\theta^+}$&
 $\delta_{cor}^{h^+}$&
 $\delta_{cor}^{N^+}$&
 $\delta_{cor}^{B^+}$&
 $\delta_{cor}^{S^+}$ &
 $\phi_{NC}/Y_+$ & $F_2$&
 $\Delta_{F_2}$& $\Delta_{F_3}$& $\Delta_{F_L}$ \\
 $(\rm GeV^2)$ & & & &
 $(\%)$ &$(\%)$ &$(\%)$ & 
 $(\%)$ &$(\%)$ &$(\%)$ &$(\%)$ &$(\%)$ &
 $(\%)$ &$(\%)$ &$(\%)$ &$(\%)$ &$(\%)$ & &
        &$(\%)$ &$(\%)$ &$(\%)$ \\ \hline \hline
$ 1500$&$0.02000$&$0.738$&$0.728$&$  6.4$&$  4.0$&$  7.5$&
$  3.8$&$  1.1$&$  0.4$&$  1.1$&$ -0.6$&$ -0.7$&$  0.0$&$  0.0$&$ -0.7$&$-$&
$0.730$ & $  -  $ & $  -  $ & $  -  $ & $  -  $ \\
$ 1500$&$0.03200$&$0.462$&$0.597$&$  5.9$&$  3.5$&$  6.8$&
$  3.3$&$  0.4$&$  1.3$&$  1.1$&$  0.0$&$ -0.5$&$  1.0$&$  0.4$&$  0.0$&$-$&
$0.598$ & $0.609$ & $  2.6$ & $ -3.1$ & $ -1.3$ \\
$ 1500$&$0.05000$&$0.295$&$0.585$&$  5.2$&$  2.9$&$  6.0$&
$  2.8$&$  0.5$&$  0.4$&$  0.8$&$  0.5$&$  0.5$&$  0.3$&$  0.4$&$  0.0$&$-$&
$0.586$ & $0.590$ & $  2.5$ & $ -2.7$ & $ -0.4$ \\
$ 1500$&$0.08000$&$0.185$&$0.427$&$  5.5$&$  2.9$&$  6.2$&
$  2.7$&$  0.3$&$  0.5$&$  1.0$&$  0.3$&$ -0.5$&$  0.6$&$  0.5$&$  0.0$&$-$&
$0.428$ & $0.427$ & $  2.5$ & $ -2.2$ & $ -0.1$ \\
$ 1500$&$0.13000$&$0.114$&$0.344$&$  6.7$&$  3.2$&$  7.4$&
$  3.1$&$  0.5$&$  0.3$&$  1.1$&$  0.6$&$ -0.6$&$ -0.3$&$  0.6$&$  0.0$&$-$&
$0.345$ & $0.343$ & $  2.2$ & $ -1.7$ & $  0.0$ \\
$ 1500$&$0.18000$&$0.082$&$0.329$&$  7.1$&$  4.5$&$  8.4$&
$  3.6$&$  1.8$&$  0.1$&$  2.7$&$  1.8$&$ -1.8$&$ -0.3$&$  0.9$&$  0.0$&$-$&
$0.330$ & $0.327$ & $  2.2$ & $ -1.3$ & $  0.0$ \\
$ 1500$&$0.25000$&$0.059$&$0.226$&$  8.0$&$  4.9$&$  9.4$&
$  4.2$&$  2.3$&$  1.1$&$  2.7$&$  2.3$&$ -1.0$&$ -0.7$&$  0.7$&$  0.0$&$-$&
$0.226$ & $0.224$ & $  2.1$ & $ -1.0$ & $  0.0$ \\
$ 1500$&$0.40000$&$0.037$&$0.104$&$ 12.1$&$  9.9$&$ 15.7$&
$  6.7$&$  3.2$&$  3.9$&$  7.3$&$  3.2$&$  1.3$&$ -3.5$&$ -5.4$&$  0.0$&$-$&
$0.104$ & $0.103$ & $  1.8$ & $ -0.6$ & $  0.0$ \\
$ 1500$&$0.65000$&$0.023$&$0.0166$&$ 30.2$&$ 19.4$&$ 35.9$&
$ 14.4$&$  8.4$&$  5.6$&$ 13.0$&$  8.5$&$ -2.7$&$ -3.9$&$ -8.7$&$  0.0$&$-$&
$0.0166$ & $0.0160$ & $  4.4$ & $ -0.4$ & $  0.0$ \\ \hline
$ 2000$&$0.03200$&$0.615$&$0.642$&$  6.4$&$  4.2$&$  7.7$&
$  3.9$&$  0.5$&$  1.5$&$  1.6$&$  0.6$&$ -0.5$&$  1.3$&$  0.5$&$  0.0$&$-$&
$0.644$ & $0.671$ & $  3.8$ & $ -5.3$ & $ -2.5$ \\
$ 2000$&$0.05000$&$0.394$&$0.472$&$  7.4$&$  4.0$&$  8.4$&
$  3.7$&$  0.9$&$  0.9$&$  1.7$&$  1.2$&$ -0.6$&$  0.8$&$  0.6$&$  0.0$&$-$&
$0.473$ & $0.481$ & $  3.8$ & $ -4.7$ & $ -0.7$ \\
$ 2000$&$0.08000$&$0.246$&$0.405$&$  6.7$&$  3.5$&$  7.6$&
$  3.2$&$  0.4$&$  0.9$&$  1.5$&$ -0.4$&$ -1.0$&$  0.8$&$  0.7$&$  0.0$&$-$&
$0.406$ & $0.408$ & $  3.5$ & $ -3.8$ & $ -0.2$ \\
$ 2000$&$0.13000$&$0.151$&$0.368$&$  7.4$&$  4.0$&$  8.4$&
$  3.6$&$  1.5$&$  0.4$&$  1.7$&$  1.6$&$ -0.8$&$  0.1$&$  0.0$&$  0.0$&$-$&
$0.369$ & $0.367$ & $  3.3$ & $ -2.8$ & $  0.0$ \\
$ 2000$&$0.18000$&$0.109$&$0.342$&$  8.3$&$  3.9$&$  9.2$&
$  3.7$&$  0.3$&$  0.6$&$  1.0$&$  0.3$&$  0.2$&$ -0.8$&$  0.5$&$  0.0$&$-$&
$0.343$ & $0.340$ & $  3.1$ & $ -2.2$ & $  0.0$ \\
$ 2000$&$0.25000$&$0.079$&$0.233$&$  9.0$&$  4.9$&$ 10.2$&
$  4.4$&$  2.1$&$  1.0$&$  2.3$&$  2.2$&$ -0.3$&$ -0.6$&$ -0.3$&$  0.0$&$-$&
$0.233$ & $0.230$ & $  3.0$ & $ -1.7$ & $  0.0$ \\
$ 2000$&$0.40000$&$0.049$&$0.108$&$ 12.4$&$ 10.1$&$ 16.0$&
$  7.5$&$  5.3$&$  2.6$&$  6.8$&$  5.3$&$ -0.9$&$ -2.4$&$ -3.3$&$  0.0$&$-$&
$0.108$ & $0.106$ & $  2.8$ & $ -1.1$ & $  0.0$ \\
$ 2000$&$0.65000$&$0.030$&$0.0294$&$ 23.6$&$ 20.5$&$ 31.3$&
$ 14.7$&$  7.4$&$  6.9$&$ 14.2$&$  7.4$&$ -1.4$&$ -3.9$&$-11.4$&$  0.0$&$-$&
$0.0294$ & $0.0287$ & $  3.2$ & $ -0.7$ & $  0.0$ \\ \hline
$ 3000$&$0.05000$&$0.591$&$0.551$&$  5.6$&$  3.9$&$  6.8$&
$  3.6$&$  0.4$&$  1.4$&$  1.5$&$  0.0$&$ -0.7$&$  1.2$&$  0.6$&$ -0.1$&$-$&
$0.553$ & $0.585$ & $  6.3$ & $ -9.9$ & $ -1.7$ \\
$ 3000$&$0.08000$&$0.369$&$0.439$&$  6.2$&$  4.0$&$  7.4$&
$  3.6$&$  0.5$&$  1.1$&$  1.8$&$  1.2$&$ -0.5$&$  0.9$&$  0.8$&$  0.0$&$-$&
$0.441$ & $0.452$ & $  6.0$ & $ -8.2$ & $ -0.4$ \\
$ 3000$&$0.13000$&$0.227$&$0.340$&$  7.3$&$  3.8$&$  8.2$&
$  3.7$&$  1.0$&$  0.5$&$  0.9$&$  0.8$&$ -0.4$&$ -0.4$&$  0.1$&$  0.0$&$-$&
$0.341$ & $0.343$ & $  5.7$ & $ -6.1$ & $ -0.1$ \\
$ 3000$&$0.18000$&$0.164$&$0.313$&$  7.8$&$  5.9$&$  9.8$&
$  5.0$&$  3.3$&$  1.0$&$  3.2$&$  3.1$&$ -0.3$&$ -0.5$&$  0.1$&$  0.0$&$-$&
$0.314$ & $0.312$ & $  5.3$ & $ -4.7$ & $  0.0$ \\
$ 3000$&$0.25000$&$0.118$&$0.195$&$  9.7$&$  4.7$&$ 10.8$&
$  4.0$&$  1.1$&$  0.7$&$  2.4$&$  2.3$&$  0.4$&$ -0.7$&$  0.0$&$  0.0$&$-$&
$0.195$ & $0.192$ & $  5.1$ & $ -3.6$ & $  0.0$ \\
$ 3000$&$0.40000$&$0.074$&$0.108$&$ 11.5$&$  6.4$&$ 13.1$&
$  5.5$&$  2.8$&$  1.4$&$  3.3$&$  2.9$&$  0.3$&$ -1.2$&$ -1.1$&$  0.0$&$-$&
$0.108$ & $0.106$ & $  4.6$ & $ -2.3$ & $  0.0$ \\
$ 3000$&$0.65000$&$0.045$&$0.0118$&$ 28.9$&$ 16.8$&$ 33.4$&
$ 12.4$&$  6.7$&$  5.4$&$ 11.4$&$  6.6$&$  1.2$&$ -3.5$&$ -8.6$&$  0.0$&$-$&
$0.0118$ & $0.0113$ & $  5.7$ & $ -1.4$ & $  0.0$ \\ \hline
$ 5000$&$0.08000$&$0.615$&$0.422$&$  7.0$&$  4.3$&$  8.2$&
$  4.1$&$  0.9$&$  1.6$&$  1.5$&$  0.1$&$ -0.6$&$  1.3$&$  0.5$&$ -0.1$&$-$&
$0.425$ & $0.475$ & $ 11.0$ & $-20.1$ & $ -1.4$ \\
$ 5000$&$0.13000$&$0.379$&$0.339$&$  8.3$&$  4.3$&$  9.3$&
$  4.3$&$  1.2$&$  1.0$&$  0.9$&$  0.6$&$ -0.2$&$  0.6$&$  0.2$&$  0.0$&$-$&
$0.341$ & $0.359$ & $ 10.4$ & $-15.2$ & $ -0.3$ \\
$ 5000$&$0.18000$&$0.273$&$0.263$&$  9.4$&$  4.7$&$ 10.5$&
$  4.7$&$  2.2$&$  0.3$&$  0.5$&$ -0.5$&$ -0.1$&$ -0.1$&$  0.0$&$  0.0$&$-$&
$0.263$ & $0.269$ & $  9.9$ & $-11.8$ & $ -0.1$ \\
$ 5000$&$0.25000$&$0.197$&$0.301$&$  9.6$&$  9.4$&$ 13.4$&
$  9.0$&$  7.5$&$  1.3$&$  2.9$&$  2.8$&$ -0.5$&$ -0.6$&$  0.3$&$  0.0$&$-$&
$0.302$ & $0.300$ & $  9.4$ & $ -8.8$ & $  0.0$ \\
$ 5000$&$0.40000$&$0.123$&$0.130$&$ 12.3$&$  6.1$&$ 13.7$&
$  5.9$&$  2.5$&$  0.3$&$  1.2$&$  1.1$&$  0.2$&$  0.3$&$  0.5$&$  0.0$&$-$&
$0.130$ & $0.127$ & $  8.3$ & $ -5.5$ & $  0.0$ \\
$ 5000$&$0.65000$&$0.076$&$0.00760$&$ 40.9$&$ 25.0$&$ 47.9$&
$ 21.3$&$ 17.1$&$  7.9$&$ 13.2$&$  8.9$&$  2.4$&$ -6.3$&$ -7.1$&$  0.0$&$-$&
$0.00761$ & $0.00721$ & $  8.9$ & $ -3.4$ & $  0.0$ \\ \hline
$ 8000$&$0.13000$&$0.606$&$0.344$&$ 11.0$&$  5.5$&$ 12.3$&
$  5.4$&$  1.8$&$  1.2$&$  1.0$&$  0.6$&$  0.3$&$  0.7$&$  0.1$&$ -0.1$&$-$&
$0.347$ & $0.419$ & $ 17.0$ & $-33.1$ & $ -0.9$ \\
$ 8000$&$0.18000$&$0.438$&$0.361$&$ 10.7$&$  6.5$&$ 12.5$&
$  6.3$&$  3.1$&$  1.7$&$  1.6$&$  0.9$&$ -1.0$&$  0.9$&$  0.1$&$  0.0$&$-$&
$0.364$ & $0.406$ & $ 15.8$ & $-26.1$ & $ -0.3$ \\
$ 8000$&$0.25000$&$0.315$&$0.224$&$ 13.1$&$  7.1$&$ 14.9$&
$  6.7$&$  3.9$&$  0.0$&$  2.5$&$  2.4$&$  0.4$&$  0.2$&$  0.7$&$  0.0$&$-$&
$0.224$ & $0.234$ & $ 15.2$ & $-19.4$ & $ -0.1$ \\
$ 8000$&$0.40000$&$0.197$&$0.0982$&$ 18.0$&$ 10.3$&$ 20.7$&
$  9.8$&$  7.2$&$  1.2$&$  3.2$&$  3.2$&$ -0.3$&$ -0.3$&$  0.0$&$  0.0$&$-$&
$0.0984$ & $0.0971$ & $ 13.4$ & $-12.0$ & $  0.0$ \\
$ 8000$&$0.65000$&$0.121$&$0.0228$&$ 28.9$&$ 28.7$&$ 40.7$&
$ 26.6$&$ 22.7$&$  6.3$&$ 11.0$&$  8.2$&$  4.1$&$ -4.7$&$ -4.0$&$  0.0$&$-$&
$0.0228$ & $0.0213$ & $ 14.5$ & $ -7.3$ & $  0.0$ \\ \hline
$12000$&$0.18000$&$0.656$&$0.233$&$ 19.5$&$  5.0$&$ 20.1$&
$  4.6$&$  2.8$&$  0.4$&$  2.0$&$ -0.9$&$ -1.8$&$  0.4$&$  0.2$&$ -0.3$&$-$&
$0.236$ & $0.320$ & $ 22.7$ & $-48.1$ & $ -0.9$ \\
$12000$&$0.25000$&$0.473$&$0.117$&$ 23.6$&$  7.3$&$ 24.7$&
$  7.1$&$  6.1$&$  1.5$&$  1.5$&$  1.0$&$  0.6$&$  0.9$&$  0.3$&$  0.0$&$-$&
$0.117$ & $0.139$ & $ 21.1$ & $-36.7$ & $ -0.3$ \\
$12000$&$0.40000$&$0.295$&$0.0424$&$ 35.4$&$ 11.8$&$ 37.3$&
$ 11.6$&$ 11.1$&$  0.2$&$  2.0$&$  1.8$&$  0.7$&$ -0.3$&$  0.1$&$  0.0$&$-$&
$0.0425$ & $0.0441$ & $ 19.1$ & $-22.6$ & $ -0.1$ \\
$12000$&$0.65000$&$0.182$&$0.0179$&$ 40.8$&$ 28.9$&$ 50.0$&
$ 27.4$&$ 26.2$&$  4.8$&$  9.0$&$  6.4$&$  5.1$&$ -3.6$&$ -1.1$&$  0.0$&$-$&
$0.0180$ & $0.0170$ & $ 19.1$ & $-13.5$ & $  0.0$ \\ \hline
$20000$&$0.25000$&$0.788$&$0.106$&$ 38.3$&$  5.2$&$ 38.7$&
$  4.8$&$  2.9$&$  0.6$&$  2.0$&$  1.1$&$ -1.5$&$  0.9$&$  0.1$&$ -0.4$&$-$&
$0.109$ & $0.192$ & $ 29.5$ & $-71.8$ & $ -1.0$ \\
$20000$&$0.40000$&$0.492$&$0.0381$&$ 50.0$&$ 11.7$&$ 51.4$&
$ 11.5$&$ 11.0$&$  1.6$&$  2.4$&$  2.1$&$  0.7$&$  0.9$&$  0.3$&$  0.0$&$-$&
$0.0385$ & $0.0490$ & $ 26.7$ & $-48.0$ & $ -0.2$ \\
$20000$&$0.65000$&$0.303$&$0.0110$&$ 70.7$&$ 38.7$&$ 80.6$&
$ 38.0$&$ 37.5$&$  1.0$&$  7.7$&$  5.9$&$  4.9$&$ -0.7$&$ -0.3$&$  0.0$&$-$&
$0.0110$ & $0.0113$ & $ 26.2$ & $-28.6$ & $  0.0$ \\ \hline
$30000$&$0.40000$&$0.738$&$0.164$&$ 46.1$&$ 13.7$&$ 48.1$&
$ 13.4$&$ 12.5$&$  0.9$&$  2.4$&$ -0.5$&$ -2.2$&$ -0.4$&$  0.2$&$ -0.9$&$-$&
$0.168$ & $0.319$ & $ 32.4$ & $-79.1$ & $ -0.5$ \\ \hline
 \end{tabular}}

 \begin{table}[htbp]
 \begin{center}
  \tiny
  \rotatebox{90}{\tncc}
 \end{center}
 \end{table}
\newcommand{\ccxs}{
 \begin{tabular}{|r|r|r||l|r||r|r|r||r|r||r|r|r|r|r||r|}
 \hline
 $Q^2$ & $x$ & $y$ & 
 ${\rm d}^2\sigma_{CC}/dxdQ^2$ & ${\phi}_{CC}$ &
 $\delta_{stat}$ & $\delta_{sys}$ & $\delta_{tot}$ &
 $\delta_{unc}$ & $\delta_{unc}^{h}$ &
 $\delta_{cor}$ & $\delta_{cor}^{V^+}$ & $\delta_{cor}^{h^+}$& 
 $\delta_{cor}^{N^+}$ & $\delta_{cor}^{B^+}$ &
 $\Delta_{CC}^{QED}$ \\
 $(\rm GeV^2)$ & & &
 $(\rm pb / \rm GeV^2)$ & &
 $(\%)$ & $(\%)$ & $(\%)$ &
 $(\%)$ & $(\%)$ & 
 $(\%)$ & $(\%)$ & $(\%)$ & $(\%)$ & $(\%)$ &
 $(\%)$ \\ \hline\hline
$  300$&$0.0130$&$0.227$&
 $0.703 \cdot 10^{ 0}$&$1.184$&
  $ 20.3$&$ 11.8$&$ 23.5$&
   $  8.2$&$  1.5$&
    $  8.6$&$  6.7$&$ -2.2$&$ -0.3$&$ -4.9$&$  0.3  $\\
$  300$&$0.0320$&$0.092$&
 $0.283 \cdot 10^{ 0}$&$1.171$&
  $ 13.7$&$  5.9$&$ 14.9$&
   $  4.6$&$  1.3$&
    $  3.8$&$  2.6$&$ -1.4$&$  1.0$&$ -2.1$&$  0.4  $\\
$  300$&$0.0800$&$0.037$&
 $0.585 \cdot 10^{-1}$&$0.606$&
  $ 19.0$&$  7.4$&$ 20.4$&
   $  6.0$&$  2.9$&
    $  4.5$&$  1.0$&$ -2.4$&$  1.6$&$ -3.2$&$  5.2  $\\ \hline
$  500$&$0.0130$&$0.379$&
 $0.570 \cdot 10^{ 0}$&$1.018$&
  $ 14.6$&$  8.5$&$ 16.9$&
   $  6.5$&$  2.3$&
    $  5.4$&$  4.9$&$ -1.4$&$  0.4$&$ -1.8$&$ -4.4  $\\
$  500$&$0.0320$&$0.154$&
 $0.189 \cdot 10^{ 0}$&$0.829$&
  $ 12.1$&$  4.8$&$ 13.0$&
   $  3.9$&$  1.2$&
    $  2.8$&$  1.8$&$ -1.1$&$  0.8$&$ -1.6$&$ -0.7  $\\
$  500$&$0.0800$&$0.062$&
 $0.465 \cdot 10^{-1}$&$0.511$&
  $ 13.4$&$  4.4$&$ 14.1$&
   $  4.0$&$  0.4$&
    $  2.0$&$  0.4$&$ -0.7$&$  1.9$&$ -0.1$&$ -0.7  $\\
$  500$&$0.1300$&$0.038$&
 $0.194 \cdot 10^{-1}$&$0.346$&
  $ 25.1$&$  7.2$&$ 26.1$&
   $  6.4$&$  2.2$&
    $  2.9$&$  0.2$&$ -1.6$&$ -2.2$&$ -0.9$&$ -3.5  $\\ \hline
$ 1000$&$0.0320$&$0.308$&
 $0.121 \cdot 10^{ 0}$&$0.609$&
  $ 10.5$&$  4.7$&$ 11.5$&
   $  3.8$&$  1.3$&
    $  2.5$&$  1.9$&$ -1.3$&$  0.4$&$ -0.8$&$ -3.1  $\\
$ 1000$&$0.0800$&$0.123$&
 $0.406 \cdot 10^{-1}$&$0.512$&
  $ 10.2$&$  3.5$&$ 10.8$&
   $  3.3$&$  0.6$&
    $  1.2$&$  0.6$&$ -0.7$&$  0.8$&$  0.0$&$ -0.4  $\\
$ 1000$&$0.1300$&$0.076$&
 $0.162 \cdot 10^{-1}$&$0.332$&
  $ 16.5$&$  6.1$&$ 17.6$&
   $  5.9$&$  0.5$&
    $  0.9$&$  0.3$&$  0.5$&$  0.7$&$  0.0$&$ -2.5  $\\
$ 1000$&$0.2500$&$0.039$&
 $0.794 \cdot 10^{-2}$&$0.313$&
  $ 23.5$&$ 16.8$&$ 28.9$&
   $ 16.1$&$  2.0$&
    $  4.9$&$  0.0$&$  2.4$&$ -4.2$&$  0.0$&$ -4.9  $\\ \hline
$ 2000$&$0.0320$&$0.615$&
 $0.762 \cdot 10^{-1}$&$0.495$&
  $  9.8$&$  4.5$&$ 10.8$&
   $  4.0$&$  0.2$&
    $  2.3$&$  2.1$&$  0.1$&$  0.8$&$ -0.2$&$ -5.6  $\\
$ 2000$&$0.0800$&$0.246$&
 $0.228 \cdot 10^{-1}$&$0.370$&
  $  9.9$&$  3.8$&$ 10.6$&
   $  3.6$&$  0.6$&
    $  1.2$&$  0.8$&$ -0.8$&$  0.4$&$  0.0$&$ -3.1  $\\
$ 2000$&$0.1300$&$0.152$&
 $0.168 \cdot 10^{-1}$&$0.442$&
  $ 11.1$&$  5.7$&$ 12.5$&
   $  5.2$&$  2.0$&
    $  2.6$&$  0.1$&$ -1.3$&$  2.2$&$  0.0$&$ -5.0  $\\
$ 2000$&$0.2500$&$0.079$&
 $0.337 \cdot 10^{-2}$&$0.171$&
  $ 18.4$&$  7.3$&$ 19.8$&
   $  6.9$&$  0.3$&
    $  2.0$&$  0.0$&$ -1.4$&$ -1.5$&$  0.0$&$-10.0  $\\ \hline
$ 3000$&$0.0800$&$0.369$&
 $0.201 \cdot 10^{-1}$&$0.407$&
  $  8.7$&$  4.3$&$  9.7$&
   $  4.2$&$  1.2$&
    $  1.4$&$  0.9$&$  0.9$&$  0.2$&$  0.0$&$ -7.5  $\\
$ 3000$&$0.1300$&$0.227$&
 $0.107 \cdot 10^{-1}$&$0.354$&
  $ 10.7$&$  4.5$&$ 11.6$&
   $  3.8$&$  1.1$&
    $  2.1$&$  0.2$&$  1.2$&$  1.7$&$  0.0$&$ -3.9  $\\
$ 3000$&$0.2500$&$0.118$&
 $0.251 \cdot 10^{-2}$&$0.159$&
  $ 16.3$&$  6.1$&$ 17.4$&
   $  6.0$&$  0.4$&
    $  1.1$&$  0.1$&$  1.0$&$ -0.4$&$  0.0$&$ -5.7  $\\
$ 3000$&$0.4000$&$0.074$&
 $0.531 \cdot 10^{-3}$&$0.054$&
  $ 37.8$&$ 17.1$&$ 41.5$&
   $ 14.2$&$  3.9$&
    $  9.6$&$  0.0$&$  1.8$&$ -9.4$&$  0.0$&$-12.6  $\\ \hline
$ 5000$&$0.0800$&$0.615$&
 $0.842 \cdot 10^{-2}$&$0.250$&
  $ 13.8$&$  8.1$&$ 16.0$&
   $  5.9$&$  1.8$&
    $  5.6$&$  1.4$&$  3.8$&$  3.8$&$  0.0$&$-13.0  $\\
$ 5000$&$0.1300$&$0.379$&
 $0.530 \cdot 10^{-2}$&$0.256$&
  $ 12.3$&$  5.6$&$ 13.5$&
   $  5.1$&$  2.5$&
    $  2.3$&$  0.4$&$  2.2$&$  0.5$&$  0.0$&$-13.8  $\\
$ 5000$&$0.2500$&$0.197$&
 $0.192 \cdot 10^{-2}$&$0.179$&
  $ 14.2$&$  5.7$&$ 15.3$&
   $  5.3$&$  3.0$&
    $  1.7$&$  0.1$&$  1.4$&$  0.9$&$  0.0$&$ -9.8  $\\
$ 5000$&$0.4000$&$0.123$&
 $0.261 \cdot 10^{-3}$&$0.039$&
  $ 33.3$&$ 10.4$&$ 34.9$&
   $  9.5$&$  4.0$&
    $  4.6$&$  0.0$&$  4.5$&$ -0.5$&$  0.0$&$ -4.6  $\\ \hline
$ 8000$&$0.1300$&$0.606$&
 $0.178 \cdot 10^{-2}$&$0.137$&
  $ 20.3$&$ 12.6$&$ 23.9$&
   $ 11.4$&$  5.1$&
    $  5.4$&$  0.6$&$  4.7$&$  2.7$&$  0.0$&$-13.0  $\\
$ 8000$&$0.2500$&$0.315$&
 $0.903 \cdot 10^{-3}$&$0.134$&
  $ 17.3$&$ 11.0$&$ 20.5$&
   $  9.5$&$  8.1$&
    $  5.8$&$  0.0$&$  5.3$&$  2.2$&$  0.0$&$ -8.6  $\\
$ 8000$&$0.4000$&$0.197$&
 $0.152 \cdot 10^{-3}$&$0.036$&
  $ 40.8$&$ 26.0$&$ 48.4$&
   $ 25.2$&$ 14.5$&
    $  6.6$&$  0.0$&$  6.6$&$  0.2$&$  0.0$&$-14.8  $\\ \hline
$15000$&$0.2500$&$0.591$&
 $0.126 \cdot 10^{-3}$&$0.041$&
  $ 37.8$&$ 16.4$&$ 41.2$&
   $ 15.0$&$ 13.6$&
    $  6.8$&$  0.3$&$  6.5$&$  1.7$&$  0.0$&$-16.0  $\\
$15000$&$0.4000$&$0.369$&
 $0.117 \cdot 10^{-3}$&$0.061$&
  $ 28.8$&$ 23.2$&$ 37.0$&
   $ 22.0$&$ 13.4$&
    $  7.6$&$  0.0$&$  7.4$&$  1.3$&$  0.0$&$-14.7  $\\ \hline
\multicolumn{16}{c}{\begin{minipage}{199mm}
 \caption{\sl \label{ccdxdq2}The CC $e^+p$ 
 double differential cross section 
 ${\rm d}^2\sigma_{CC}/{\rm d}x{\rm d}Q^2$
 and the structure function term $\phi_{CC}$,
 shown with statistical ($\delta_{stat}$),
 systematic ($\delta_{sys}$) and
 total ($\delta_{tot}$) errors.
 Also shown are the total uncorrelated systematic error
 ($\delta_{unc}$) and its contribution from 
 the hadronic energy error ($\delta_{unc}^{h}$).
 The effect of the other uncorrelated
 systematic errors is included in ($\delta_{unc}$). 
 In addition the correlated systematic error
 ($\delta_{cor}$) and its contributions from a
 positive variation of one standard deviation of the
 error due to the cuts against photoproduction
 ($\delta_{cor}^{V^+}$), of the hadronic
 energy error ($\delta_{cor}^{h^+}$), of the error
 due to noise subtraction ($\delta_{cor}^{N^+}$)
 and of the error due to background subtraction
 ($\delta_{cor}^{B^+}$) are given.
 The normalisation uncertainty of $1.5\%$ is
 not included in the errors.
 The last column gives the correction for QED radiative 
 effects ($\Delta^{QED}_{CC}$).}
\end{minipage}}
 \end{tabular}}

 \begin{table}[htb]
 \begin{center}
 \footnotesize
 \rotatebox{90}{\ccxs}
 \end{center}
 \end{table}
 \begin{table}[htb]
 \begin{center}
 \tiny
 \begin{tabular}{|r|c||c|r|r||r|r|r||r|r|r|r|r|r|r|}
 \hline
 $Q^2$  &$x$ & $\tilde{\sigma}_{NC}$ &
 $\delta_{tot}$ & $\delta_{stat}$ & $\delta_{unc}$ &
 $\delta_{unc}^{E}$ &
 $\delta_{unc}^{h}$&
 $\delta_{cor}$ &
 $\delta_{cor}^{E^+}$ &
 $\delta_{cor}^{\theta^+}$&
 $\delta_{cor}^{h^+}$&
 $\delta_{cor}^{N^+}$&
 $\delta_{cor}^{B^+}$&
 $\delta_{cor}^{S^+}$ \\
 $(\rm GeV^2)$ & & &
 $(\%)$ &$(\%)$ &$(\%)$ &$(\%)$ &$(\%)$ &$(\%)$ &
 $(\%)$ &$(\%)$ &$(\%)$ &$(\%)$ &$(\%)$ &$(\%)$ 
 \\\hline\hline
$  100$&$0.00130$&$1.368$&$  7.0$&$  4.0$&$  5.5$&$  1.5$&$  0.2$&$  1.2$&$ -0.7$&$ -1.0$&$  0.5$&$  0.4$&$  -$&$ -0.9$\\
$  100$&$0.00200$&$1.293$&$  5.3$&$  3.4$&$  3.3$&$  0.5$&$  0.3$&$  2.4$&$  0.5$&$ -1.8$&$  0.9$&$  0.6$&$ -1.1$&$  -$\\ \hline
$  120$&$0.00160$&$1.342$&$  6.6$&$  4.2$&$  4.8$&$  0.1$&$  0.2$&$  1.7$&$ -0.3$&$ -1.5$&$  0.6$&$  0.3$&$  -$&$ -0.7$\\
$  120$&$0.00200$&$1.325$&$  5.0$&$  3.3$&$  3.2$&$  0.5$&$  0.2$&$  2.1$&$ -0.4$&$ -1.4$&$ -0.3$&$  0.3$&$ -1.4$&$  -$\\
$  120$&$0.00320$&$1.198$&$  4.8$&$  3.1$&$  3.3$&$  1.1$&$  0.4$&$  1.9$&$  0.8$&$ -1.7$&$  0.5$&$  0.5$&$ -0.3$&$  -$\\ \hline
$  150$&$0.00200$&$1.339$&$  6.7$&$  4.4$&$  4.8$&$  0.9$&$  0.2$&$  1.2$&$  0.8$&$ -0.8$&$  0.5$&$  0.4$&$  -$&$ -0.6$\\ \hline
$  200$&$0.00260$&$1.188$&$  7.1$&$  4.9$&$  4.8$&$  0.1$&$  0.4$&$  1.9$&$  0.4$&$ -1.7$&$  0.6$&$  0.4$&$  -$&$ -0.6$\\ \hline
$  250$&$0.00330$&$1.126$&$  7.9$&$  5.7$&$  4.9$&$  0.6$&$  0.2$&$  2.1$&$  1.0$&$ -2.0$&$  0.5$&$  0.3$&$  -$&$ -0.7$\\ \hline
$  300$&$0.00390$&$1.068$&$  8.0$&$  6.1$&$  4.9$&$  0.2$&$  0.2$&$  1.4$&$ -0.4$&$ -1.3$&$  0.4$&$  0.3$&$  -$&$ -0.7$\\ \hline
$  400$&$0.00530$&$1.101$&$  8.3$&$  6.4$&$  5.1$&$  0.7$&$  0.3$&$  1.5$&$ -0.6$&$ -1.4$&$  0.3$&$  0.3$&$  -$&$ -0.6$\\ \hline
$  500$&$0.00660$&$1.099$&$  8.7$&$  6.9$&$  5.2$&$  0.2$&$  0.3$&$  1.0$&$ -0.3$&$ -0.8$&$  0.5$&$  0.4$&$  -$&$ -0.2$\\ \hline
$  650$&$0.00850$&$1.056$&$  9.9$&$  8.2$&$  5.6$&$  1.0$&$  0.0$&$  0.5$&$ -1.0$&$ -0.2$&$ -0.2$&$  0.2$&$  -$&$ -0.4$\\ \hline
$  800$&$0.01050$&$0.938$&$ 10.9$&$  9.2$&$  5.9$&$  0.4$&$  0.3$&$  1.1$&$ -0.6$&$ -1.1$&$  0.3$&$  0.0$&$  -$&$ -0.1$\\ \hline
 \end{tabular}
 \end{center}
 \caption[RESULT]
 {\sl \label{ncextra} The NC $e^-p$
  reduced cross section
  $\tilde{\sigma}_{NC}(x,Q^2)$ from the high-$y$ analysis,
  shown with statistical ($\delta_{stat}$) and
  total ($\delta_{tot}$) errors.
  Also shown are the total uncorrelated systematic 
  ($\delta_{unc}$) error and two of its contributions: 
  the electron energy error ($\delta_{unc}^{E}$) 
  and the hadronic energy error ($\delta_{unc}^{h}$). 
  The effect of the other uncorrelated
  systematic errors is included in ($\delta_{unc}$). 
  In addition the correlated systematic error
  ($\delta_{cor}$) and its contributions from a
  positive variation of one 
  standard deviation of the
  electron energy error ($\delta_{cor}^{E^+}$), of
  the polar electron angle error
  ($\delta_{cor}^{\theta^+}$), of the hadronic
  energy error ($\delta_{cor}^{h^+}$), of the error
  due to noise subtraction ($\delta_{cor}^{N^+}$),
  of the error due to background subtraction
  ($\delta_{cor}^{B^+}$) and of the error
  due to charge symmetry background subtraction
  ($\delta_{cor}^{S^+}$) are given.
  The normalisation uncertainty of $1.8\%$ is
  not included in the errors. All $e^- p$ data not previously
  reported in~\cite{h1elec} are given, including the   
  new high $y$ data and three data points at $Q^2=100$ and 
  $120{\rm\,GeV}^2$ from the nominal analysis phase space.}
 \end{table}
 \begin{table}[htb]
 \begin{center}
 \begin{tabular}{|c|c|c|c|c|c|c|c|}
 \hline
 $Q^2$  &$x$ & $y$ & $\phi_{NC}$ & $F_L$ &
 $\delta_{stat}$ & $\delta_{sys}$ & $\delta_{tot}$ \\
 $(\rm GeV^2)$ & & & & & & & \\ \hline\hline
 \multicolumn{8}{|c|}{$e^-p$ data} \\
 \hline
$  110$&$ 0.00144$&$ 0.75$&$1.440$&$0.298$&$ 0.074$&$ 0.133$&$ 0.154$\\
$  175$&$ 0.00230$&$ 0.75$&$1.346$&$0.298$&$ 0.077$&$ 0.113$&$ 0.139$\\
$  280$&$ 0.00368$&$ 0.75$&$1.162$&$0.390$&$ 0.085$&$ 0.099$&$ 0.132$\\
$  450$&$ 0.00591$&$ 0.75$&$1.164$&$0.117$&$ 0.097$&$ 0.101$&$ 0.140$\\
$  700$&$ 0.00919$&$ 0.75$&$1.072$&$0.042$&$ 0.117$&$ 0.098$&$ 0.153$\\ \hline
 \multicolumn{8}{|c|}{$e^+p$ data}\\
 \hline
$  110$&$ 0.00144$&$ 0.75$&$1.518$&$0.198$&$ 0.032$&$ 0.083$&$ 0.092$\\
$  175$&$ 0.00230$&$ 0.75$&$1.426$&$0.164$&$ 0.038$&$ 0.064$&$ 0.076$\\
$  280$&$ 0.00368$&$ 0.75$&$1.292$&$0.171$&$ 0.041$&$ 0.057$&$ 0.072$\\
$  450$&$ 0.00591$&$ 0.75$&$1.163$&$0.133$&$ 0.045$&$ 0.052$&$ 0.070$\\
$  700$&$ 0.00919$&$ 0.75$&$1.037$&$0.096$&$ 0.053$&$ 0.062$&$ 0.082$\\ \hline
 \end{tabular}
 \end{center}
 \caption[RESULT]
 {\sl \label{fltab} The NC structure function term $\phi_{NC}(x,Q^2)$
 and the structure function $F_L$, shown
 with its statistical ($\delta_{stat}$),
 systematic ($\delta_{sys}$) and total 
 ($\delta_{tot}$) absolute error. 
 The total error includes a contribution
 arising from the model uncertainties in the calculated $\tilde{F_2}$. 
 These are obtained by varying
 the assumptions of the H1 Low $y$ QCD fit as listed in table~\ref{tabmodunc}.
 The luminosity uncertainties of
 the $e^+p$ and $e^-p$ data sets
 are included in the systematic error.}
 \end{table}
 \begin{table}[htb]
 \begin{center}
 \begin{tabular}{|r|c|r|c|c|c|}
 \hline
\multicolumn{1}{|c|}{$Q^2$}  & $x$   & \multicolumn{1}{|c|}{\raisebox{-0.6mm}{$x\tilde{F}_3$}} & $\delta_{stat}$ & $\delta_{sys}$ &
   $\delta_{tot}$ \\
 $(\rm GeV^2)$  &  &           &         &         &         \\ \hline\hline
 $ 1500$ & $0.020$ & $ 0.052$  & $0.036$ & $0.025$ & $0.044$ \\
 $ 1500$ & $0.032$ & $ 0.074$  & $0.032$ & $0.026$ & $0.042$ \\
 $ 1500$ & $0.050$ & $ 0.076$  & $0.039$ & $0.028$ & $0.048$ \\
 $ 1500$ & $0.080$ & $ 0.067$  & $0.050$ & $0.035$ & $0.061$ \\\hline
 $ 5000$ & $0.050$ & $ 0.088$  & $0.037$ & $0.024$ & $0.044$ \\
 $ 5000$ & $0.080$ & $ 0.150$  & $0.031$ & $0.020$ & $0.037$ \\
 $ 5000$ & $0.130$ & $ 0.160$  & $0.036$ & $0.023$ & $0.043$ \\
 $ 5000$ & $0.180$ & $ 0.099$  & $0.041$ & $0.025$ & $0.048$ \\
 $ 5000$ & $0.250$ & $ 0.089$  & $0.049$ & $0.039$ & $0.062$ \\
 $ 5000$ & $0.400$ & $ 0.027$  & $0.045$ & $0.034$ & $0.057$ \\
 $ 5000$ & $0.650$ & $-0.008$  & $0.015$ & $0.011$ & $0.019$ \\\hline
 $12000$ & $0.180$ & $ 0.149$  & $0.077$ & $0.021$ & $0.080$ \\
 $12000$ & $0.250$ & $ 0.113$  & $0.053$ & $0.017$ & $0.056$ \\
 $12000$ & $0.400$ & $ 0.035$  & $0.038$ & $0.019$ & $0.043$ \\
 $12000$ & $0.650$ & $-0.007$  & $0.015$ & $0.011$ & $0.018$ \\\hline\hline
 \multicolumn{1}{|c|}{$Q^2$}  & $x$   & \multicolumn{1}{|c|}{\raisebox{-0.6mm}{$xF_3^{\gamma Z}$}} & $\delta_{stat}$ & $\delta_{sys}$ &
   $\delta_{tot}$ \\
 $(\rm GeV^2)$  &  &           &         &         &         \\ \hline
 $ 1500$ & $0.026$ & $ 0.59$  & $0.22$ & $0.17$ & $0.28$ \\
 $ 1500$ & $0.050$ & $ 0.38$  & $0.13$ & $0.09$ & $0.16$ \\
 $ 1500$ & $0.080$ & $ 0.57$  & $0.12$ & $0.08$ & $0.14$ \\
 $ 1500$ & $0.130$ & $ 0.61$  & $0.14$ & $0.09$ & $0.16$ \\
 $ 1500$ & $0.180$ & $ 0.37$  & $0.12$ & $0.06$ & $0.13$ \\
 $ 1500$ & $0.250$ & $ 0.29$  & $0.11$ & $0.05$ & $0.12$ \\
 $ 1500$ & $0.400$ & $ 0.09$  & $0.08$ & $0.05$ & $0.09$ \\
 $ 1500$ & $0.650$ & $-0.02$  & $0.03$ & $0.02$ & $0.04$ \\\hline
 \end{tabular}
 \end{center}
 \caption[RESULT]
 {\sl \label{tab:xf3} The upper part of the table shows the generalised 
 structure function $x\tilde{F}_3$ with statistical ($\delta_{stat}$),
 systematic ($\delta_{sys}$) and total
 ($\delta_{tot}$) absolute errors. The luminosity uncertainties of the
 $e^+p$ and $e^-p$ data are included in the systematic error. 
 The lower part of the table shows the structure
 function $xF_3^{\gamma Z}$ obtained by averaging over different $Q^2$ values
 and transforming to a $Q^2$ value at $1\,500{\rm\,GeV}^2$.}
 \end{table}

\end{document}